\documentclass[11pt,english,onecolumn,draftcls]{IEEEtran}
\usepackage[T1]{fontenc}
\usepackage[latin9]{inputenc}
\usepackage{xcolor}
\usepackage{pdfcolmk}
\usepackage{float}
\usepackage{amsmath}
\usepackage{amsthm}
\usepackage{amssymb}
\usepackage{graphicx}
\PassOptionsToPackage{normalem}{ulem}
\usepackage{ulem}

\makeatletter

\providecommand{\tabularnewline}{\\}
\floatstyle{ruled}
\newfloat{algorithm}{tbp}{loa}
\providecommand{\algorithmname}{Algorithm}
\floatname{algorithm}{\protect\algorithmname}
\providecolor{lyxadded}{rgb}{0,0,1}
\providecolor{lyxdeleted}{rgb}{1,0,0}
  \theoremstyle{plain}
  \newtheorem{thm}{\protect\theoremname}
  \theoremstyle{plain}
  \newtheorem{lem}{\protect\lemmaname}
  \theoremstyle{remark}
  \newtheorem{rem}{\protect\remarkname}

\usepackage{cite}
\usepackage{times}
\usepackage{subfig}

\makeatother

\usepackage{babel}
\providecommand{\lemmaname}{Lemma}
\providecommand{\remarkname}{Remark}
\providecommand{\theoremname}{Theorem}

\begin{document}

\title{Cache-induced Hierarchical Cooperation in Wireless Device-to-Device
Caching Networks}

\author{{\normalsize{}An Liu,$^{1}$ }\textit{\normalsize{}Member IEEE}{\normalsize{},
Vincent Lau,$^{1}$}\textit{\normalsize{} Fellow IEEE}{\normalsize{}
and Giuseppe Caire,$^{2}$ }\textit{\normalsize{}Fellow IEEE}{\normalsize{}\\$^{1}$Department
of Electronic and Computer Engineering, Hong Kong University of Science
and Technology\\$^{2}$Department of Telecommunication Systems, Technical
University of Berlin}}
\maketitle
\begin{abstract}
We consider a wireless device-to-device (D2D) caching network where
$n$ nodes are placed on a regular grid of area $A\left(n\right)$.
Each node caches $L_{C}F$ (coded) bits from a library of size $LF$
bits, where $L$ is the number of files and $F$ is the size of each
file. Each node requests a file from the library independently according
to a popularity distribution. Under a commonly used ``physical model''
and Zipf popularity distribution, we characterize the optimal per-node
capacity scaling law for \textit{extended networks} (i.e., $A\left(n\right)=n$).
Moreover, we propose a \textit{cache-induced} \textit{hierarchical
cooperation} scheme and associated cache content placement optimization
algorithm to achieve the optimal per-node capacity scaling law. When
the path loss exponent $\alpha<3$, the optimal per-node capacity
scaling law achieved by the cache-induced hierarchical cooperation
can be significantly better than that achieved by the existing state-of-the-art
schemes. To the best of our knowledge, this is the first work that
completely characterizes the per-node capacity scaling law for wireless
caching networks under the physical model and Zipf distribution with
an arbitrary skewness parameter $\tau$. While scaling law analysis
yields clean results, it may not accurately reflect the throughput
performance of a large network with a finite number of nodes. Therefore,
we also analyze the throughput of the proposed cache-induced hierarchical
cooperation for networks of practical size. The analysis and simulations
verify that cache-induced hierarchical cooperation can also achieve
a large throughput gain over the cache-assisted multihop scheme for
networks of practical size.
\end{abstract}


\section{Introduction}

An increase of 1000x in wireless data traffic is expected in the near
future. More than 50\% of this will be generated by high-definition
video and content delivery applications. Many recent works have shown
that wireless caching is one of the most promising solutions to handle
the high traffic load caused by content delivery applications \cite{Caire_INFOCOM12_femtocache,Niesen_TIT12_FLcaching,Liu_TWC15arxiv_adhoccaching}.
By exploiting the fact that content is \textquotedblleft cachable\textquotedblright ,
wireless nodes can cache some popular content during off-peak hours
(\textit{cache initialization phase}), in order to reduce the traffic
rate at peak hours (\textit{content delivery phase}). Early works
focused on caching at the network side, such as at base stations (BSs).
Recently, however, caching at the user/device side has also gained
increasing interest, due to the number of wireless devices increasing
faster than the number of BSs, and wireless device storage being arguably
the cheapest and most rapidly growing network resource. It has been
shown in \cite{Caire_arxiv13_D2Dcaching,Caire_arxiv13_d2dcachingtradeoff,Altieri_arxiv14_d2dcaching,Jeon_ICC15_D2Dcaching}
that combining wireless device caching with short-range device-to-device
(D2D) communications can significantly improve the throughput of wireless
networks. Although many efficient wireless caching schemes have been
proposed, the fundamental limit of such \textit{wireless D2D caching
networks} and the associated optimal caching scheme remains an open
problem. In this paper, we will provide a partial solution to this
open problem.

\subsection{Related Work}

\subsubsection{Capacity Scaling Law in Wireless Ad Hoc Networks\label{subsec:Capacity-Scaling-Law-adhoc}}

It is extremely hard to characterize the exact capacity of general
wireless networks. For large wireless networks, scaling laws provide
a useful way to characterize the behavior of the capacity order. The
capacity scaling law of wireless ad hoc networks was first studied
by Gupta and Kumar in the seminal paper \cite{GuptaKumar}, where
they showed that in a large wireless ad hoc networks with $n$ randomly
located nodes, the aggregate throughput of the classical multihop
scheme scales at most as $\Theta\left(\sqrt{n}\right)$ under a \textit{protocol
model}. Since then, a number of works \cite{Xie_liangliang_IT04_network_information_capacity,Jovicic_TIT2004_TCahoc,Kumar_TIT06_TCadhoc}
have studied the information theoretic capacity scaling law under
a more realistic \textit{physical model} that includes distance-dependent
propagation path-loss, fading, Gaussian noise, and signal interference.
In this case, the capacity scaling law depends on whether the network
is \textquotedblleft extended\textquotedblright{} (constant node density,
with the network area growing as $\Theta\left(n\right)$), or \textquotedblleft dense\textquotedblright{}
(constant network area, with the node density growing as $\Theta\left(n\right)$).
Specifically, it was shown in \cite{Tse_IT07_CapscalingHMIMO} that
under a physical model with path loss exponent $\alpha\geq2$, the
total network capacity of a \textit{dense network} scales as $\Theta\left(n\right)$
and that of an \textit{extended network} scales as $\Theta\left(n^{2-\frac{\min\left(3,\alpha\right)}{2}}\right)$,
both of which are orders better than the $\Theta\left(\sqrt{n}\right)$
scaling law achieved by the classical multihop scheme. Moreover, this
capacity scaling is achieved by \textit{hierarchical cooperation},
with the number of hierarchical stages going to infinity. In \cite{Niesen_TIT09_CSadhoc},
the authors studied the capacity scaling in ad hoc networks with arbitrary
node placement, and the capacity regions of ad hoc networks with the
more complicated unicast or multicast traffic model was studied in
\cite{Niesen_TIT10_CSadhoc}.

Note that the results in \cite{Tse_IT07_CapscalingHMIMO,Niesen_TIT09_CSadhoc,Niesen_TIT10_CSadhoc}
depended heavily on the physical channel model, which assumes independent
fading coefficients between different nodes. In contrast, the authors
in \cite{Franceschetti_TIT09_Maxwellscaling} showed that the capacity
of a wireless network with area $\mathcal{A}$ is fundamentally limited
by $\Theta\left(\frac{\sqrt{\mathcal{A}}}{\lambda}\right)$ using
Maxwell\textquoteright s equations, where $\lambda$ is the carrier
frequency. The results in \cite{Franceschetti_TIT09_Maxwellscaling}
imply that for practical dense networks, the assumption of independent
fading coefficients may only be valid when $n\leq\frac{\sqrt{\mathcal{A}}}{\lambda}=\Theta\left(\frac{1}{\lambda}\right)$.
Since $\Theta\left(\frac{1}{\lambda}\right)$ is usually not large
enough to be considered as an asymptotic regime, the scaling law for
dense networks is less interesting in practice, as pointed out in
\cite{Caire_TIT2015_HMIMOimp}. For extended networks, $\mathcal{A}$
scales linearly with $n$, and thus the scaling law analysis is more
relevant in practice. Therefore, in this paper, we will only study
scaling laws for extended networks. For clarity, we will assume rich
scattering and focus on the case with independent fading coefficients.
However, we will also discuss the extension of the scaling law results
to the case when the assumption of independent fading coefficients
is invalid and $\frac{\sqrt{\mathcal{A}}}{\lambda}$ becomes the limiting
factor of the capacity.

\subsubsection{Capacity Scaling Law in Caching Networks}

\cite{Gitzenis_TIT13_wirelesscache} studied the joint optimization
of cache content replication and routing in a \textit{regular network}
and identified the throughput scaling laws for various regimes. Single-hop
device-to-device (D2D) caching networks, where the content delivery
scheme is restricted to single-hop transmission, were considered in
\cite{Caire_arxiv13_D2Dcaching,Caire_arxiv13_d2dcachingtradeoff}.
Under a Zipf popularity distribution \cite{Yamakami_PDCAT06_Zipflaw}
with skewness parameter less than one, and the protocol model, it
was shown in \cite{Caire_arxiv13_D2Dcaching,Caire_arxiv13_d2dcachingtradeoff}
that the per-node capacity scales as $\Theta\left(L_{C}/L\right)$,
where $L_{C}$ is the number of files that each node can cache (cache
capacity in the unit of file size) and $L$ is the total number of
files in the content library. Multi-hop D2D caching networks with
the protocol model were considered in \cite{Jeon_ICC15_D2Dcaching}.
By allowing multihop transmission, the per-node capacity scales as
$\Theta\left(\sqrt{L_{C}/L}\right)$ when the popularity distribution
has the \textquotedblleft heavy tail\textquotedblright{} property,
which is much better than the single-hop case. 

In \cite{Niesen_TIT12_FLcaching}, the authors studied a different
caching network topology, where a single transmitter serves $n$ user
nodes through a common noiseless link of fixed capacity (bottleneck
link). Coded caching schemes were proposed for this scenario to create
coded multicast gain. Specifically, in the cache initialization phase,
each file is partitioned into packets and each node stores subsets
of packets from each file. In the content delivery phase, the BS can
compute a multicast network-coded message (transmitted via the common
link) such that each node can decode its own requested file from the
multicast message and its cached file packets (side information).
Under the worst-case arbitrary demands model, the per-node throughput
scaling is again given by $\Theta\left(L_{C}/L\right)$, which is
the same scaling law as achieved by single-hop D2D caching networks.
A number of extensions under different user demands and network structures
can be found in \cite{Caire_GlobalSIP2014_codedcaching,Hachem_ISIT14_Codedcaching,Niesen_TIT2016_codedcaching}.

\subsubsection{Physical Layer (PHY) Caching}

A key feature of wireless networks is that interference can be handled
at the physical layer (PHY) beyond the simple exclusion principle
built into the previously mentioned protocol model. In particular,
caching can also be exploited to mitigate interference and enable
cooperative transmission at the PHY. For example, in cellular networks,
when the user requested data exist in the BS cache (cache hits), they
induce \emph{dynamic side information} to the BSs, which can be further
exploited to enhance the capacity of the radio interface. The concept
of \textit{cache-induced opportunistic MIMO cooperation}, or \textit{PHY
caching}, was first introduced in \cite{Liu_TSP14_CacheRelay,Liu_TSP13_CacheIFN}
to achieve significant spectral efficiency gain without consuming
BS backhaul. Since then, there have been many works on PHY caching,
and they can be classified into two major classes, as discussed below. 

\textbf{High-SNR and fixed-size network regimes:} These works focus
on the degrees of freedom (DoF), i.e., the coefficient of the $O(\log\textrm{SNR})$
leading term of the network sum capacity as SNR grows, but the network
has a fixed number of nodes. For example, \cite{Wei_TSP15_cacheDoFrelay,Wei_Globecom15_DoFcachedIFC}
studied the average sum DoF (averaged over the user demands) for relay
and interference channels with BS caching, respectively, under some
achievable scheme. On the other hand, \cite{naderializadeh2016fundamental}
studied the max-min sum DoF (i.e., maximizing the worst-case sum DoF
over the user demands) of one-hop interference networks with caching
at both the transmitters and receivers, and the impact of caching
on the DoF of a Gaussian vector broadcast channel with delayed channel
state information at the transmitter (CSIT) was also investigated
in \cite{Zhang_2016fundamentalCSITFeedback_arxiv}. 

\textbf{Large network and fixed SNR regimes:} These works focus on
studying the capacity/throughput scaling laws as the number of nodes
grows, but with fixed SNR. In \cite{Liu_ToN16_capscalingCaching},
PHY caching was used to exploit both the \textit{cache-induced MIMO
cooperation gain} and the \textit{cache-assisted multihopping gain}
(i.e., reducing the number of hops from the source to the destination)
in backhaul-limited multi-hop wireless networks, and the throughput
scaling laws achievable by PHY caching were identified for extended
networks under Zipf popularity distributions (see also \cite{Liu_TWC15arxiv_adhoccaching}).
It was shown in \cite{Liu_TWC15arxiv_adhoccaching,Liu_ToN16_capscalingCaching}
that by exploiting the cache-induced MIMO cooperation, PHY caching
can achieve significant throughput gain over conventional caching,
which purely exploits cache-assisted multihopping gain. However, exploiting
the cache-induced MIMO cooperation does not provide order gain in
terms of throughput scaling laws.

\subsection{Contributions}

\begin{table}
\begin{centering}
{\footnotesize{}}%
\begin{tabular}{|l|l|c|c|}
\hline 
 &  & {\footnotesize{}Protocol model} & {\footnotesize{}Physical model extended network}\tabularnewline
\hline 
{\footnotesize{}Without cache} & {\footnotesize{}Capacity scaling} & {\footnotesize{}$\Theta\left(n^{-\frac{1}{2}}\right)$ \cite{GuptaKumar}} & {\footnotesize{}$\Theta\left(n^{1-\frac{\min\left(3,\alpha\right)}{2}}\right)$
\cite{Tse_IT07_CapscalingHMIMO}}\tabularnewline
\cline{2-4} 
 & {\footnotesize{}Achievable scheme} & {\footnotesize{}Multihop} & {\footnotesize{}H-Coop. \cite{Tse_IT07_CapscalingHMIMO}}\tabularnewline
\hline 
{\footnotesize{}With cache} & {\footnotesize{}Capacity scaling} & {\footnotesize{}$\Theta\left(\sqrt{L_{C}/L}\right)$ \cite{Jeon_ICC15_D2Dcaching}} & {\footnotesize{}{*}$\Theta\left(\left(L_{C}/L\right)^{\frac{\min\left(3,\alpha\right)}{2}-1}\right)$}\tabularnewline
\cline{2-4} 
{\footnotesize{}(heavy tail } & {\footnotesize{}Achievable scheme} & {\footnotesize{}Random caching} & {\footnotesize{}{*}Cache-induced H-Coop.}\tabularnewline
{\footnotesize{}popularity)} &  & {\footnotesize{}+ Multihop \cite{Jeon_ICC15_D2Dcaching}} & \tabularnewline
\hline 
{\footnotesize{}With cache} & {\footnotesize{}Capacity scaling} & {\footnotesize{}Unknown} & {\footnotesize{}{*}Known for Zipf}\tabularnewline
\cline{2-4} 
{\footnotesize{}(general popularity)} & {\footnotesize{}Achievable scheme} & {\footnotesize{}Unknown} & {\footnotesize{}{*}Known for Zipf}\tabularnewline
\hline 
\end{tabular}
\par\end{centering}{\footnotesize \par}
\caption{\label{tab:contri}{\small{}Summary of the }\textit{\small{}per node
capacity scaling laws}{\small{} for wireless D2D networks with and
without caching, where ``H-Coop.'' stands for hierarchical cooperation.
The contribution of this paper is highlighted with a star symbol.}}
\end{table}
As discussed above, capacity scaling laws have been obtained under
the protocol model for one-hop and multihop wireless caching networks.
For multihop wireless caching networks under the physical model, \textit{achievable}
scaling laws have been obtained and cache-induced MIMO cooperation
has been shown to provide gains in terms of throughput (but not in
terms of scaling laws). However, the question of the capacity scaling
laws of wireless caching networks under the physical model has been
unanswered so far. In this paper, we provide an answer to this open
question under the assumption of Zipf popularity distribution. Table
\ref{tab:contri} summarizes the existing capacity scaling law results
for wireless D2D networks with and without caching, as well as the
scaling law results from our work (highlighted in blue).

In this paper, we address the fundamental capacity scaling in extended
wireless D2D caching networks under the physical model, and propose
an associated order-optimal caching and content delivery scheme. With
respect to the previous work on cache-induced MIMO cooperation, we
shall design a more advanced cooperation scheme that can achieve an
order gain in the throughput scaling law. As explained in Section
\ref{subsec:Capacity-Scaling-Law-adhoc}, the capacity scaling law
is less interesting for dense networks, and thus we will only focus
on extended networks for the scaling law analysis. While scaling law
analysis yields clean results, it cannot accurately reflect how a
large network with a finite number of nodes really performs in terms
of throughput. For example, as shown in the simulations in \cite{Caire_TIT2015_HMIMOimp},
the original hierarchical cooperation scheme in \cite{Tse_IT07_CapscalingHMIMO}
performs even worse than the multihop scheme for networks of practical
size. Therefore, in this paper, we will also analyze the throughput
of the proposed caching and content delivery scheme to verify its
performance gain for such networks. The main contributions of the
paper are summarized below.
\begin{itemize}
\item \textbf{Cache-induced hierarchical cooperation:} In this paper, we
combine the ideas of PHY caching (cache-induced MIMO cooperation)
and hierarchical cooperation, and propose a novel caching and content
delivery scheme called cache-induced hierarchical cooperation, which
can achieve both a higher scaling law in extended networks and huge
throughput gain in networks of practical size.
\item \textbf{Cache content placement optimization:} We propose a low complexity
cache content placement algorithm to optimize the parameters of the
cache-induced hierarchical cooperation scheme, and establish the order
optimality of the proposed algorithm.
\item \textbf{Throughput analysis:} We analyze the throughput performance
of the proposed cache-induced hierarchical cooperation, and show that
it can achieve significant throughput gain over conventional caching,
which purely exploits cache-assisted multihopping gain.
\item \textbf{Capacity scaling laws in extended wireless D2D caching networks:}
For the extended network model under a Zipf popularity distribution,
we derive both the achievable throughput scaling laws of the proposed
cache-induced hierarchical cooperation and an information-theoretic
upper bound of the throughput scaling law. The scaling laws of the
achievability and converse coincide, so that we can establish the
capacity scaling law for the Zipf popularity distribution with the
general skewness parameter $\tau$. For the case of a \textquotedblleft heavy
tail\textquotedblright{} Zipf popularity distribution (i.e., the skewness
parameter $\tau\leq1$), the per node capacity scales as $\Theta\left(\left(L_{C}/L\right)^{\frac{\min\left(3,\alpha\right)}{2}-1}\right)$.
When $\alpha<3$ and $L_{C}/L\ll1$, this per node capacity scaling
law is much better than the $\Theta\left(\left(L_{C}/L\right)^{1/2}\right)$
per node capacity scaling law of the cache-assisted multihop scheme
under both the protocol model \cite{Jeon_ICC15_D2Dcaching} and physical
model \cite{Liu_TWC15arxiv_adhoccaching}. 
\end{itemize}

\subsection{Paper Organization}

In Section \ref{sec:System-Model}, we introduce the architecture
of wireless D2D caching networks and the channel model. In Section
\ref{sec:Exist-Achievable-Schemes}, we discuss some preliminary results
on the improved hierarchical cooperation scheme in \cite{Caire_TIT2015_HMIMOimp}
and the classical multihop scheme, which are designed for wireless
ad-hoc/D2D networks without caching. In Section \ref{sec:Order-wise-Optimal-Control}
and \ref{sec:Cache-Content-Placement}, we describe the proposed cache-induced
hierarchical cooperation scheme and the associated cache content placement
optimization algorithm, respectively. The throughput performance of
the proposed scheme is analyzed and compared in Section \ref{sec:Achievable-Throughput-and}.
The achievable scaling law of the cache-induced hierarchical cooperation
and the converse proof are given in Section \ref{sec:Scaling-Laws-in}
for extended networks. The conclusion is given in Section \ref{sec:Conclusion}.

\section{System Model\label{sec:System-Model}}

\subsection{Wireless Device-to-Device Caching Networks\label{subsec:Wireless-Device-to-Device-Cachin}}

Consider a wireless D2D caching network with $n$ nodes placed on
a regular grid of area $A\left(n\right)$. For clarity, we focus on
networks with $n=4^{M}$ nodes, where $M$ is some positive integer,
and let $V\left(n\right)$ denote the set of all nodes in the network.
The results can be easily generalized to the case when $M$ is not
an integer without affecting the first-order performance.

In a wireless D2D caching network, the nodes request data (e.g., music
or video) from a content library $\mathcal{L}=\left\{ W_{1},W_{2},...,W_{L}\right\} $
of $L=\left|\mathcal{L}\right|$ files (information messages), where
$W_{l}$ are drawn at random and independently with a uniform distribution
over a message set $\mathbb{F}_{2}^{F}$ (binary strings of length
$F$). Each node has a cache of size $FL_{C}$ bits, which can be
used to store a portion of the content files to serve the requests
generated by the nodes in the network. We assume that $L_{C}<L$ to
avoid the trivial case when every node has enough cache capacity to
store the whole content library $\mathcal{L}$. Furthermore, we assume
$nL_{C}>L$ so that there is at least one complete copy of each content
file in the caches of the entire network. 

There are two phases during the operation of a wireless D2D caching
network, namely the\textit{ cache initialization phase} and the \textit{content
delivery phase}. 

In the cache initialization phase, each node caches a portion of the
(possibly encoded) content files. In general, the \textit{caching
scheme} is defined as a collection of $n$ mappings $\mathcal{B}_{i}:\mathbb{F}_{2}^{FL}\rightarrow\mathbb{F}_{2}^{FL_{C}},i=1,...,n$
from the content library $\mathcal{L}$ to the content $B_{i}=\mathcal{B}_{i}\left(\mathcal{L}\right)$
cached at node $i$. Since the popularity of content files change
very slowly (e.g., new movies are usually posted on a weekly or monthly
timescale), the cache update overhead in the cache initialization
phase is usually small. This is a reasonable assumption widely used
in the literature \cite{Gitzenis_TIT13_wirelesscache,Caire_arxiv13_D2Dcaching,Caire_arxiv13_d2dcachingtradeoff,Niesen_TIT12_FLcaching,Liu_TWC15arxiv_adhoccaching,Liu_ToN16_capscalingCaching}.

In the content delivery phase, time is divided into time slots and
each node independently requests the $l$-th content file with probability
$p_{l}$, where probability mass function $\mathbf{p}=\left[p_{1},...,p_{L}\right]$
represents the popularity of the content files. Without loss of generality,
we assume $p_{1}\geq p_{2}\cdots\geq p_{L}$. Each node requests files
one after another. When a requested file is delivered to node $i$,
node $i$ will request the next file immediately according to the
popularity distribution $\mathbf{p}$. 

Let $l_{i}\left(t\right)$ denote the content file requested by node
$i$ at time slot $t$, and let $\boldsymbol{l}\left(t\right)=\left[l_{1}\left(t\right),...,l_{n}\left(t\right)\right]^{T}$
denote the user request vector (URV). Let $t_{i}^{j}$ denote the
time slot when $l_{i}\left(t\right)$ changes for the $j$-th time.
In other words, node $i$ starts to request file $l_{i}\left(t_{i}^{j}\right)$
at time slot $t_{i}^{j}$, and the delivery of file $l_{i}\left(t_{i}^{j}\right)$
to node $i$ is finished at time slot $t_{i}^{j+1}-1$. If the content
$\mathcal{B}_{i}\left(\mathcal{L}\right)$ cached at node $i$ is
sufficient to decode the requested content file $l_{i}\left(t_{i}^{j}\right)$,
node $i$ can obtain the requested file $l_{i}\left(t_{i}^{j}\right)$
immediately. Otherwise, node $i$ has to obtain more information about
the content file $l_{i}\left(t_{i}^{j}\right)$ from the other nodes
in the network. Specifically, at time slot $t\in\left[t_{i}^{j},...,t_{i}^{j+1}-1\right]$,
each node $i^{'}\neq i$ generates an \textit{information message
$U_{i,i^{'}}\left(t\right)=\mathcal{U}_{i,i^{'}}\left(B_{i^{'}},t\right)$
}for node $i$ using a \textit{content delivery} \textit{encoder}
$\mathcal{U}_{i,i^{'}}\left(\cdot,t\right):\mathbb{F}_{2}^{FL_{C}}\rightarrow\mathbb{F}_{2}^{\left|U_{i,i^{'}}\left(t\right)\right|}$\footnote{Note that $U_{i,i^{'}}\left(t\right)$ can be empty, i.e., node $i^{'}$
does not generate any information message for node $i$ at time slot
$t$.}. Let $U_{i}^{j}=\cup_{i^{'}\neq i}\cup_{t\in\left[t_{i}^{j},...,t_{i}^{j+1}-1\right]}U_{i,i^{'}}\left(t\right)$
denote the aggregate information message for the $j$-th request of
node $i$. The content delivery scheme treats the aggregate information
messages of different users as independent messages and delivers each
aggregate information message to the desired node. To be more specific,
the content delivery scheme ensures that node $i$ can successfully
receive the aggregate information message $U_{i}^{j}$ within the
time window $\left[t_{i}^{j},...,t_{i}^{j+1}-1\right]$ for any $i$
and $j$. Note that $t_{i}^{j}$ is a random variable depending on
the specific content delivery scheme, the random URV process $\boldsymbol{l}\left(t\right)$,
and other underlying random processes in the network, such as the
fading channel and noise. When node $i$ obtains $U_{i}^{j}$ at time
$t_{i}^{j+1}-1$, node $i$ will apply a decoding function $\hat{W}_{l_{i}}=\phi_{i}^{j}\left(U_{i}^{j},B_{i}\right)$
to obtain the estimated file $\hat{W}_{l_{i}}$, where $\phi_{i}^{j}:\mathbb{F}_{2}^{\left|U_{i}^{j}\right|}\times\mathbb{F}_{2}^{FL_{C}}\rightarrow\mathbb{F}_{2}^{F}$.
A content delivery scheme is \textit{feasible} if
\[
\lim_{F\rightarrow\infty}\Pr\left[\phi_{i}^{j}\left(U_{i}^{j},B_{i}\right)\neq W_{l_{i}}\right]=0,\forall i,j.
\]
Fano's inequality implies that a necessary condition for a content
delivery scheme to be feasible is
\begin{equation}
H\left(W_{l_{i}}|U_{i}^{j},B_{i}\right)\leq\varepsilon_{F}F,\label{eq:decodcons}
\end{equation}
where $\varepsilon_{F}$ is a vanishing quantity as $F\rightarrow\infty$.

Similar to \cite{Gitzenis_TIT13_wirelesscache,Liu_TWC15arxiv_adhoccaching},
we assume a symmetric traffic model where all users have the same
\textit{average throughput requirement} $R$ (averaged over all possible
realizations of user requests). To be more specific, the average data
rate of node $i$ is defined as $R_{i}=\frac{F}{\overline{T}_{i}}$,
where 
\[
\overline{T}_{i}=\lim_{J\rightarrow\infty}\frac{1}{J}\sum_{j=1}^{J}\mathbb{E}\left[t_{i}^{j+1}-t_{i}^{j}\right]=\lim_{J\rightarrow\infty}\frac{1}{J}\mathbb{E}\left[t_{i}^{J+1}-t_{i}^{1}\right]
\]
is the average delivery time of one file to node $i$. A \textit{symmetric
per node throughput} $R$ is achievable if there exists a feasible
caching and content delivery scheme with $F\rightarrow\infty$, such
that $R_{i}\geq R,\forall i$. 

\subsection{Wireless Channel Model}

We use a similar channel model to that in \cite{Tse_IT07_CapscalingHMIMO,Niesen_TIT10_CSadhoc}.
The channel coefficient between a transmitter node $j$ and a receiver
node $i$ is
\[
h_{i,j}=\left(r_{i,j}\right)^{-\alpha/2}\exp\left(\sqrt{-1}\theta_{i,j}\right),
\]
where $r_{i,j}$ is the distance between node $j$ and $i$, $\theta_{i,j}$
is the random phase with uniform distribution on $(0,2\pi]$, and
$\alpha>2$ is the path loss exponent. At each node, the received
signal is also corrupted by a circularly symmetric Gaussian noise
with zero mean and unit variance. 

\section{Preliminaries On Hierarchical Cooperation and Classical Multihop
Schemes\label{sec:Exist-Achievable-Schemes}}

In the proposed cache-induced hierarchical cooperation scheme in Section
\ref{sec:Order-wise-Optimal-Control}, there are two physical layer
(PHY) transmission modes, namely, the \textit{hierarchical cooperation}
mode and \textit{multihop }mode. These two PHY transmission modes
are based on the hierarchical cooperation scheme in \cite{Caire_TIT2015_HMIMOimp}
and the classical multihop scheme, respectively. The original hierarchical
cooperation and multihop schemes are designed for wireless ad-hoc/D2D
networks without caching. In this case, the per node throughput $R$
depends on the \textit{traffic pattern}, i.e., the number of source-destination
pairs and the locations of each source-destination pair. In \cite{Caire_TIT2015_HMIMOimp},
the throughput performance of the hierarchical cooperation and multihop
schemes are analyzed and compared for a wireless D2D network with
$n$ nodes under \textit{uniform permutation traffic}, where the network
consists of $n$ source-destination pairs with the same throughput
requirement $R$, such that each node is both a source and a destination,
and pairs are selected at random over the set of $n$-permutations
$\pi$ that do not fix any elements (i.e., for which $\pi\left(i\right)\neq i$
for all $i=1,...,n$). In this section, we review some preliminary
results from \cite{Caire_TIT2015_HMIMOimp}, which will be useful
in the later sections.

\subsection{Throughput Performance of Hierarchical Cooperation under Uniform
Permutation Traffic}

We first describe the hierarchical cooperation scheme for wireless
D2D networks with $A\left(n\right)=1$. The hierarchical cooperation
is based on a three-phase cooperative transmission scheme. The basic
hierarchical cooperation scheme was first proposed in \cite{Tse_IT07_CapscalingHMIMO}.
In such a scheme, the network is first divided into $n/N$ clusters
of $N$ nodes each and then the following three phases are used to
achieve cooperation gain.
\begin{itemize}
\item \textbf{Phase 1 (Information Dissemination):} Each source distributes
$N$ distinct sub-packets of its message to the $N$ neighboring nodes
in the same cluster. One transmission is active per each cluster,
in a round robin fashion, and clusters are active simultaneously in
order to achieve some spatial spectrum reuse. The inter-cluster interference
is controlled by the reuse factor $T_{r}$, i.e., each cluster has
one transmission opportunity every $T_{r}^{2}$ time slots. 
\item \textbf{Phase 2 (Long-Range MIMO Transmission):} One cluster at a
time is active, and when a cluster is active it operates as a single
$N$-antenna MIMO transmitter, sending $N$ independently encoded
data streams to a destination cluster. Each node in the cooperative
receiving cluster stores its own received signal.
\item \textbf{Phase 3 (Cooperative Reception):} All receivers in each cluster
share their own received and quantized signals so that each destination
in the cluster decodes its intended message on the basis of the (quantized)
$N$-dimensional observation. Each destination performs joint typical
decoding to obtain its own desired message based on the quantized
signals.
\end{itemize}

The basic hierarchical cooperation scheme employs the above three-phase
cooperative transmission scheme as a recursive building block applied
for local communication of a higher stage, i.e., at a larger space
scale in the network. This scheme was improved in \cite{Ozgur_TIT2010_HMIMOimp,Caire_TIT2015_HMIMOimp}.
Specifically, \cite{Ozgur_TIT2010_HMIMOimp} proposed an improvement
where the local communication phase is formulated as a network multiple
access problem instead of being decomposed into a number of unicast
network problems. \cite{Caire_TIT2015_HMIMOimp} further improved
the throughput performance by using more efficient TDMA scheduling.
In this paper, we will use the hierarchical cooperation ``method
4'' from \cite{Caire_TIT2015_HMIMOimp}, with both improvements,
as a building block for the proposed cache-induced hierarchical cooperation.
For convenience, we will call ``method 4'' from \cite{Caire_TIT2015_HMIMOimp}
the \textit{improved hierarchical cooperation} scheme, and its throughput
performance is summarized in the following theorem.
\begin{thm}
[Per node throughput of hierarchical cooperation]Consider a wireless
D2D network of size $n$ and area $A\left(n\right)=1$ under uniform
permutation traffic. Suppose each node has a sufficiently large transmit
power with a uniform bound $P_{max}$ that does not scale with $n$.
The improved hierarchical cooperation scheme with $s$ stages achieves
a per node throughput of 
\[
R_{H}^{(s)}\left(n,P_{I}\right)=\begin{cases}
\log\left(1+\frac{\textrm{SNR}}{1+P_{I}}\right)\frac{n^{-\frac{1}{2}}}{2\sqrt{2}T_{r}} & s=1\\
R_{c}\left(\alpha,P_{I}\right)\frac{n^{\frac{-1}{s+1}}}{\left(1+s\right)T_{r}^{\frac{2s}{s+1}}\left(3\cdot2^{s-1}\right)^{\frac{s}{2\left(s+1\right)}}}, & s\geq2
\end{cases},
\]
where
\begin{align*}
\textrm{SNR} & =2^{2\left(3+\alpha/\ln2\right)}\\
T_{r} & =\left\lceil \sqrt{\textrm{SNR}}^{1/\alpha}+1\right\rceil \\
P_{I} & =\sum_{i=1}^{\sqrt{n}}8i\textrm{SNR}\left(T_{r}i-1\right)^{-\alpha}.
\end{align*}
and $R_{c}\left(\alpha,P_{I}\right)$ is determined in Section III-A
in \cite{Caire_TIT2015_HMIMOimp}.
\end{thm}

Note that $R_{c}\left(\alpha,P_{I}\right)\leq\log\left(1+\frac{\textrm{SNR}}{1+P_{I}}\right)$,
and please refer to Section III-A in \cite{Caire_TIT2015_HMIMOimp}
for the details about how to determine the exact value of $R_{c}\left(\alpha,P_{I}\right)$.
Let $s_{n}^{\star}=\textrm{argmax}_{s}R_{H}^{(s)}\left(n,P_{I}\right)$
denote the optimal number of stages. There is no closed form for $s_{n}^{\star}$.
However, $s_{n}^{\star}$ can be easily found by a simple one dimensional
search. Furthermore, it is shown in \cite{Shen_TIT2009_HMIMO} that
$s_{n}^{\star}=\Theta\left(\sqrt{\ln n}\right)$.

Now we use the method in \cite{Tse_IT07_CapscalingHMIMO} to extend
the above hierarchical cooperation scheme from $A\left(n\right)=1$
to an arbitrary $A\left(n\right)\geq\Theta\left(1\right)$. Compared
to networks with $A\left(n\right)=1$, the distance between nodes
in networks with an arbitrary $A\left(n\right)$ is scaled by a factor
of $\sqrt{A\left(n\right)}$, and hence for the same transmit powers,
the received powers are all scaled by a factor of $A\left(n\right)^{-\alpha/2}$.
The hierarchical scheme for fixed peak power per node ($O\left(1\right)$
power per node) yields an \textit{average} power per node of $O\left(1/n\right)$
since nodes are active only a fraction of $O\left(1/n\right)$ the
time \cite{Tse_IT07_CapscalingHMIMO}. For a network with arbitrary
$A\left(n\right)$, we need to scale the peak power up by a factor
$A\left(n\right)^{\alpha/2}$ in order to compensate for the path
loss. Imposing an average power per node $O\left(1\right)$, this
yields that we can operate the network under the hierarchical cooperation
scheme for a fraction of time $\min\left(nA\left(n\right)^{-\alpha/2},1\right)$.
In this way, the hierarchical cooperation scheme for arbitrary $A\left(n\right)$
can achieve a per node throughput of 
\[
\widetilde{R}_{H}^{(s)}\left(n,P_{I}\right)=R_{H}^{(s)}\left(n,P_{I}\right)\min\left(nA\left(n\right)^{-\alpha/2},1\right).
\]

\subsection{Throughput Performance of Multihop Scheme under Uniform Permutation
Traffic}

The multihop scheme is a classical communication architecture that
has been widely used in practice. In this scheme, for a given source-destination
pair, a routing path is first formed from the source to the destination.
Then, on each routing path, packets are relayed from node to node.
On each link of the routing path, each packet is fully decoded using
conventional single-user decoding with all interference treated as
noise.

In \cite{Caire_TIT2015_HMIMOimp}, the performance of the multihop
scheme is compared with that of the hierarchical cooperation scheme
for dense wireless D2D networks, under the following assumptions.
The routing between each source-destination pair is to first proceed
horizontally and then vertically in the network grid. Distance-dependent
power control is applied and the interference is controlled by the
reuse factor $T_{r}$, chosen to enforce the optimality condition
of treating interference as noise (TIN) as $T_{r}=\left\lceil \sqrt{\textrm{SNR}}^{1/\alpha}+1\right\rceil $.
Under these assumptions, the per node throughput for uniform permutation
traffic is given by
\begin{align*}
R_{M}\left(n,P_{I}\right) & =\log\left(1+\frac{\textrm{SNR}}{1+P_{I}}\right)\frac{n^{-\frac{1}{2}}}{\left\lceil \sqrt{\textrm{SNR}}^{1/\alpha}+1\right\rceil ^{2}},\\
\textrm{SNR} & =2^{2\left(3+\alpha/\ln4\right)}.
\end{align*}

\subsection{Extension to Per Cluster Uniform Permutation Traffic\label{subsec:Extension-to-PerClusterUT}}

\begin{figure}
\begin{centering}
\textsf{\includegraphics[clip,width=70mm]{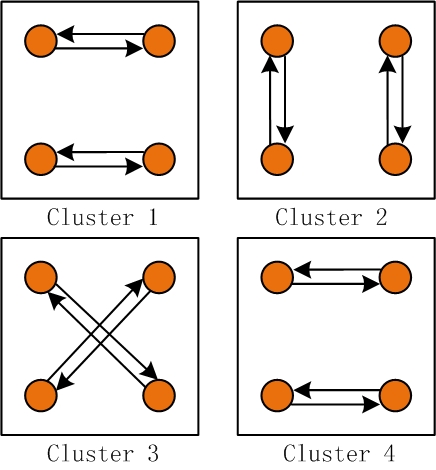}}
\par\end{centering}
\caption{\label{fig:ClusterUtraffic}{\small{}An illustration of per cluster
uniform permutation traffic over clusters of size $N=4$ in a network
of size $n=16$.}}
\end{figure}
In the proposed cache-induced hierarchical cooperation scheme, the
traffic induced by the requests of all nodes will be grouped into
\textit{per cluster uniform permutation traffic} over clusters (sub-networks)
of different sizes. The above hierarchical cooperation scheme or multihop
scheme can be used to handle per cluster uniform permutation traffic
over the $n/N$ non-overlapping clusters with the same cluster size
$N$, where there is uniform permutation traffic within each cluster
but there is no traffic among clusters, as illustrated in Fig. \ref{fig:ClusterUtraffic}.
Specifically, each cluster of size $N$ simultaneously employs the
hierarchical cooperation scheme or multihop scheme to serve the uniform
permutation traffic within the cluster. Note that there is no need
to apply TDMA among clusters to control the inter-cluster interference
because the TDMA scheme within each cluster with reuse factor $T_{r}$
already guarantees that the received power of the interference is
upper bounded by $P_{I}=\sum_{i=1}^{\sqrt{n}}8i\textrm{SNR}\left(T_{r}i-1\right)^{-\alpha}$.
A similar idea is also used in \cite{Caire_TIT2015_HMIMOimp} to improve
the TDMA scheduling for hierarchical cooperation. The choice of hierarchical
cooperation or multihop scheme depends on which will achieve higher
throughput. If $\widetilde{R}_{H}^{(s_{N}^{\star})}\left(N,P_{I}\right)>R_{M}\left(N,P_{I}\right)$,
we will use the hierarchical cooperation scheme; otherwise, we will
use the multihop scheme. In this case, the achievable per node throughput
is given by 
\begin{equation}
R_{u}\left(N\right)=\max\left(\widetilde{R}_{H}^{(s_{N}^{\star})}\left(N,P_{I}\right),R_{M}\left(N,P_{I}\right)\right).\label{eq:Ru}
\end{equation}

\section{Cache-induced Hierarchical Cooperation\label{sec:Order-wise-Optimal-Control}}

In this section, we elaborate the proposed achievable scheme, called
cache-induced hierarchical cooperation, which works for both dense
and extended wireless D2D caching networks.

\subsection{Key Components of the Cache-induced Hierarchical Cooperation Scheme}

\begin{figure}
\begin{centering}
\textsf{\includegraphics[clip,width=90mm]{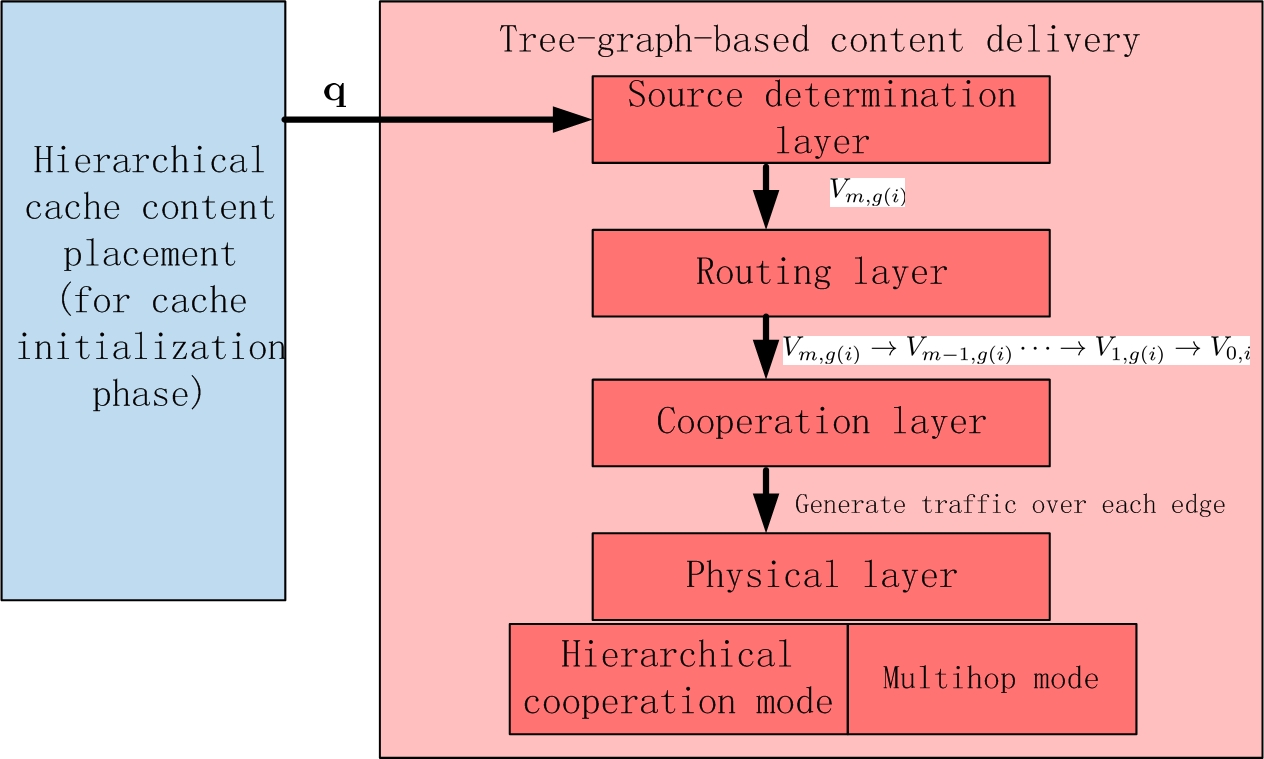}}
\par\end{centering}
\caption{\label{fig:components}{\small{}Components of the cache-induced hierarchical
cooperation and their inter-relationship.}}
\end{figure}
The components of the proposed cache-induced hierarchical cooperation
scheme and their inter-relationship are illustrated in Fig. \ref{fig:components}.
There are two major components: the \textit{hierarchical cache content
placement}, working in the cache initiation phase. and the\textit{
tree-graph-based content delivery}, working in the content delivery
phase. The hierarchical cache content placement decides how to distribute
the content files into caches of different nodes (or mathematically
decides the $n$ mappings $\mathcal{B}_{i},\forall i$). The tree-graph-based
content delivery exploits the cached content at each node to serve
the user requests, and it consists of four layers: the source determination
layer, routing layer, cooperation layer and physical layer. In this
content delivery scheme, the original network is abstracted as a tree
graph and the source determination and routing are based on this graph.
Specifically, each node is a leaf node in the tree graph and the set
of source nodes for a leaf node is an internal node in the tree. The
routing layer routes messages between the source nodes and destination
node. The cooperation layer provides this tree abstraction to the
routing layer by appropriately concentrating traffic over the network.
Finally, the PHY implements this concentration of messages in the
wireless network based on two PHY transmission modes, namely, \textit{hierarchical
cooperation} mode and multihop\textit{ }mode. The details of the components
are elaborated in the following subsections.

\subsection{Hierarchical Cache Content Placement\label{subsec:MDS-Cache-Encoding}}

\begin{figure}
\begin{centering}
\textsf{\includegraphics[clip,width=85mm]{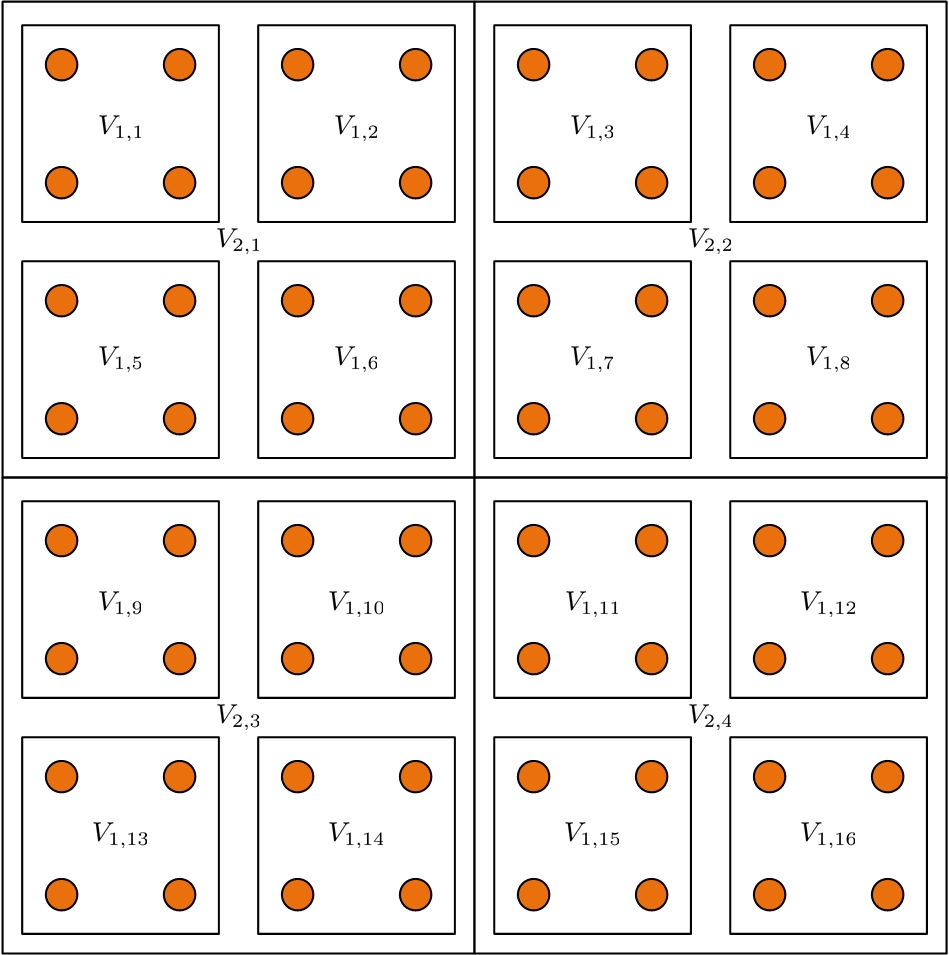}}
\par\end{centering}
\caption{\label{fig:levelsclusters}{\small{}Illustration of clusters at different
levels for a network with $n=64$ nodes.}}
\end{figure}
In the proposed hierarchical cache content placement, nodes are partitioned
into clusters of different levels. In the $m$-th level, $A\left(n\right)$
is partitioned into $4^{M-m}$ squares of equal size, as illustrated
in Fig. \ref{fig:levelsclusters}. Then the $4^{m}$ nodes in the
same square form a cluster in the $m$-th level. Let $V_{m,i}\subseteq V\left(n\right)$
be the $i$-th cluster in the $m$-th level for $i\in\left\{ 1,...,4^{M-m}\right\} $.
In the hierarchical cache content placement, all nodes cache the same
number of $q_{l}F$ bits for the $l$-th file. Moreover, $q_{l}$
can only take values from the discrete set $\left\{ 0,\frac{1}{4^{M}},\frac{1}{4^{M-1}},...,\frac{1}{4},1\right\} $.
If $q_{l}=\frac{1}{4^{m}}$, during the cache initiation phase, the
$l$-th file will be equally distributed over the nodes in $V_{m,i}$
for any $i\in\left\{ 1,...,4^{M-m}\right\} $. In other words, each
node in $V_{m,i}$ caches a portion of the $4^{-m}F$ bits for file
$l$ such that the $l$-th file can be reconstructed by collecting
all portions from the caches of nodes in $V_{m,i}$. For convenience,
we say that file $l$ is cached at the $m$-th level if $q_{l}=\frac{1}{4^{m}}$.
Such hierarchical cache content placement with parameter $q_{l}\in\left\{ 0,\frac{1}{4^{M}},\frac{1}{4^{M-1}},...,\frac{1}{4},1\right\} $
can also be specified by another set of parameters $\mathbf{x}=\left[x_{0},x_{1},...,x_{M}\right]\in\mathbb{Z}_{+}^{M+1}$,
where 
\[
x_{m}=\sum_{l=1}^{L}1\left(q_{l}=4^{-m}\right)
\]
is the number of files cached at the $m$-th level, and $1\left(\cdot\right)$
is the indication function. Note that $\mathbf{x}$ must satisfy the
constraint $\sum_{m=0}^{M}x_{m}=L$ so that there is at least one
complete copy of each content file in the caches of the entire network.
Moreover, $\mathbf{x}$ must also satisfy the cache size constraint
$\sum_{m=0}^{M}x_{m}4^{-m}\leq L_{C}$. Clearly, if file $l$ is more
popular than file $l^{'}$ (i.e., $p_{l}\geq p_{l^{'}}$), file $l$
should be replicated more frequently than file $l^{'}$. Therefore,
without loss of optimality, we let $q_{1}\geq q_{2}\cdots\geq q_{L}$.
In other words, the more popular files are cached at lower levels
and the less popular files are cached at higher levels.

\subsection{Capacitated Graph for Content Delivery}

\begin{figure}
\begin{centering}
\textsf{\includegraphics[clip,width=100mm]{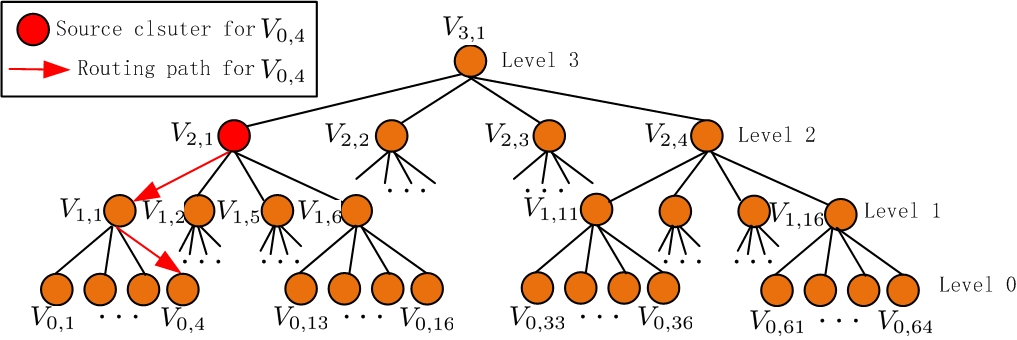}}
\par\end{centering}
\caption{\label{fig:capacitatedgraph}{\small{}Illustration of the capacitated
graph $\mathcal{G}$ for the network in Fig. \ref{fig:levelsclusters}
with $n=64$ nodes. Suppose node $V_{0,4}$ requests a file cached
at the first level. Then the set of source nodes is $V_{2,1}$ (red
node) and the routing path from the source cluster $V_{2,1}$ to the
destination $V_{0,4}$ is illustrated with red arrows.}}
\end{figure}

For a given hierarchical cache content placement with parameter $\mathbf{x}$,
the content delivery scheme is based on a capacitated graph $\mathcal{G}$,
which is similar to the communication schemes considered in \cite{Niesen_TIT10_CSadhoc,Niesen_IT12_caching}.
Specifically, the D2D networks is represented by a tree graph $\mathcal{G}$
whose leaf nodes are the nodes in $V\left(n\right)$ and whose internal
nodes are node clusters. There are $M+1$ levels in the tree graph
$\mathcal{G}$, where the lowest level is called the $0$-th level,
the next lowest level is called the first level, and the highest level
is called the $M$-th level. With a slight abuse of notation, let
$V_{0,i}\subseteq V\left(n\right)$ also denote the $i$-th leaf node
at the $0$-th level of $\mathcal{G}$, which represents the $i$-th
node in the network. For $m\in\left\{ 1,...,M\right\} $, let $V_{m,i}$
also denote the $i$-th internal node at the $m$-th level of $\mathcal{G}$.
There are only edges between the nodes in adjacent levels. For $m\in\left\{ 2,...,M\right\} $,
there is an edge between an internal node $V_{m-1,j}$ and an internal
node $V_{m,i}$ if $V_{m-1,j}\subseteq V_{m,i}$. Similarly, there
is an edge between a leaf node $V_{0,j}$ and an internal node $V_{1,i}$
at the first level if $V_{0,j}\subseteq V_{1,i}$. An example of the
capacitated graph $\mathcal{G}$ is given in Fig. \ref{fig:capacitatedgraph}.

\subsubsection{Source Determination Layer}

Let $V_{m,g\left(i\right)}$ denote the internal node at the $m$-th
level in $\mathcal{G}$ that contains the leaf node $V_{0,i}$ (i.e.,
$V_{0,i}\subseteq V_{m,g\left(i\right)}$). Then for a leaf node $V_{0,i}$
requesting the $l$-th content cached at the $m$-th level (i.e.,
$q_{l}=4^{-m}$), the set of source nodes is given by $V_{m,g\left(i\right)}$,
which is the cluster at the $m$-th level that contains $V_{0,i}$,
as illustrated in Fig. \ref{fig:capacitatedgraph} for {\small{}$V_{0,4}$}.
Since the cache of each node in $V_{m,g\left(i\right)}$ stores a
different portion of the $4^{-m}F$ bits of the $l$-th file, the
leaf node $V_{0,i}$ can reconstruct a complete copy of the $l$-th
file by collecting all the portions (\textit{subfiles}) from the nodes
in $V_{m,g\left(i\right)}.$

\subsubsection{Routing Layer}

When a leaf node $V_{0,i}$ requests the $l$-th content cached at
the $m$-th level, the requested content is sent from the source set
$V_{m,g\left(i\right)}$ to the destination $V_{0,i}$ via the path
$V_{m,g\left(i\right)}\rightarrow V_{m-1,g\left(i\right)}\cdots\rightarrow V_{1,g\left(i\right)}\rightarrow V_{0,i}$
in the capacitated graph $\mathcal{G}$, as illustrated in Fig. \ref{fig:capacitatedgraph}
for $V_{0,4}$. This corresponds to the concentration of the content
to smaller and smaller clusters until it finally concentrates to the
single leaf node $V_{0,i}$ that requested the content. 

\subsubsection{Cooperation Layer}

\begin{figure}
\begin{centering}
\textsf{\includegraphics[clip,width=150mm]{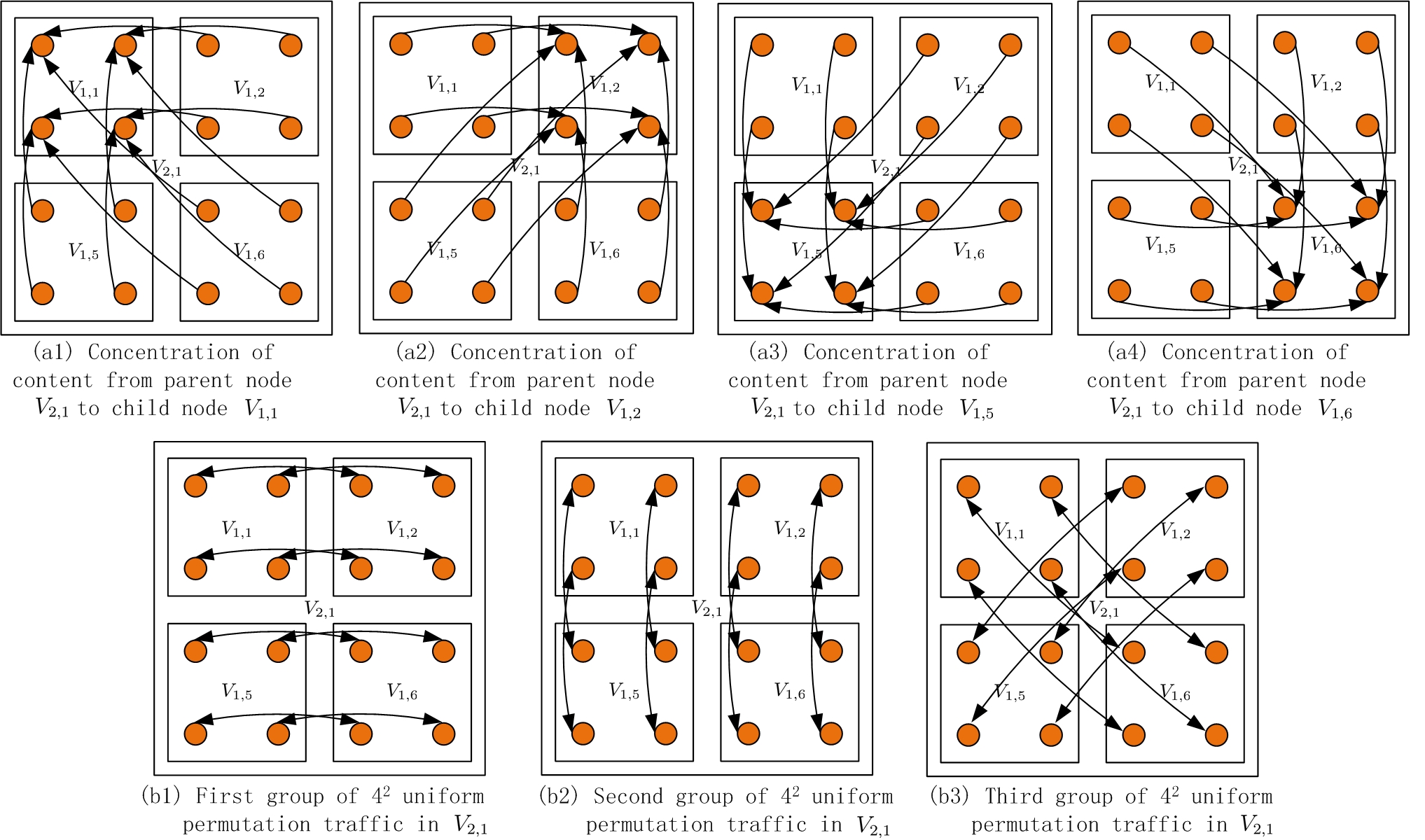}}
\par\end{centering}
\caption{\label{fig:trafficpattern}{\small{}Subfigures (a1)-(a4) illustrate
the concentration of content from parent node $V_{2,1}$ to child
nodes $V_{1,1},V_{1,2},V_{1,5}$ and $V_{1,6}$, respectively, for
the network in Fig. \ref{fig:levelsclusters}. The PHY partitions
all the $3\times4^{2}=48$ subfile transmissions induced by the concentration
of content from parent node $V_{2,1}$ to its child nodes into three
groups of uniform permutation traffic. At each time, the PHY schedules
one group for transmission using either the hierarchical cooperation
or multihop scheme, as illustrated in Subfigures (b1)-(b3). }}
\end{figure}

To send information along an edge from a parent node to a child node,
the routing layer calls upon the cooperation layer. Specifically,
suppose the routing layer calls the cooperation layer to send a message
from a parent node $V_{m,i}$ to a child node $V_{m-1,j}$. Assume
each node in $V_{m,i}$ has access to a distinct $4^{-m}$ fraction
of the message to be sent. Then each node in $V_{m,i}\backslash V_{m-1,j}$
sends its part of the message to a node in $V_{m-1,j}$ such that
after the transmission, each node in $V_{m-1,j}$ will have access
to a distinct $4^{-(m-1)}$ fraction of the message, as illustrated
in Fig. \ref{fig:trafficpattern} for $m=2$.

\subsubsection{Physical Layer}

The PHY groups the traffic induced by the cooperation layer into per
cluster uniform permutation traffic within different clusters at different
levels so that we can use the existing hierarchical cooperation scheme
or multihop scheme described in Section \ref{sec:Exist-Achievable-Schemes}
as building blocks to handle the traffic induced by the cooperation
layer. Specifically, the choice of PHY transmission mode for level
$m$ with cluster size $4^{m}$ depends on which PHY mode achieves
a higher throughput, as described in Section \ref{subsec:Extension-to-PerClusterUT}.

To achieve this, the PHY needs to properly partition the available
resources between different levels and different clusters, and schedule
the transmissions. Specifically, the resource partitioning and scheduling
at different levels/clusters are elaborated below.

The PHY time shares between the transmissions of $M_{b}$ active levels,
where $M_{b}=\max_{m}\:\textrm{s.t. }x_{m}>0$. (Note that all levels
higher than $M_{b}$ are inactive since no content files are cached
at these levels.) Note that $M_{b}\geq1$ since $L_{C}<L$. For simplicity,
in our achievability strategy, we choose to serve the levels in a
round robin manner with the same fraction of time per level. This
turns out to be sufficient in terms of scaling laws. 

Within the $m$-th level for $m>0$, there is no communication between
clusters at the $m$-th level and the communications within each cluster
of the $m$-th level occur simultaneously to achieve spatial reuse
gain. This is exactly the per cluster uniform permutation traffic
with cluster size $N=4^{m}$ described in Section \ref{subsec:Extension-to-PerClusterUT}.
Therefore, we can use the hierarchical cooperation or multihop scheme
described in Section \ref{subsec:Extension-to-PerClusterUT} to handle
the traffic at the $m$-th level.

Within the $i$-th cluster at the $m$-th level $V_{m,i}$ for $m>0$,
there are a total number of $3\times4^{m}$ subfile transmissions
that need to be scheduled since each node in $V_{m,i}$ needs to collect
a portion of the message from the other three nodes in $V_{m,i}$,
as illustrated in Fig. \ref{fig:trafficpattern}. At each time, the
PHY can schedule $4^{m}$ subfile transmissions into a group of uniform
permutation traffic for the nodes in $V_{m,i}$. Therefore, the PHY
needs to further time share between the transmissions of the three
groups of uniform permutation traffic, as illustrated in Fig. \ref{fig:trafficpattern}.

As a result, the per traffic rate at the $m$-th level for $m>0$
can be calculated as in the following lemma.
\begin{lem}
[Per traffic rate at different levels]For any cluster $V_{m,i},i\in\left\{ 1,...,4^{M-m}\right\} $
at the $m$-th level, a per node rate of $R_{m}$ is achievable from
one node in $V_{m,i}$ to another node in $V_{m,i}$, where 
\[
R_{m}=\frac{R_{u}\left(4^{m}\right)}{3M_{b}},m=1,...,M_{b}.
\]
$R_{u}\left(N\right)$ is the per node rate for a regular grid network
with $n$ nodes under per cluster uniform permutation traffic with
cluster size $N$, as given in (\ref{eq:Ru}).
\end{lem}

Since there are a total number of $4^{m}$ effective transmissions
from a parent node $\nu_{m,i}$ to a child node $\nu_{m-1,j}$, the
edge between $\nu_{m,i}$ and $\nu_{m-1,j}$ can provide an achievable
throughput of $C_{m}=4^{m}R_{m}$.

\section{Cache Content Placement Optimization\label{sec:Cache-Content-Placement}}

In this section, we aim at finding the optimal cache content placement
parameter $\mathbf{x}$ to maximize the per node throughput $R$.
We first derive the per node throughput $R$ for given cache content
placement parameter $\mathbf{x}$ and formulate the cache content
placement optimization problem. Then we propose a low-complexity cache
content placement algorithm. 

\subsection{Problem Formulation}

We first analyze the total average traffic rate over an edge $e_{m,i,j}$
between a parent node $\nu_{m,i}$ to a child node $\nu_{m-1,j}$
at the $m$-th level, when the per node throughput requirement is
$R$. Whenever a user in $V_{m-1,i}$ requests a file that is cached
at the $m^{'}$-th level with $m^{'}>m-1$, it will induce a traffic
rate of $R$ on the edge $e_{m,i,j}$. Under the hierarchical cache
content placement scheme, files with indices $\left\{ \sum_{i=0}^{m-1}x_{i}+1,\sum_{i=0}^{m-1}x_{i}+2,...,L\right\} $
are cached at levels higher than the $\left(m-1\right)$-th level.
Therefore, for given cache content placement parameter $\mathbf{x}$
and per node rate requirement $R$, the total average traffic rate
over the edge $e_{m,i,j}$ is $4^{m-1}\sum_{l=\sum_{i=0}^{m-1}x_{i}+1}^{L}p_{l}R$.
Clearly, a per node throughput $R$ is achievable if and only if the
induced total average traffic rate does not exceed the capacity of
the edges at all levels. Hence, for given cache content placement
parameter $\mathbf{x}$, the maximum achievable per node throughput
is $\max R,\:\textrm{s.t. }\sum_{l=\sum_{i=0}^{m-1}x_{i}+1}^{L}p_{l}R\leq C_{m}4^{-\left(m-1\right)}$.
Consequently, the cache content placement optimization problem to
maximize the per node throughput can be formulated as
\begin{eqnarray}
\max_{\mathbf{x}\in\mathbb{Z}_{+}^{m},R} & R\label{eq:cachePmain}\\
\textrm{s.t. } & \sum_{l=\sum_{i=0}^{m-1}x_{i}+1}^{L}p_{l}R\leq C_{m}4^{-\left(m-1\right)}, & m=1,...,M\label{eq:rateconmain}\\
 & \sum_{m=0}^{M}x_{m}4^{-m}\leq L_{C},\label{eq:cachesizeconmain}\\
 & \sum_{m=0}^{M}x_{m}=L,\label{eq:cacheallLmain}
\end{eqnarray}
where the second constraint is the cache size constraint. Note that
for convenience, we have extended the definition of $C_{m}$ from
$m\in\left\{ 1,...,M_{b}\right\} $ to $m\in\left\{ 1,...,M\right\} $,
where $C_{m}=\frac{4^{m}R_{u}\left(4^{m}\right)}{3M_{b}},\forall m\in\left\{ M_{b}+1,...,M\right\} $.
Since $\sum_{l=\sum_{i=0}^{m-1}x_{i}+1}^{L}p_{l}R=0$ for $m>M_{b}$,
Constraint (\ref{eq:rateconmain}) is equivalent to $\sum_{l=\sum_{i=0}^{m-1}x_{i}+1}^{L}p_{l}R\leq C_{m}4^{-\left(m-1\right)},m=1,...,M_{b}$,
which is the link capacity constraint for the $M_{b}$ active levels.

When $nL_{C}<L$, the condition $\sum_{m=0}^{M}x_{m}=L$ can never
be satisfied. In this case, Problem (\ref{eq:cachePmain}) is infeasible,
which indicates that no cache content placement scheme can guarantee
a non-zero rate for all users since the entire network cannot cache
a complete copy for every file. Since we have assumed that $nL_{C}\geq L$
to avoid such a degenerate case, Problem (\ref{eq:cachePmain}) is
always feasible.

The cache content placement optimization problem in (\ref{eq:cachePmain})
is an integer optimization problem, and an explicit (closed-form)
solution amenable to the order-optimality analysis of the resulting
throughput scaling law seems difficult to obtain. In the next section,
we propose a low-complexity algorithm which can find an order-optimal
solution for (\ref{eq:cachePmain}).

\subsection{Low-Complexity Cache Content Placement Algorithm}

The capacity $C_{m}$ of the edge $e_{m,i,j}$ depends on the number
of active levels $M_{b}$, which is a complicated function of $x_{m}$.
Clearly, $C_{m}$ can be bounded as $\frac{1}{M}\overline{C}_{m}4^{\left(m-1\right)}\leq C_{m}\leq\overline{C}_{m}4^{\left(m-1\right)}$,
where
\begin{equation}
\overline{C}_{m}=\frac{4R_{u}\left(4^{m}\right)}{3},m=1,...,M.\label{eq:Cmbar}
\end{equation}
Define $f\left(x\right)=\left(\left\lceil x\right\rceil -x\right)p_{\left\lfloor x\right\rfloor }+\sum_{l=\left\lceil x\right\rceil }^{L}p_{l},x\in\left[1,L+1\right]$.
Then we have $f\left(\sum_{i=0}^{m-1}x_{i}+1\right)=\sum_{l=\sum_{i=0}^{m-1}x_{i}+1}^{L}p_{l}$
for $\sum_{i=0}^{m-1}x_{i}+1\in\mathbb{Z}_{++}$. Consider the following
simplified cache content placement optimization problem:
\begin{eqnarray}
\max_{\mathbf{x}\in\mathbb{R}_{+}^{m},R} & R\label{eq:cachePmain-1}\\
\textrm{s.t. } & f\left(\sum_{i=0}^{m-1}x_{i}+1\right)R\leq\overline{C}_{m}, & m=1,...,M\nonumber \\
 & \sum_{m=0}^{M}x_{m}4^{-m}\leq L_{C}, & \sum_{m=0}^{M}x_{m}=L,\nonumber 
\end{eqnarray}
where we have replaced $C_{m}$ with its upper bound $\overline{C}_{m}4^{\left(m-1\right)}$
and relaxed the integer optimization variables $\mathbf{x}$ to real
variables. Note that the optimal solution of (\ref{eq:cachePmain-1})
would be the same if we were to replace $C_{m}$ with its lower bound
$\overline{C}_{m}4^{\left(m-1\right)}/M$, although the optimal objective
value would be scaled by $1/M$. 

Based on the above analysis, the low-complexity cache content placement
algorithm first solves the optimal solution of the simplified problem
in (\ref{eq:cachePmain-1}), and then (approximately) projects the
solution to the feasible set of the original problem in (\ref{eq:cachePmain}).
In general, the function $f\left(x\right)$ depends on the content
popularity distributions $p_{1},...,p_{L}$ and may not be convex.
Therefore, Problem (\ref{eq:cachePmain-1}) may not be convex. In
the following theorem, we prove that the optimal solution of Problem
(\ref{eq:cachePmain-1}) must satisfy certain sufficient and necessary
optimality conditions, from which a low complexity algorithm can be
derived.
\begin{thm}
[Optimality Condition of (\ref{eq:cachePmain-1})]\label{thm:Optimality-Condition-of}$\left(\mathbf{x}^{*},R^{*}\right)$
is the optimal solution of Problem (\ref{eq:cachePmain-1}) if and
only if
\begin{eqnarray}
f\left(\sum_{i=m^{*}}^{m-1}x_{i}^{*}+1\right)R^{*} & = & \overline{C}_{m},m=m^{*}+1,...,M,\label{eq:Requ}\\
R^{*} & \leq & \overline{C}_{m^{*}},\nonumber \\
\sum_{m=0}^{M}x_{m}^{*} & = & L,\label{eq:sumL}\\
\sum_{m=0}^{M}x_{m}^{*}4^{-m} & = & L_{C},\label{eq:cachesize}
\end{eqnarray}
where $m^{*}=\min m\:\textrm{s.t. }x_{m}^{*}>0$, and $\overline{C}_{0}=+\infty$
when $m^{*}=0$. 
\end{thm}

Please refer to Appendix \ref{subsec:Proof-of-TheoremOptCond} for
the proof.

From the optimality condition (\ref{eq:Requ}) in Theorem \ref{thm:Optimality-Condition-of},
the optimal cache content placement is to balance the traffic loading
of the active levels. In Theorem \ref{thm:Optimality-Condition-of},
$m^{*}$ is the lowest level at which a file can be cached and it
depends on the cache size. The larger the cache size, the smaller
$m^{*}$ is. For example, when $L_{C}=L/n$, which is the minimum
cache size to make the problem feasible, we have $m^{*}=M$, i.e.,
we can only cache all files at the highest level $M$. On the other
hand, when $L_{C}=L$, we have $m^{*}=0$; i.e., the cache size is
enough to cache all files at the lowest level. As $L_{C}$ increases
from $L/n$ to $L$, $m^{*}$ decreases from $M$ to $0$. Motivated
by this observation, we propose a bisection algorithm to find $m^{*}$
and the optimal solution $\mathbf{x}^{*}$. 

Specifically, for a given $m^{*}$, the solution of (\ref{eq:Requ})
and (\ref{eq:sumL}) is

\begin{eqnarray}
x_{m^{*}}^{*} & = & f^{-1}\left(\frac{\overline{C}_{m^{*}+1}}{R}\right)-1,\nonumber \\
x_{m}^{*} & = & f^{-1}\left(\frac{\overline{C}_{m+1}}{R}\right)-f^{-1}\left(\frac{\overline{C}_{m}}{R}\right),m=m^{*}+1,...,M-1\nonumber \\
x_{M}^{*} & = & L-f^{-1}\left(\frac{\overline{C}_{M}}{R}\right)+1.\label{eq:solRequ}
\end{eqnarray}
Substituting (\ref{eq:solRequ}) into (\ref{eq:cachesize}), we have
\begin{equation}
L_{m^{*}}\left(R\right)\triangleq\sum_{m=m^{*}+1}^{M}\frac{3f^{-1}\left(\frac{\overline{C}_{m}}{R}\right)}{4^{m}}+\frac{L+1}{4^{M}}-\frac{1}{4^{m^{*}}}=L_{C}.\label{eq:solR}
\end{equation}
If we can find a solution $R^{*}$ of $L_{m^{*}}\left(R\right)=L_{C}$
for $R\in\left(\overline{C}_{m^{*}+1},\overline{C}_{m^{*}}\right]$,
then $m^{*}$ and $R^{*}$ satisfy all the optimality conditions in
Theorem \ref{thm:Optimality-Condition-of}.

For a given $m^{*}$, if $L_{m^{*}}\left(\overline{C}_{m^{*}+1}\right)\geq L_{C}$,
it implies the cache size is insufficient to cache files at level
$m^{*}$ and thus $m^{*}$ should be increased. If $L_{m^{*}}\left(\overline{C}_{m^{*}}\right)<L_{C}$,
it implies the cache size is still sufficient to cache files at the
lower level and thus $m^{*}$ should be decreased. If $L_{m^{*}}\left(\overline{C}_{m^{*}+1}\right)<L_{C}$
and $L_{m^{*}}\left(\overline{C}_{m^{*}}\right)\geq L_{C}$, (\ref{eq:solR})
must have a unique solution $R^{*}$ for $R\in\left(\overline{C}_{m^{*}+1},\overline{C}_{m^{*}}\right]$
because $L_{m^{*}}\left(R\right)$ is a strictly increasing function
of $R$. In this case, $\left(\mathbf{x}^{*},R^{*}\right)$ is the
optimal solution of Problem (\ref{eq:cachePmain-1}) according to
Theorem \ref{thm:Optimality-Condition-of}. 

Based on the above analysis, the overall cache content placement algorithm
is summarized in Algorithm \ref{alg:cachebisec}. In Algorithm \ref{alg:cachebisec},
Step 1 is the bisection algorithm to find $m^{*}$ and the optimal
solution $\mathbf{x}^{*}$ of Problem (\ref{eq:cachePmain-1}). After
step 1, we can determine the cache allocated to each level, e.g.,
the optimal cache allocated to level $m$ (i.e., the amount of the
cache used to store the files cached at the $m$-th level) is $x_{m}^{*}4^{-m}F$.
However, such a cache allocation scheme may not be feasible because
$x_{m}^{*}$ may not be integer. Therefore, step 2 is to find a feasible
solution $\mathbf{x}^{o}$ of (\ref{eq:cachePmain}) that is close
to $\mathbf{x}^{*}$ (or equivalently, find a cache allocation scheme
such that the cache allocated to the $m$-th level is close to $x_{m}^{*}4^{-m}F$
and can be divided by $4^{-m}F$ ). Specifically, when $m=0$, the
available cache size is $x_{0}^{*}4^{-0}F$ and thus the cache allocated
to the $0$-th level is $\left\lfloor x_{0}^{*}\right\rfloor 4^{-0}F$.
Correspondingly, the number of files stored at the $0$-th level is
$x_{0}^{o}=\left\lfloor x_{0}^{*}\right\rfloor $. The released cache
size from the $0$-th level is $b_{0}F=\left(x_{0}^{*}4^{-0}-x_{0}^{o}4^{-0}\right)F$.
When $m=1$, the available cache size (including the cache size released
from the $0$-th level) is $\left(x_{1}^{*}4^{-1}F+b_{0}F\right)$,
and thus the cache allocated to the $1$-th level is $\left\lfloor \left(x_{1}^{*}4^{-1}F+b_{0}F\right)/\left(4^{-1}F\right)\right\rfloor 4^{-1}F=\left\lfloor x_{1}^{*}+b_{0}4^{1}\right\rfloor 4^{-1}F$.
Correspondingly, the number of files stored at the $1$-th level is
$x_{1}^{o}=\left\lfloor x_{1}^{*}+b_{0}4^{1}\right\rfloor $. The
released cache size from the first level is $b_{1}=\left(x_{1}^{*}4^{-1}+b_{0}-x_{1}^{o}4^{-1}\right)F$.
Similarly, when $m>1$, the available cache size (including the cache
size released from the $\left(m-1\right)$-th level) is $\left(x_{m}^{*}4^{-m}F+b_{m-1}F\right)$
and thus the cache allocated to the $m$-th level is $\left\lfloor \left(x_{m}^{*}4^{-m}F+b_{m-1}F\right)/\left(4^{-m}F\right)\right\rfloor 4^{-1}F=\left\lfloor x_{m}^{*}+b_{m-1}4^{m}\right\rfloor 4^{-m}F$.
Correspondingly, the number of files stored at the $m$-th level is
$x_{m}^{o}=\left\lfloor x_{m}^{*}+b_{m-1}4^{m}\right\rfloor $. The
released cache size from the $m$-th level is $b_{m}=\left(x_{m}^{*}4^{-m}+b_{m-1}-x_{m}^{o}4^{-m}\right)F$.
The above cache allocation process continues until all $L$ files
have been cached. Finally, steps 2c - 2g are to balance the traffic
loading of different levels.

Despite various relaxations and approximations, we will show that
the proposed low-complexity cache content placement algorithm is order
optimal in Section \ref{sec:Achievable-Throughput-and}; i.e., it
achieves the same order of throughput as the optimal solution of the
cache content placement optimization problem in (\ref{eq:cachePmain}).

\begin{algorithm}
\caption{\label{alg:cachebisec}Cache content placement Algorithm}
\textbf{\small{}Step 1 (Bisection for solving}{\small{} }\textbf{\small{}(\ref{eq:cachePmain-1})):}{\small{}
Let $m_{L}=0$. If $L_{0}\left(\overline{C}_{1}\right)<L_{C}$, let
$m^{*}=m_{L}$ and goto Step 1c.}{\small \par}

{\small{}Let $m_{H}=M$. If $L4^{-M}\geq L_{C}$, let $m^{*}=m_{H}$,
$x_{m}^{*}=0,\forall m<m^{*}$, $x_{M}^{*}=L$ and goto Step 2; otherwise
let $m^{*}=\left\lfloor \frac{m_{L}+m_{H}}{2}\right\rfloor $.}{\small \par}

{\small{}\ \ \ \ }\textbf{\small{}1a: }{\small{}If $L_{m^{*}}\left(\overline{C}_{m^{*}+1}\right)\geq L_{C}$,
let $m_{L}=m^{*}$; elseif $L_{m^{*}}\left(\overline{C}_{m^{*}}\right)<L_{C}$,
let $m_{H}=m^{*}$; else goto Step 1c.}{\small \par}

{\small{}\ \ \ \ }\textbf{\small{}1b:}{\small{} If $m_{H}-m_{L}=1$,
let $m^{*}=m_{H}$ and goto Step 1c; otherwise let $m^{*}=\left\lfloor \frac{m_{L}+m_{H}}{2}\right\rfloor $
and goto Step 1a.}{\small \par}

{\small{}\ \ \ \ }\textbf{\small{}1c: }{\small{}Let $x_{m}^{*}=0,\forall m<m^{*}$
and $\left\{ x_{m}^{*},m=m^{*},...,M\right\} $ be the solution of
(\ref{eq:Requ}) and (\ref{eq:sumL}) as given in (\ref{eq:solRequ}).}{\small \par}

\textbf{\small{}Step 2 (Find a feasible solution close to }{\small{}$\mathbf{x}^{*}$}\textbf{\small{}):}{\small{}
Let $b_{m^{*}-1}=0$, $x_{m}^{o}=0,m=0,...,m^{*}-1$ and $m=m^{*}$.}{\small \par}

{\small{}\ \ \ \ }\textbf{\small{}2a:}{\small{} Let $x_{m}^{o}=\left\lfloor x_{m}^{*}+b_{m-1}4^{m}\right\rfloor $.
$b_{m}=x_{m}^{*}4^{-m}+b_{m-1}-x_{m}^{o}4^{-m}$.}{\small \par}

{\small{}\ \ \ \ }\textbf{\small{}2b:}{\small{} If $\sum_{i=0}^{m}x_{i}^{o}\geq L$,
let $x_{m}^{o}=L-\sum_{i=0}^{m-1}x_{i}^{o}$ and goto Step 2c; otherwise
let $m=m+1$ and goto Step 2a.}{\small \par}

{\small{}\ \ \ \ }\textbf{\small{}2c: }{\small{}Let $M^{\circ}=\textrm{argmax}_{m}x_{m}^{\circ},\:\textrm{s.t. }x_{m}^{\circ}>0$
and $m^{\circ}=\textrm{argmin}_{m}x_{m}^{\circ},\:\textrm{s.t. }x_{m}^{\circ}>1$.
If $M^{\circ}-m^{\circ}\leq2$, goto Step 3. }{\small \par}

{\small{}\ \ \ \ }\textbf{\small{}2d:}{\small{} Let}\textbf{\small{}
$x_{m^{\circ}}^{'}=x_{m^{\circ}}^{\circ}-1$}{\small{}, $x_{m^{\circ}+1}^{'}=x_{m^{\circ}+1}^{\circ}+1$
and $x_{m}^{'}=x_{m}^{\circ},\forall m\notin\left\{ m^{\circ},m^{\circ}+1\right\} $.}\textbf{\small{} }{\small \par}

{\small{}\ \ \ \ }\textbf{\small{}2e:}{\small{} Let $M^{'}=\textrm{argmax}_{m}x_{m}^{'},\:\textrm{s.t. }x_{m}^{'}>0$.}{\small \par}

{\small{}\ \ \ \ }\textbf{\small{}2f:}{\small{} While $M^{'}>m^{\circ}+1$
and $x_{M^{'}}^{'}>0$ and $\sum_{m=0}^{M}x_{m}^{'}4^{-m}-4^{-M^{'}}+4^{-(m^{\circ}+1)}\leq L_{C}$ }{\small \par}

{\small{}\ \ \ \ \ \ \ \ Let $x_{M^{'}}^{'}=x_{M^{'}}^{'}-1$,
$x_{m^{\circ}+1}^{'}=x_{m^{\circ}+1}^{'}+1$, }{\small \par}

{\small{}\ \ \ \ \ \ \ \ Let $M^{'}=\textrm{argmax}_{m}x_{m}^{'},\:\textrm{s.t. }x_{m}^{'}>0$.}{\small \par}

{\small{}\ \ \ \ }\textbf{\small{}2g:}{\small{} Let $R^{'}=\min_{m\in\left\{ m^{\circ},...,M^{'}\right\} }\overline{C}_{m}/f\left(\sum_{i=m^{\circ}}^{m-1}x_{i}^{'}+1\right)$
and $R^{\circ}=\min_{m\in\left\{ m^{\circ},...,M^{\circ}\right\} }\overline{C}_{m}/f\left(\sum_{i=m^{\circ}}^{m-1}x_{i}^{\circ}+1\right)$.
If $R^{'}>R^{\circ}$, let $x_{m}^{\circ}=x_{m}^{'},\forall m$, goto
Step 2c. }{\small \par}

\textbf{\small{}Step 3 (Termination):}{\small{} Output $\mathbf{x}^{o}$.}{\small \par}
\end{algorithm}

\begin{rem}
In Appendix \ref{subsec:Reformulation-of-the}, we reformulate Problem
(\ref{eq:cachePmain}) with fixed $R$ as a zero-one linear programming
(ZOLP) feasibility problem. Based on the ZOLP reformulation in (\ref{eq:ZOLP}),
it is possible to find the optimal solution of Problem (\ref{eq:cachePmain})
by a bisection search over $R$, where for each fixed $R$, the ZOLP
feasibility problem (\ref{eq:ZOLP}) is solved using standard ZOLP
solvers. Although it is difficult to analyze the performance of the
optimal solution, the ZOLP reformulation in (\ref{eq:ZOLP}) is elegant
and may potentially achieve a better throughput performance. Readers
interested in algorithm design may refer to Appendix \ref{subsec:Reformulation-of-the}
for the details.
\end{rem}

\section{Throughput Performance of Cache-induced Hierarchical Cooperation\label{sec:Achievable-Throughput-and} }

For general content popularity distributions, it is very difficult
to analyze the performance of the proposed cache-induced hierarchical
cooperation scheme with the cache content placement parameter $\mathbf{x}$
determined by Algorithm \ref{alg:cachebisec}. In this section, we
assume the content popularity follows Zipf distribution \cite{Yamakami_PDCAT06_Zipflaw}
and analyze the throughput performance of the proposed scheme. Under
the Zipf popularity distribution, the probability of requesting the
$l$-th file is given by
\begin{equation}
p_{l}=\frac{1}{Z_{\tau}\left(L\right)}l^{-\tau},l=1,...,L,\label{eq:zipp}
\end{equation}
where $\tau$ is the \textit{popularity skewness parameter} and $Z_{\tau}\left(L\right)=\sum_{l=1}^{L}l^{-\tau}$
is a normalization factor.

\subsection{Throughput Bounds under Zipf Popularity Distribution}

In this subsection, we derive the upper and lower bounds for the throughput
of the proposed scheme under the Zipf popularity distribution. To
achieve this, we first give upper and lower bounds for the $\overline{C}_{m}$'s
in (\ref{eq:Cmbar}).
\begin{lem}
\label{lem:edgecapbound}The upper and lower bounds of $\overline{C}_{m}$
can be expressed in a unified form as $c_{n}4^{-m\gamma_{n}}$ for
some coefficient $c_{n}$ and $\gamma_{n}$ that depends on $n=4^{M}$.
Specifically, for a wireless D2D network with $A\left(n\right)=n^{\kappa}$
nodes, with $\kappa\geq0$, $\overline{C}_{m}$ can be bounded as
$c_{n}^{L}\left(\kappa\right)4^{-m\gamma_{n}^{L}\left(\kappa\right)}\leq\overline{C}_{m}\leq c_{n}^{U}\left(\kappa\right)4^{-m\gamma_{n}^{U}\left(\kappa\right)}$,
where 
\begin{align*}
\gamma_{n}^{U}\left(\kappa\right) & =\min\left(\frac{1}{2s_{M}+1}+\left(\frac{\alpha\kappa}{2}-1\right)^{+},\frac{1}{2}\right),\\
c_{n}^{U}\left(\kappa\right) & =\frac{2}{T_{r}3^{\frac{5}{4}}}\log\left(1+\frac{\textrm{SNR}}{1+P_{I}}\right),s_{M}=\sqrt{M\ln4},\\
\gamma_{n}^{L}\left(\kappa\right) & =\min\left(\frac{1}{s_{M}+1}+\left(\frac{\alpha\kappa}{2}-1\right)^{+},\frac{1}{2}\right),\\
c_{n}^{L}\left(\kappa\right) & =\frac{4R_{c}\left(\alpha,P_{I}\right)}{3\left(1+s_{M}\right)T_{r}^{2}\left(3\cdot2^{s_{M}-1}\right)^{\frac{s_{M}}{2\left(s_{M}+1\right)}}}.
\end{align*}
\end{lem}

The proof follows straightforwardly from the definition of $\overline{C}_{m}$
and the results in Section \ref{sec:Exist-Achievable-Schemes}. The
detailed derivations are omitted for conciseness.

One key challenge to derive the throughput lower bound is to quantify
the throughput loss due to various relaxations and approximations
in Algorithm \ref{alg:cachebisec}. This challenge is addressed in
the following lemma. 
\begin{lem}
\label{lem:Performance-loss-Integer}Let $\left(\mathbf{x}^{*},R^{*}\right)$
denote the optimal solution of the relaxed cache content placement
optimization problem in (\ref{eq:cachePmain-1}) obtained in Step
1 of Algorithm \ref{alg:cachebisec}, and $\mathbf{x}^{o}$ denote
the feasible cache content placement parameter found by Step 2 of
Algorithm \ref{alg:cachebisec}. Then $\left(\mathbf{x}^{o},\frac{1}{M\left(1+2^{\tau}\right)}R^{*}\right)$
must be a feasible solution of the original cache content placement
optimization problem in (\ref{eq:cachePmain}).
\end{lem}

Please refer to Appendix \ref{subsec:Proof-of-Lemmaloss} for the
proof.

Clearly, the optimal objective of Problem (\ref{eq:cachePmain-1})
provides an upper bound for the throughput achievable with the proposed
cache-induced hierarchical cooperation. However, it is highly non-trivial
to obtain the closed-form expression for the optimal objective of
Problem (\ref{eq:cachePmain-1}) since there is no closed-form solution
for Problem (\ref{eq:cachePmain-1}). To overcome this challenge,
we first derive closed-form upper and lower bounds $R_{U}$ and $R_{L}$
for the optimal objective of Problem (\ref{eq:cachePmain-1}). Then
$R_{U}$ and $\frac{1}{M\left(1+2^{\tau}\right)}R_{L}$ provide an
upper bound and a lower bound for the achievable throughput, respectively.
The detailed analysis is given in Appendix \ref{subsec:Proof-of-Theoremtpbounds},
and the final results are summarized in the following theorem.
\begin{thm}
[Throughput Bounds]\label{thm:thp-bounds}Consider a wireless D2D
network with area $A\left(n\right)=n^{\kappa}$, $\kappa\geq0$. Let
$R^{\star}$ and $R^{\circ}$ denote the per node throughput achieved
by the cache-induced hierarchical cooperation scheme with the optimal
cache content placement parameter $\mathbf{x}^{\star}$ (i.e., the
optimal solution of (\ref{eq:cachePmain})) and with the low-complexity
cache content placement solution $\mathbf{x}^{\circ}$in Algorithm
\ref{alg:cachebisec}, respectively. Then both $R^{\star}$ and $R^{\circ}$
are lower bounded as $R^{\star}\geq R^{\circ}\geq\frac{1}{M\left(1+2^{\tau}\right)}R_{L}$,
where 
\[
R_{L}=\begin{cases}
\overline{R}_{L|\tau<1} & \tau\in\left[0,1\right)\\
\left(\frac{\left(e^{2}L-L-1\right)4^{-M}+L_{C}}{4\left(e^{2}L-1\right)}\right)^{\gamma_{n}}c_{n} & \tau=1\\
\left(\frac{4^{\frac{1+\gamma_{n}-\tau}{\tau-1}}-1}{4^{\frac{\gamma_{n}+\tau-1}{\tau-1}}-4}\right)^{\gamma_{n}}\frac{c_{n}L_{C}^{\gamma_{n}}L^{\tau-1-\gamma_{n}}}{\tau} & \tau\in\left(1,\gamma_{n}+1\right)\\
\left(3\log_{4}L+4\right)^{-\gamma_{n}}\frac{c_{n}}{\tau}L_{C}^{\tau-1} & \tau=\gamma_{n}+1\\
\frac{c_{n}L_{C}^{\tau-1}}{\left(\frac{3\tau^{\frac{1}{\tau-1}}4^{\frac{\gamma_{n}+1-\tau}{\tau-1}}}{1-4^{\frac{\gamma_{n}+1-\tau}{\tau-1}}}+4\tau^{\frac{1}{\gamma_{n}}}\right)^{\tau-1}} & \tau>\gamma_{n}+1
\end{cases}
\]
\[
R_{L|\tau<1}=c_{n}\min\left(\left(\frac{4^{\gamma_{n}+1}-1}{4^{\gamma_{n}+2}-16}\frac{L_{C}}{L}\right)^{\gamma_{n}},\frac{3\left(1-\frac{L_{C}}{L}\right)^{-1}}{4^{\gamma_{n}+1}-1}\right),
\]
$c_{n}=c_{n}^{L}\left(\kappa\right)$ and $\gamma_{n}=\gamma_{n}^{L}\left(\kappa\right)$.
Moreover, both $R^{\star}$ and $R^{\circ}$ are upper bounded as
$R_{U}\geq R^{\star}\geq R^{\circ}$, where 

\[
R_{U}=\begin{cases}
\frac{c_{n}}{1-\tau}\left(\frac{4^{\gamma_{n}+1}-1}{4^{\gamma_{n}}-1}\right)^{\gamma_{n}}\left(\frac{L_{C}}{L}\right)^{\gamma_{n}} & \tau\in\left[0,1\right)\\
c_{n}\left(\frac{4e\left(4^{M}L_{C}+L+1\right)}{3\cdot4^{M}L^{1-\frac{1}{\ln L}}}\right)^{\gamma_{n}}\ln L & \tau=1\\
\frac{L_{C}^{\gamma_{n}}}{\left(\frac{L^{\frac{1+\gamma_{n}-\tau}{\gamma_{n}}}\left(\frac{\tau}{c_{n}}\right)^{\frac{1}{\gamma_{n}}}}{2^{\tau-1}}-4c_{n}^{-\frac{1}{\gamma_{n}}}\right)^{\gamma_{n}}} & \tau\in\left(1,\gamma_{n}+1\right)\\
\frac{\left(\tau-L^{1-\tau}\right)c_{n}4^{-\gamma_{n}}}{\left(L_{C}+1\right)^{1-\tau}-\left(L+1\right)^{1-\tau}} & \tau\geq\gamma_{n}+1
\end{cases}
\]
$c_{n}=c_{n}^{U}\left(\kappa\right)$ and $\gamma_{n}=\gamma_{n}^{U}\left(\kappa\right)$.
\end{thm}

\subsection{Comparison With Cache-assisted Multihop Scheme}

In this subsection, we compare the per node throughput of the cache-induced
hierarchical cooperation scheme with that of a cache-assisted multihop
scheme, which only has multihop PHY mode. Following a similar analysis,
it can be shown that the lower and upper bounds of the per node throughput
of the cache-assisted multihop scheme is given in the same form as
$R_{L}$ and $R_{U}$ in Theorem \ref{thm:thp-bounds}, but with different
coefficients $c_{n}=c_{n}^{M}$ and $\gamma_{n}=\gamma_{n}^{M}$,
where 
\[
\gamma_{n}^{M}=\frac{1}{2},c_{n}^{M}=\frac{4}{3T_{r}^{2}}\log\left(1+\frac{\textrm{SNR}}{1+P_{I}}\right).
\]
Note that both $R_{L}$ and $R_{U}$ increase with $c_{n}$ and decrease
with $\gamma_{n}$, where $\gamma_{n}$ determines the scaling of
the throughput bounds w.r.t. $n$, as will be shown later in Theorem
\ref{thm:Achievable-Scaling-Law}, and $c_{n}$ determines the constant
coefficients in the scaling law. Both $\gamma_{n}^{U}\left(\kappa\right)$
and $\gamma_{n}^{L}\left(\kappa\right)$ of the proposed scheme are
smaller than the $\gamma_{n}^{M}=1/2$ of the cache-assisted multihop
scheme. As a result, the proposed scheme has huge throughput gain
over the cache-assisted multihop scheme, especially when $\kappa$
is smaller (i.e., denser networks), as will be shown in the simulations.

\begin{figure}
\begin{centering}
\includegraphics[width=85mm]{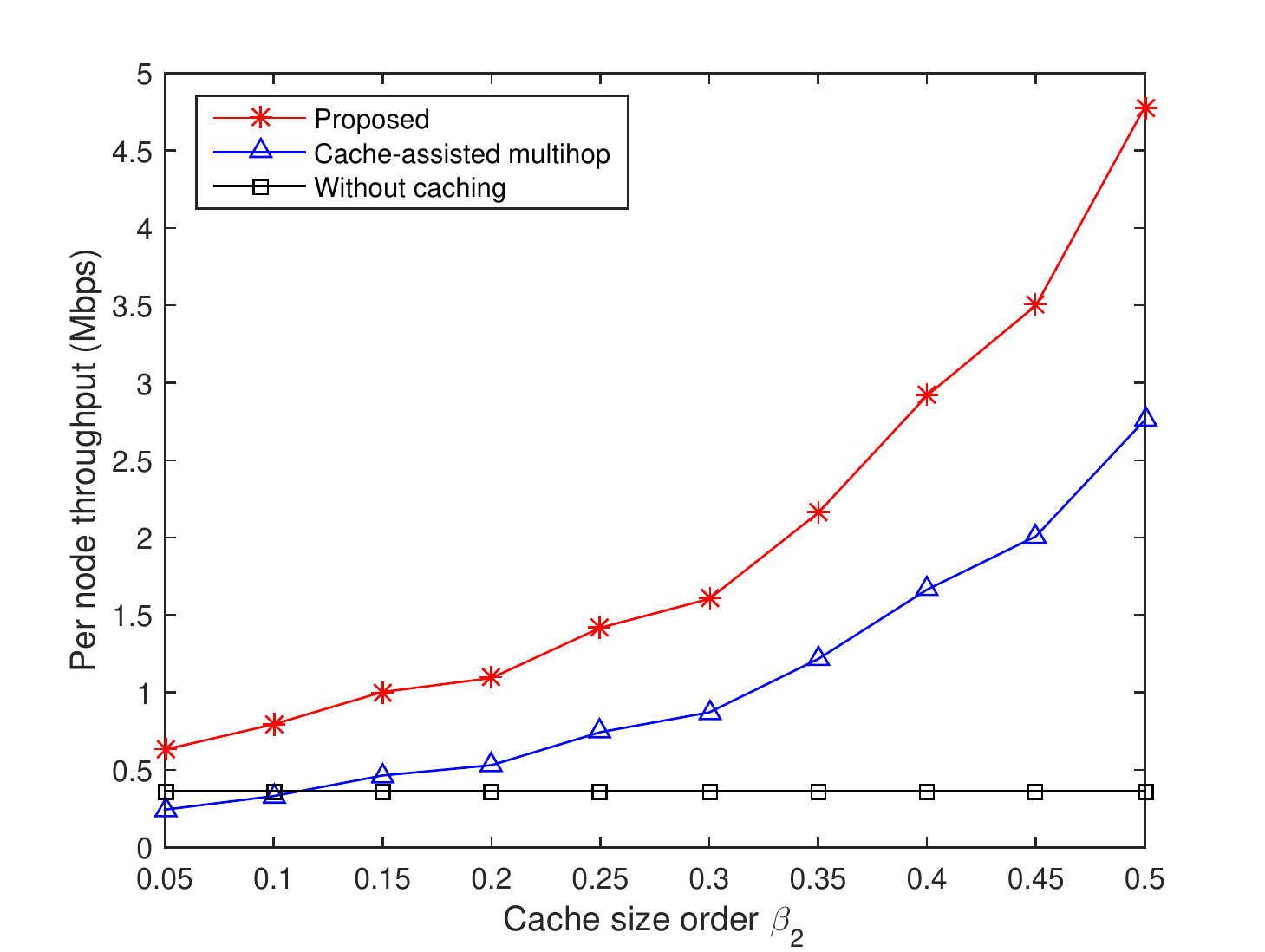}
\par\end{centering}
\caption{\label{fig:VarLC}{\small{}Per node throughput versus the cache size
order $\beta_{2}$ for a dense network with $n=4^{9}$ nodes. The
content popularity skewness is $\tau=1$ and the path loss exponent
is $\alpha=4$.}}
\end{figure}

\begin{figure}
\begin{centering}
\includegraphics[width=85mm]{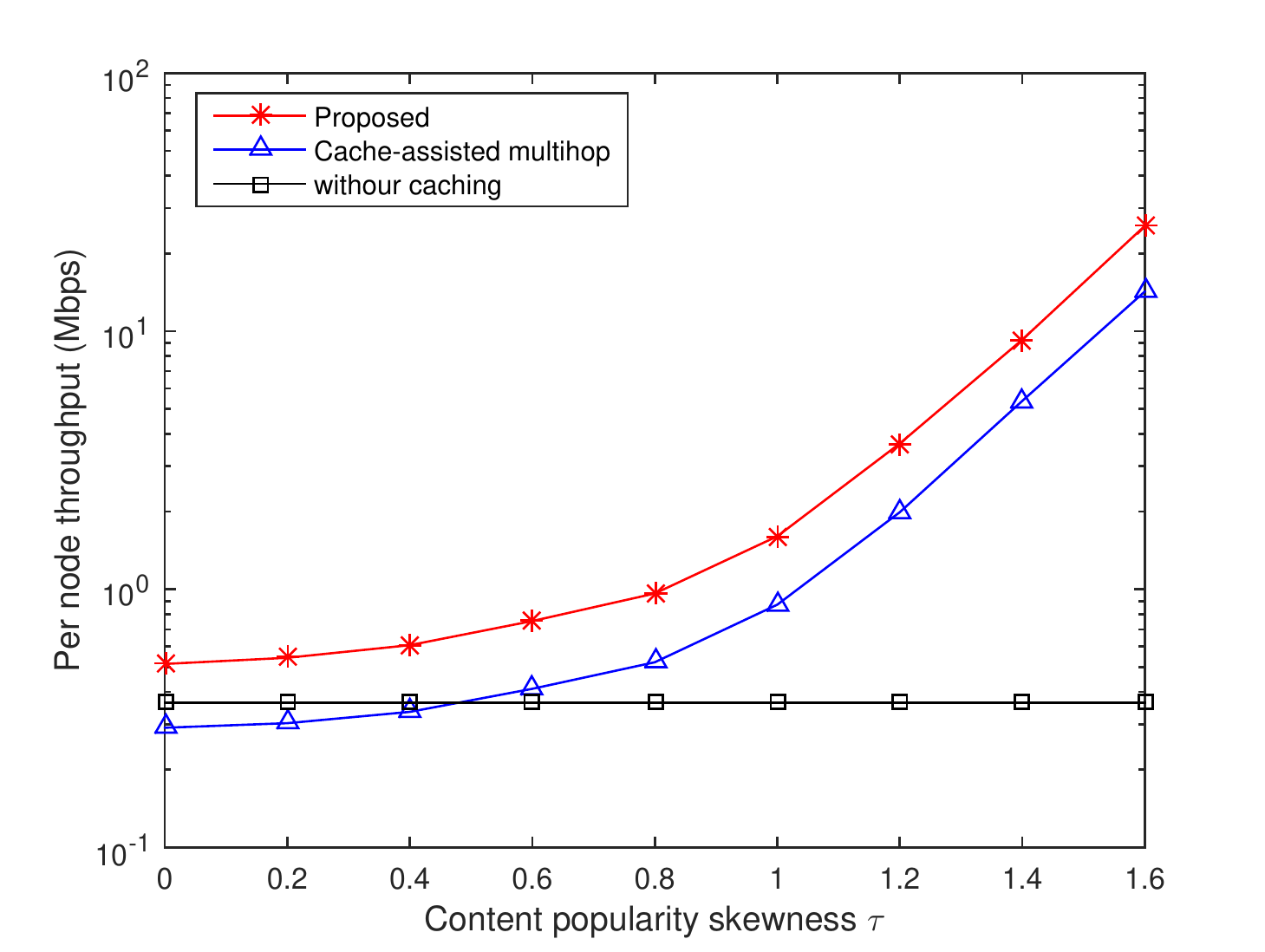}
\par\end{centering}
\caption{\label{fig:Vartau}{\small{}Per node throughput versus the content
popularity skewness $\tau$ for a dense network with $n=4^{9}$ nodes.
The cache size order is $\beta_{2}=0.3$, and the path loss exponent
is $\alpha=4$.}}
\end{figure}

\begin{figure}
\begin{centering}
\includegraphics[width=85mm]{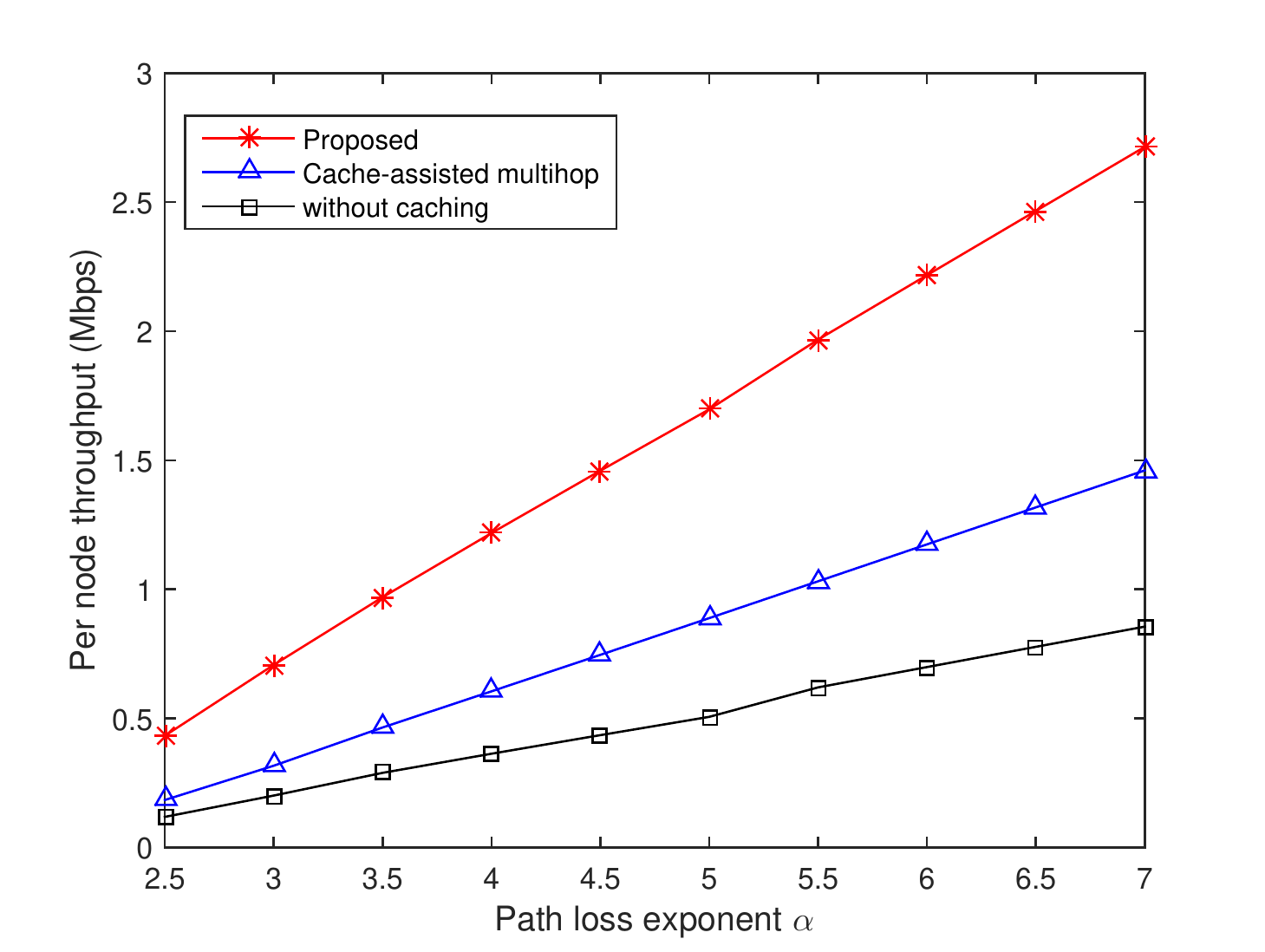}
\par\end{centering}
\caption{\label{fig:Varalf}{\small{}Per node throughput versus the path loss
exponent $\alpha$ for a dense network with $n=4^{9}$ nodes. The
cache size order is $\beta_{2}=0.3$, and the content popularity skewness
$\tau=1$.}}
\end{figure}

In Figs. \ref{fig:VarLC} - \ref{fig:Varalf}, we illustrate the throughput
gain of the proposed cache-induced hierarchical cooperation for a
dense wireless D2D network with $n=4^{9}$ nodes, area $A\left(n\right)=1$,
and $200$ MHz system bandwidth. There are $L=\left\lfloor n^{\beta_{1}}\right\rfloor $
content files on the content server, and the cache capacity at each
node is $L_{C}=n^{\beta_{2}}$, where $\beta_{1}=0.9$ and $\beta_{2}\in\left[0,\beta_{1}\right]$.
The throughput of the\textit{ network without caching} is also given
for comparison. The network without caching refers to arbitrary (random
uniform permutation) source-destination traffic, as in \cite{Tse_IT07_CapscalingHMIMO},
using the improved hierarchical cooperation in \cite{Caire_TIT2015_HMIMOimp}.
This comparison is just to give an idea of the advantage of caching
when the demands are restricted to being in a given library of messages,
rather than random source-destination traffic.

In Fig. \ref{fig:VarLC}, we plot the per node throughput versus the
cache size order $\beta_{2}$. The total number of content files is
$L=n^{0.9}$, and the content popularity skewness $\tau$ is fixed
as 1. It can be seen that the throughput of both the proposed scheme
and cache-assisted multihop scheme increases with the cache size order
$\beta_{2}$. Moreover, the proposed scheme achieves large throughput
gain over the two baseline schemes. 

We then simulate the case when the BS cache size is much smaller than
the total content size. In Fig. \ref{fig:Vartau}, we plot the per
node throughput versus the content popularity skewness $\tau$. The
total number of content files is $L=n^{0.9}$, and the cache size
at each node is fixed as $L_{C}=n^{0.3}$. The results in Fig. \ref{fig:Vartau}
show that the throughput of both the proposed scheme and cache-assisted
multihop scheme increases with the content popularity skewness $\tau$.
Again, the proposed hierarchical cooperation achieves a large throughput
gain over the two baseline schemes.

In Fig. \ref{fig:Varalf}, we plot the per node throughput versus
the path loss exponent $\alpha$. It can be seen that the throughput
gain of the proposed scheme increases with the path loss exponent
$\alpha$ for dense networks.

\section{Scaling Laws in Extended Networks\label{sec:Scaling-Laws-in}}

\subsection{System Scaling Regime}

In order to study the throughput scaling of extended wireless D2D
caching networks (i.e., $A\left(n\right)=n$) for asymptotically large
$n$, we consider that $L$ and $L_{C}$ scale with $n$ according
to the following functions:
\[
L=a_{1}n^{\beta_{1}}\textrm{ and }L_{C}=a_{2}n^{\beta_{2}},
\]
where $\beta_{1},a_{1},a_{2}>0$ and $\beta_{2}\in\left[0,\beta_{1}\right]$.
When $\beta_{1}=\beta_{2}$, we assume $a_{1}>a_{2}$ to avoid the
trivial case when each node has enough cache capacity to store the
entire library $\mathcal{L}$. Moreover, since $nL_{C}>L$, we have
$\beta_{1}-\beta_{2}\leq1$ and when $\beta_{1}-\beta_{2}=1$, we
have $a_{1}\leq a_{2}$. Note that a similar scaling regime was also
considered in \cite{Caire_TIT2015_HMIMOimp}. 

Depending on the relative caching capacity $\frac{L_{C}}{L}$ at each
node, the entire parameter space can be partitioned into two regimes
as follows:
\begin{itemize}
\item Regime I: $\beta_{1}-\beta_{2}=0,a_{1}>a_{2}$.
\item Regime II: $\beta_{1}-\beta_{2}\in\left(0,1\right)$, or $\beta_{1}-\beta_{2}=1,a_{1}\leq a_{2}$.
\end{itemize}

\subsection{Throughput Scaling Laws of Cache-induced Hierarchical Cooperation}

In this subsection, we obtain the throughput scaling laws of the proposed
scheme. From the throughput bounds in Theorem \ref{thm:thp-bounds},
we can obtain the following achievable throughput scaling law.
\begin{thm}
[Achievable Scaling Law in Extended Networks]\label{thm:Achievable-Scaling-Law}For
extended networks with $A\left(n\right)=n$, the achievable throughput
$R^{\star}$ of the cache-induced hierarchical cooperation satisfies
the following scaling law. In Regime I, we have 
\[
R^{\star}=\begin{cases}
\Omega\left(1\right), & \tau\in\left[0,1\right]\\
\Omega\left(n^{\beta_{2}\left(\tau-1\right)}\right), & \tau>1
\end{cases},
\]
where $R^{\star}=\Omega\left(n^{\eta}\right)$ means that the order
of $R^{\star}$ is no less than $n^{\eta}$: $n^{\eta}/R^{\star}=O\left(1\right)$.
In Regime II, the achievable throughput scaling law depends on the
popularity skewness parameter $\tau$, summarized as follows:
\[
R^{\star}=\begin{cases}
\Omega\left(n^{\left(\beta_{2}-\beta_{1}\right)\left(\frac{\min\left(3,\alpha\right)}{2}-1\right)-\epsilon_{\alpha}}\right) & \tau\in\left[0,1\right]\\
\Omega\left(n^{\beta_{1}\left(\tau-\frac{\min\left(3,\alpha\right)}{2}\right)+\beta_{2}\left(\frac{\min\left(3,\alpha\right)}{2}-1\right)-\epsilon_{\alpha}}\right) & \tau\in\left(1,\frac{\min\left(3,\alpha\right)}{2}\right]\\
\Omega\left(n^{\beta_{2}\left(\tau-1\right)-\epsilon_{\alpha}}\right) & \tau>\frac{\min\left(3,\alpha\right)}{2}
\end{cases},
\]
where $\epsilon_{\alpha}=\Theta\left(\frac{1}{\sqrt{\log n}}\right)\rightarrow0$
as $n\rightarrow\infty$ for $\alpha\in\left(2,3\right)$, and $\epsilon_{\alpha}=0$
for $\alpha\geq3$. Moreover, we have $R^{\circ}=\Theta\left(R^{\star}\right)$.
\end{thm}

Theorem \ref{thm:thp-bounds} also establishes the order optimality
of the proposed low-complexity cache content placement algorithm (Algorithm
\ref{alg:cachebisec}).

In the following, we compare the achievable scaling law of the proposed
cache-induced hierarchical cooperation with that of the following
two baseline schemes: the cache-assisted multihop scheme and PHY caching
in \cite{Liu_TWC15arxiv_adhoccaching}. \cite{Liu_TWC15arxiv_adhoccaching}
only studied the achievable scaling law of the PHY caching for the
special case of $\beta_{2}=0$. However, following a similar analysis
to that in this paper, we can extend the achievable scaling law in
\cite{Liu_TWC15arxiv_adhoccaching} to the more general case considered
in this paper, as summarized in the following theorem.
\begin{thm}
[Achievable Scaling Law of Baseline Schemes]\label{thm:Achievable-Scaling-Law-BL}For
extended networks, the achievable throughput $R_{PHY}$ of the PHY
caching scheme in \cite{Liu_TWC15arxiv_adhoccaching} satisfies the
following scaling law. In Regime I, we have 
\[
R_{PHY}=\begin{cases}
\Omega\left(1\right) & \tau\in\left[0,1\right]\\
\Omega\left(n^{\beta_{2}\left(\tau-1\right)}\right) & \tau>1
\end{cases}.
\]
In Regime II, the achievable throughput scaling law depends on the
popularity skewness parameter $\tau$, summarized as follows:
\[
R_{PHY}=\begin{cases}
\Omega\left(n^{\frac{\beta_{2}-\beta_{1}}{2}-\epsilon}\right) & \tau\in\left[0,1\right]\\
\Omega\left(n^{\beta_{1}\left(\tau-\frac{3}{2}\right)+\frac{\beta_{2}}{2}-\epsilon}\right) & \tau\in\left(1,\frac{3}{2}\right]\\
\Omega\left(n^{\beta_{2}\left(\tau-1\right)-\epsilon}\right) & \tau>\frac{3}{2}
\end{cases},
\]
where $\epsilon>0$ is arbitrarily small. Moreover, the achievable
throughput of the cache-assisted multihop scheme $R_{M}$ satisfies
the same scaling law as that of $R_{PHY}$, i.e., $R_{M}=\Theta\left(R_{PHY}\right)$.
\end{thm}

For the special case of Regime I or Regime II with $\alpha\geq3$,
Theorem \ref{thm:Achievable-Scaling-Law} reduces to the achievable
scaling law of the two baseline schemes in Theorem \ref{thm:Achievable-Scaling-Law-BL}.
In Regime II with $\alpha<3$, the scaling law achieved by the cache-induced
hierarchical cooperation in Theorem \ref{thm:Achievable-Scaling-Law}
is better than that achieved by the two baseline schemes.

\begin{figure}
\begin{centering}
\includegraphics[width=85mm]{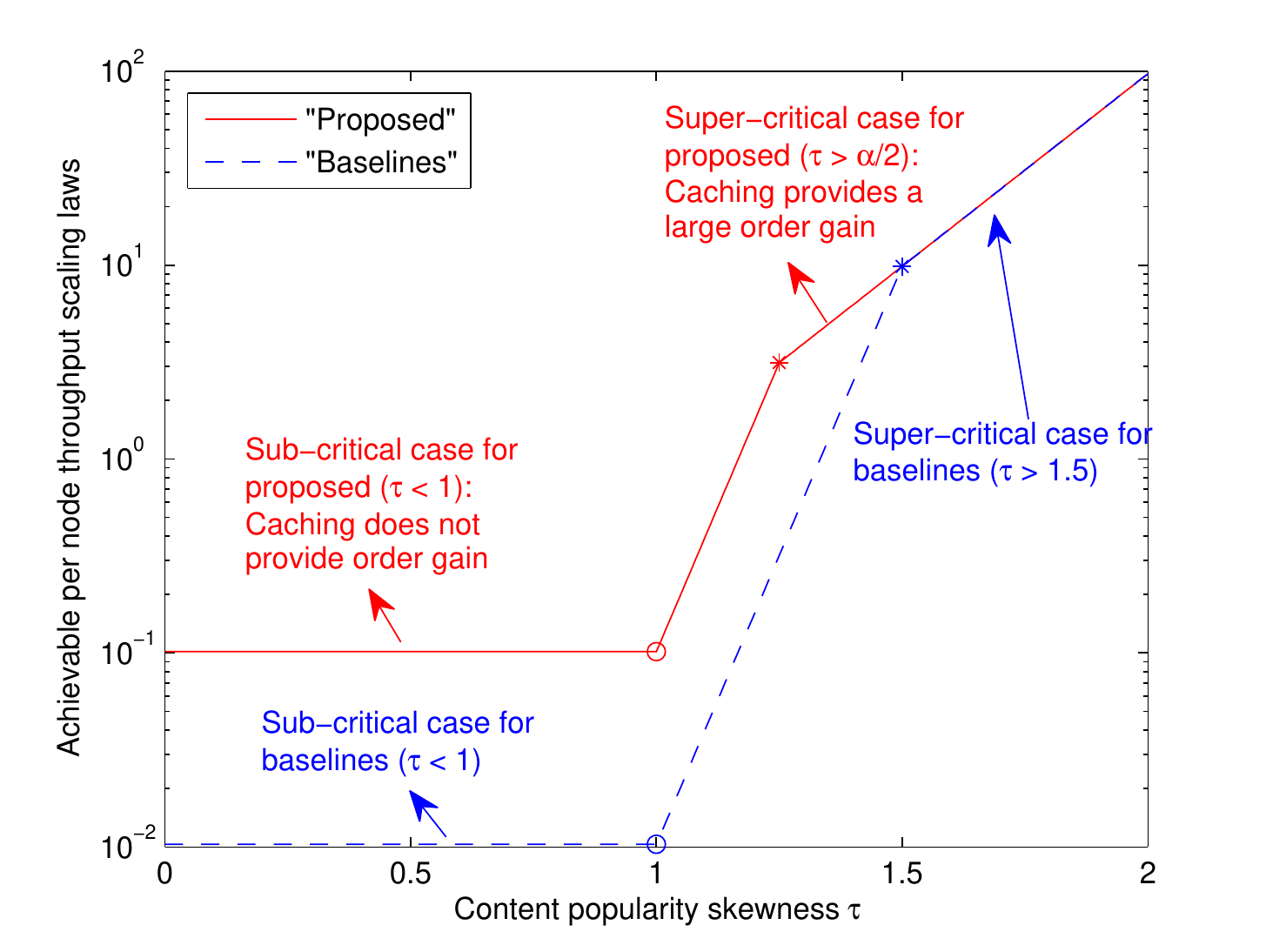}
\par\end{centering}
\caption{\label{fig:sub-super-critical}{\small{}Illustration of achievable
throughput scaling laws in Theorem \ref{thm:Achievable-Scaling-Law}
for cache-induced hierarchical cooperation (solid curve) and Theorem
\ref{thm:Achievable-Scaling-Law-BL} for baseline schemes (dashed
curve). The number of nodes in the extended network is $n=4^{11}$.
The total content size order $\beta_{1}=0.9$ and the cache size order
is $\beta_{2}=0.3$ (i.e., Regime II). The path loss exponent is $\alpha=2.5$.
In the figure, the circle and star symbols indicate the first and
second critical popularity skewness points $\tau_{a}$ and $\tau_{b}$,
respectively.}}
\end{figure}

As shown in Fig. \ref{fig:sub-super-critical}, the achievable scaling
laws of all schemes exhibit some phase transition phenomena as the
content popularity skewness $\tau$ increases. Specifically, in Regime
II, there are two \textit{critical popularity skewness} points: $\tau=\tau_{a}$
and $\tau=\tau_{b}$, where $\tau_{a}=1$, $\tau_{b}=1.5$ for the
baseline schemes and $\tau_{a}=1$, $\tau_{b}=\frac{\min\left(3,\alpha\right)}{2}$
for the cache-induced hierarchical cooperation scheme. For the \textit{sub-critical
case} when $\tau<\tau_{a}$, the per node throughput scales with $n$
as $\Omega\left(n^{\eta_{a}}\right)$ with a smaller order $\eta_{a}$.
For example, when $\beta_{1}-\beta_{2}=1$ (i.e., $L_{C}\ll L$),
$\eta_{a}=1-\frac{\min\left(3,\alpha\right)}{2}$ for the cache-induced
hierarchical cooperation, which is the same as the scaling law of
the hierarchical cooperation without caching; and $\eta_{a}=-0.5$
for the baseline schemes, which is the same as the Gupta\textendash Kumar
law \cite{GuptaKumar}. Therefore, when $\tau<\tau_{a}$ and $\beta_{1}-\beta_{2}=1$,
caching does not provide order gain. For the \textit{super-critical
case} when $\tau>\tau_{b}$, on the other hand, the per node throughput
scales with $n$ as $\Omega\left(n^{\eta_{b}}\right)$ with a much
larger order $\eta_{b}$. For example, when $\beta_{1}-\beta_{2}=1$
(i.e., $L_{C}\ll L$), we still have $\eta_{b}=\beta_{2}\left(\tau-1\right)>0$
for all schemes. In this case, caching provides a large order gain
even when $L_{C}\ll L$. 

From Fig. \ref{fig:sub-super-critical}, there are two advantages
of the proposed cache-induced hierarchical cooperation over the PHY
caching. First, when $\alpha<3$, the second critical popularity skewness
point $\tau_{b}$ of the cache-induced hierarchical cooperation is
smaller than that of the baseline schemes. This implies that the cache-induced
hierarchical cooperation can enjoy the large order gain $\eta_{b}=\beta_{2}\left(\tau-1\right)$
under weaker conditions on the popularity distribution. Second, when
$\alpha<3$, the cache-induced hierarchical cooperation can achieve
a better scaling law for $\tau<1.5$.

\subsection{Main Converse Results}

In this section, we establish upper bounds on the throughput scaling
laws. The main converse results are summarized in the following theorem.
\begin{thm}
[Upper Bound of Scaling Laws in Extended Networks]\label{thm:Converse-Scaling-Law}In
an extended wireless D2D caching network, the per node throughput
$R$ of any feasible content delivery scheme must satisfy the following
scaling laws. In Regime I, we have 
\[
R=\begin{cases}
O\left(n^{\epsilon}\right), & \tau\in\left[0,1\right]\\
O\left(n^{\beta_{2}\left(\tau-1\right)+\epsilon}\right), & \tau>1
\end{cases}.
\]
In Regime II, we have
\[
R=\begin{cases}
O\left(n^{\left(\beta_{2}-\beta_{1}\right)\left(\frac{\min\left(3,\alpha\right)}{2}-1\right)+\epsilon}\right), & \tau\in\left[0,1\right]\\
O\left(n^{\beta_{1}\left(\tau-\frac{\min\left(3,\alpha\right)}{2}\right)+\beta_{2}\left(\frac{\min\left(3,\alpha\right)}{2}-1\right)+\epsilon}\right), & \tau\in\left(1,\frac{\min\left(3,\alpha\right)}{2}\right]\\
O\left(n^{\beta_{2}\left(\tau-1\right)+\epsilon}\right), & \tau>\frac{\min\left(3,\alpha\right)}{2}
\end{cases},
\]
where $\epsilon>0$ is arbitrarily small. 
\end{thm}

Please refer to Section \ref{subsec:Converse-Proof} for the proof. 

In both Regime I and Regime II, the multiplicative gap between the
achievable per node throughput in Theorem \ref{thm:Achievable-Scaling-Law}
and its upper bound in Theorem \ref{thm:Converse-Scaling-Law} is
within $n^{\epsilon}$ for $\epsilon>0$ that can be arbitrarily small
as $n\rightarrow\infty$. Therefore, the throughput scaling law depicted
in Theorem \ref{thm:Achievable-Scaling-Law} is order-optimal in the
information theoretic sense for the Zipf popularity distribution. 

\subsection{Converse Proof\label{subsec:Converse-Proof}}

\subsubsection{Regime II with $\tau>\frac{\min\left(3,\alpha\right)}{2}$\label{subsec:Regime-II-large-tau}}

The converse result for this case can be proved by considering the
cut set bound between a reference node $i$ and the rest of the network
as follows. Let $q_{l}=I\left(B_{i};W_{l}\right)/F$. Since $B_{i}$
is a function of $W_{1},...,W_{L}$, we have
\begin{align}
H\left(B_{i}\right) & =H\left(B_{i}\right)-H\left(B_{i}|W_{1},...,W_{L}\right)\nonumber \\
 & =I\left(B_{i};W_{1},...,W_{L}\right)\nonumber \\
 & \overset{\textrm{a}}{=}\sum_{l=1}^{L}I\left(B_{i};W_{l}|W_{1},...,W_{l-1}\right)\nonumber \\
 & \overset{\textrm{b}}{=}\sum_{l=1}^{L}I\left(B_{i},W_{1},...,W_{l-1};W_{l}\right)\nonumber \\
 & \geq\sum_{l=1}^{L}I\left(B_{i};W_{l}\right)=\sum_{l=1}^{L}q_{l}F,\label{eq:HBi}
\end{align}
where (\ref{eq:HBi}-a) follows from the chain rule and (\ref{eq:HBi}-b)
follows from the fact that the messages $W_{1},...,W_{L}$ are mutually
independent. Hence, the $q_{l}$'s must satisfy the cache capacity
constraint $\sum_{l=1}^{L}q_{l}F\leq H\left(B_{i}\right)\leq L_{C}F$. 

Under any feasible content delivery scheme, the amount of information
$u_{i}^{j}$ transmitted from the rest of the network to node $i$
during the time window $\left[t_{i}^{j},...,t_{i}^{j+1}-1\right]$
must satisfy
\begin{align}
u_{i}^{j} & \geq H\left(U_{i}^{j}|B_{i}\right)\label{eq:Keyc1}\\
 & =H\left(W_{l_{i}},U_{i}^{j}|B_{i}\right)-H\left(W_{l_{i}}|U_{i}^{j},B_{i}\right)\nonumber \\
 & \geq H\left(W_{l_{i}}|B_{i}\right)-\varepsilon_{F}F\label{eq:keyFepson}\\
 & =H\left(W_{l_{i}}\right)-I\left(W_{l_{i}};B_{i}\right)-\varepsilon_{F}F\nonumber \\
 & =F\left(1-q_{l_{i}}-\varepsilon_{F}\right),\nonumber 
\end{align}
where $\varepsilon_{F}\rightarrow0$ as $F\rightarrow\infty$, (\ref{eq:Keyc1})
is to ensure that node $i$ can successfully receive the aggregate
information message $U_{i}^{j}$, and (\ref{eq:keyFepson}) follows
from the necessary condition in (\ref{eq:decodcons}). As a result,
for given $q_{l}$'s and per node rate requirement $R$, the total
average traffic rate over the cut from the rest of the network to
node $i$ is $\left(\sum_{l=1}^{L}p_{l}\left(1-q_{l}\right)-\varepsilon_{F}\right)R$.
Clearly, a per node throughput $R$ is achievable only if the induced
total average traffic rate does not exceed the sum capacity $\Gamma_{i}$
of the MISO channel between the rest of the network and node $i$,
which is upper bounded by $K\log n$ for some constant $K$ \cite{Tse_IT07_CapscalingHMIMO}.
Hence, the achievable per node throughput is upper bounded by 
\[
R\leq\max_{\left\{ q_{l}\right\} }\frac{\Gamma_{i}}{\sum_{l=1}^{L}p_{l}\left(1-q_{l}\right)-\varepsilon_{F}},\:\textrm{s.t. }\sum_{l=1}^{L}q_{l}\leq\left\lceil L_{C}\right\rceil .
\]
It is easy to see that the optimal $q_{l}$'s to maximize the achievable
per node throughput upper bound is to cache the most popular $\left\lceil L_{C}\right\rceil $
files, i.e., $q_{l}^{\star}=1,l=1,...,\left\lceil L_{C}\right\rceil $
and $q_{l}^{\star}=0,l>\left\lceil L_{C}\right\rceil $. As a result,
we have
\begin{equation}
R\leq\frac{\Gamma_{i}}{\sum_{l=1}^{L}p_{l}\left(1-q_{l}^{\star}\right)-\varepsilon_{F}}=\frac{\Gamma_{i}}{\sum_{l=\left\lceil L_{C}\right\rceil +1}^{L}p_{l}^{\star}-\varepsilon_{F}}.\label{eq:RregimeII-1}
\end{equation}
In Regime II with $\tau>\frac{\min\left(3,\alpha\right)}{2}$, we
have 
\begin{equation}
\sum_{l=\left\lceil L_{C}\right\rceil +1}^{L}p_{l}^{\star}=\Omega\left(n^{\beta_{2}\left(1-\tau\right)}\right).\label{eq:pltar}
\end{equation}
Finally, it follows from $\Gamma_{i}\leq K\log n$, (\ref{eq:RregimeII-1}),
and (\ref{eq:pltar}) that $R=O\left(n^{\beta_{2}\left(\tau-1\right)+\epsilon}\right)$
as $F\rightarrow\infty$.

\subsubsection{Regime II with $\tau\in\left[0,\frac{\min\left(3,\alpha\right)}{2}\right]$
or Regime I}

\begin{figure}
\begin{centering}
\textsf{\includegraphics[clip,width=85mm]{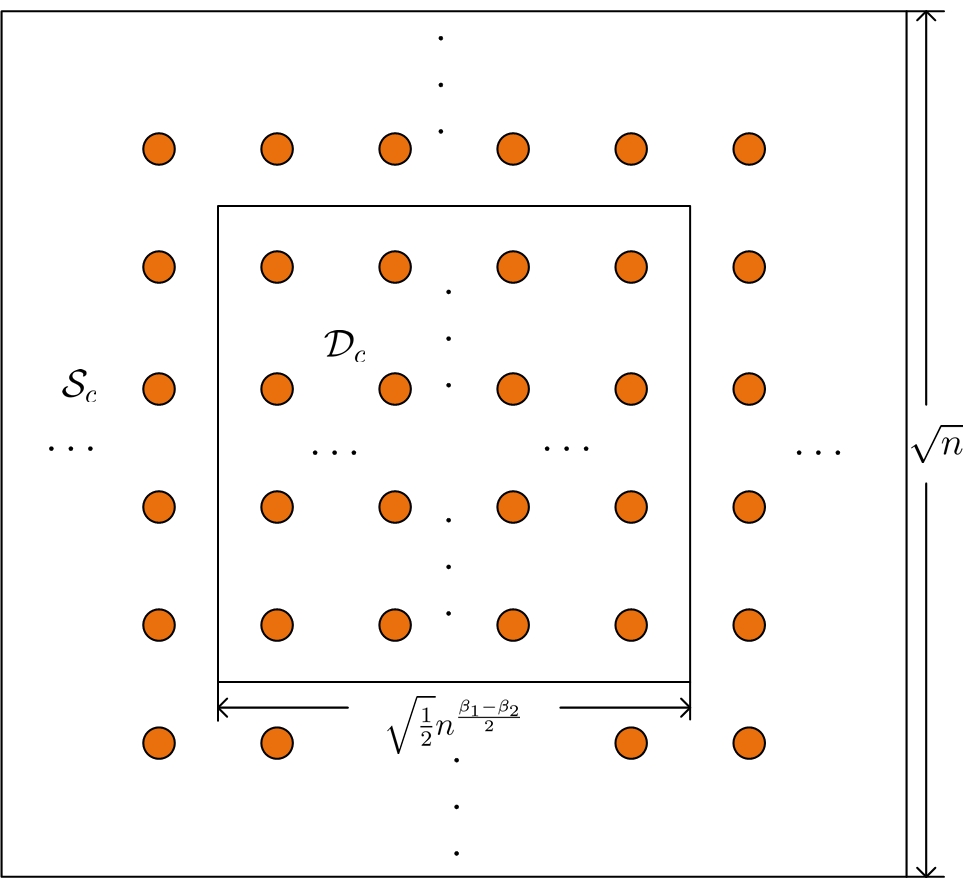}}
\par\end{centering}
\caption{\label{fig:referencesquare}{\small{}Illustration of reference square
to construct the cut set bound in Lemma \ref{lem:The-sum-capacity-c}. }}
\end{figure}
Draw a reference square at the center of the network with side length
$\sqrt{\frac{1}{2}}n^{\frac{\beta_{1}-\beta_{2}}{2}}$, as illustrated
in Fig. \ref{fig:referencesquare}. Let $\mathcal{S}_{c}$ denote
the set of nodes outside the reference square and $\mathcal{D}_{c}$
denote the set of nodes inside the reference square. The converse
result for this case can be proved by considering the cut set bound
between $\mathcal{S}_{c}$ and $\mathcal{D}_{c}$ as follows. Let
$q_{l}=I\left(\cup_{i\in\mathcal{D}_{c}}B_{i};W_{l}\right)/F$. Following
similar analysis to that in Section \ref{subsec:Regime-II-large-tau},
the $q_{l}$'s must satisfy the total cache capacity constraint $\sum_{l=1}^{L}q_{l}\leq\left|\mathcal{D}_{c}\right|L_{C}$,
where $\left|\mathcal{D}_{c}\right|=\frac{1}{2}n^{\beta_{1}-\beta_{2}}$.
Moreover, under any feasible content delivery scheme, the amount of
information $u_{i}^{j}$ transmitted from the nodes in $\mathcal{S}_{c}$
to node $i$ during the time window $\left[t_{i}^{j},...,t_{i}^{j+1}-1\right]$
must satisfy $u_{i}^{j}\geq\left(1-q_{l_{i}}-\varepsilon_{F}\right)F$.
As a result, for given $q_{l}$'s and per node rate requirement $R$,
the total average traffic rate over the cut from $\mathcal{S}_{c}$
to $\mathcal{D}_{c}$ is $\left|\mathcal{D}_{c}\right|\left(\sum_{l=1}^{L}p_{l}\left(1-q_{l}\right)-\varepsilon_{F}\right)R$.
Clearly, a per node throughput $R$ is achievable only if the induced
total average traffic rate does not exceed the sum capacity $\Gamma_{c}$
of the MIMO channel between the $\mathcal{S}_{c}$ and $\mathcal{D}_{c}$.
Hence, the achievable per node throughput is upper bounded by 
\[
R\leq\max_{\left\{ q_{l}\right\} }\frac{\Gamma_{c}}{\left|\mathcal{D}_{c}\right|\left(\sum_{l=1}^{L}p_{l}\left(1-q_{l}\right)-\varepsilon_{F}\right)},\:\textrm{s.t. }\sum_{l=1}^{L}q_{l}\leq\left\lceil \left|\mathcal{D}_{c}\right|L_{C}\right\rceil .
\]
It is easy to see that the optimal $q_{l}$'s to maximize the achievable
per node throughput upper bound is to cache the most popular $\left\lceil \left|\mathcal{D}_{c}\right|L_{C}\right\rceil $
files, i.e., $q_{l}^{\star}=1,l=1,...,\left\lceil \left|\mathcal{D}_{c}\right|L_{C}\right\rceil $
and $q_{l}^{\star}=0,l>\left\lceil \left|\mathcal{D}_{c}\right|L_{C}\right\rceil $.
As a result, we have
\begin{equation}
R\leq\frac{\Gamma_{c}}{p_{c}\left|\mathcal{D}_{c}\right|},\label{eq:RregimeII}
\end{equation}
where $p_{c}\triangleq\sum_{l=1}^{L}p_{l}\left(1-q_{l}^{\star}\right)=\sum_{l=\left\lceil \left|\mathcal{D}_{c}\right|L_{C}\right\rceil +1}^{L}p_{l}$
satisfies
\[
p_{c}=\begin{cases}
\Omega\left(1\right), & \tau\in\left[0,1\right)\\
\Omega\left(\frac{1}{\log n}\right) & \tau=1\\
\Omega\left(n^{\beta_{1}\left(1-\tau\right)}\right), & \tau>1
\end{cases}.
\]

The following lemma bounds the sum capacity $\Gamma_{c}$ between
the $\mathcal{S}_{c}$ and $\mathcal{D}_{c}$.
\begin{lem}
\label{lem:The-sum-capacity-c}The sum capacity $\Gamma_{c}$ of the
MIMO channel between the $\mathcal{S}_{c}$ and $\mathcal{D}_{c}$
is bounded as
\begin{equation}
\Gamma_{c}\leq\Theta\left(n^{\left(\beta_{1}-\beta_{2}\right)\left(2-\frac{\min\left(3,\alpha\right)}{2}\right)+\epsilon}\right),\label{eq:gamc}
\end{equation}
where $\epsilon<0$ is arbitrarily small.
\end{lem}

Please refer to Appendix \ref{subsec:Proof-of-Lemmagamc} for the
proof.

Substituting the upper bound of $\Gamma_{c}$ in (\ref{eq:gamc})
into (\ref{eq:RregimeII}) and letting $F\rightarrow\infty$, we obtain
the desired results in Theorem \ref{thm:Converse-Scaling-Law} for
Regime II with $\tau\in\left[0,\frac{\min\left(3,\alpha\right)}{2}\right]$
as well as Regime I.

\section{Conclusion\label{sec:Conclusion}}

In this paper, we combine wireless device caching and hierarchical
cooperation to significantly improve the capacity of wireless D2D
networks. Specifically, we propose a cache-induced hierarchical cooperation
scheme where the network is abstracted as a tree graph with each virtual
node in the graph representing a cluster of nodes in the network,
and the content files are cached at different levels of the tree graph
according to their popularities. The PHY has two possible modes: hierarchical
cooperation mode or multihop mode, depending on which mode yields
a better throughput. The corresponding optimal cache content placement
is formulated as an integer programming problem. We propose a low-complexity
cache content placement algorithm to solve the integer programming
problem and bound the gap w.r.t. the optimal solution. Then we analyze
the throughput performance of the cache-induced hierarchical cooperation
scheme and show that the proposed scheme achieves significant throughput
gain over the cache-assisted multihop scheme, which only supports
multihop mode in the PHY. Finally, for extended networks under Zipf
popularity distribution, we establish the per node capacity scaling
law by showing that the multiplicative gap between the achievable
per node throughput of the cache-induced hierarchical cooperation
and an upper bound of the per-node throughput is within $n^{\epsilon}$
for $\epsilon>0$ that can be arbitrarily small. When the path loss
exponent $\alpha<3$, the optimal per-node capacity scaling law in
this paper can be significantly better than that achieved by the existing
state-of-the-art schemes. To the best of our knowledge, this is the
first work that completely characterizes the per-node capacity scaling
law for wireless caching networks under the physical model and Zipf
distribution.

For clarity, we have assumed an independent phase fading channel model.
However, the achievable throughput analysis in Theorem \ref{thm:thp-bounds},
and the capacity scaling law for extended networks in Theorem \ref{thm:Achievable-Scaling-Law}
and \ref{thm:Converse-Scaling-Law} can be readily extended to the
more general PHY model. For an arbitrary PHY model, Theorem \ref{thm:thp-bounds}
and \ref{thm:Converse-Scaling-Law} still hold if we replace the coefficients
$c_{n}$ and $\gamma_{n}$ in Theorem \ref{thm:thp-bounds} (achievable
throughput bounds) and the cut set bounds of the sum capacities $\Gamma_{i}$
and $\Gamma_{c}$ in the converse proof in Section \ref{subsec:Converse-Proof}
with proper expressions under the specific PHY model. For example,
for an extended network under the free propagation model with $\alpha=2$,
the results in \cite{Chung_TIT12_CapScaling} show that $\gamma_{n}=1-\log\left(\frac{\sqrt{n}}{\lambda}\right)/\log n$,
$\Gamma_{i}=O\left(\log n\right)$ and $\Gamma_{c}=\Theta\left(\left(\frac{\sqrt{n}}{\lambda}\right)^{\beta_{1}-\beta_{2}}\right)$
as $n\rightarrow\infty$, and thus the capacity scaling law in Regime
II is given by
\[
R=\begin{cases}
\Theta\left(\lambda^{\beta_{2}-\beta_{1}}n^{\frac{\beta_{2}-\beta_{1}}{2}}\right) & \tau\in\left[0,1\right]\\
\Theta\left(\lambda^{\beta_{2}-\beta_{1}}n^{\beta_{1}\left(\tau-\frac{3}{2}\right)+\frac{\beta_{2}}{2}}\right) & \tau\in\left(1,\frac{\min\left(3,\alpha\right)}{2}\right]\\
\Theta\left(n^{\beta_{2}\left(\tau-1\right)}\right), & \tau>\frac{\min\left(3,\alpha\right)}{2}
\end{cases}.
\]
In Regime I, the capacity scaling law is still given by Theorem \ref{thm:Achievable-Scaling-Law}
and \ref{thm:Converse-Scaling-Law}. 

\appendix

\subsection{Proof of Theorem \ref{thm:Optimality-Condition-of} \label{subsec:Proof-of-TheoremOptCond}}

First, we show that if $\left(\mathbf{x}^{*},R^{*}\right)$ is the
optimal solution of Problem (\ref{eq:cachePmain-1}), it must satisfy
the conditions in Theorem \ref{thm:Optimality-Condition-of}. The
conditions $R\leq\overline{C}_{m^{*}}$ and (\ref{eq:sumL}-\ref{eq:cachesize})
are the constraints in Problem (\ref{eq:cachePmain-1}). Therefore,
we only need to prove that $\left(\mathbf{x}^{*},R^{*}\right)$ satisfy
the first condition in (\ref{eq:Requ}). Suppose there exist $m^{'}\in\left\{ m^{*}+1,...,M\right\} $
such that $f\left(\sum_{i=m^{*}}^{m^{'}-1}x_{i}^{*}+1\right)R^{*}<\overline{C}_{m^{'}}$.
Then we can find another feasible solution $\mathbf{x}$ to strictly
improve the objective function $R$ as follows. Let $x_{m^{'}}=x_{m^{'}}^{*}+\varepsilon$,
where $\varepsilon>0$ is a sufficiently small number. Let $x_{m^{'}-1}=x_{m^{'}-1}^{*}-\varepsilon$,
$x_{i}=x_{i}^{*}-\varepsilon^{'},\forall i\notin\left\{ 0,1,...,m^{*}-1\right\} \cup\left\{ m^{'}-1,m^{'}\right\} $,
and $x_{0}=x_{0}+\sum_{i\notin\left\{ 0,1,...,m^{*}-1\right\} \cup\left\{ m^{'}-1,m^{'}\right\} }\varepsilon^{'}$,
where $\varepsilon^{'}>0$ is a sufficiently small number compared
to $\varepsilon$. It can be verified that $\sum_{i=m^{*}}^{m-1}x_{i}>\sum_{i=m^{*}}^{m-1}x_{i}^{*},\forall m\in\left\{ m^{*},...,M\right\} \backslash\left\{ m^{'}\right\} $
and $\sum_{i=m^{*}}^{m^{'}-1}x_{i}>\sum_{i=m^{*}}^{m^{'}-1}x_{i}^{*}-\varepsilon$.
Since $f\left(x\right)$ is a strictly decreasing function of $x$
with a bounded derivative, for sufficiently small $\varepsilon$,
we have $f\left(\sum_{i=m^{*}}^{m-1}x_{i}+1\right)R^{*}<f\left(\sum_{i=m^{*}}^{m-1}x_{i}^{*}+1\right)R^{*}\leq\overline{C}_{m},\forall m\in\left\{ m^{*},...,M\right\} \backslash m^{'}$
and $f\left(\sum_{i=m^{*}}^{m^{'}-1}x_{i}+1\right)R^{*}<f\left(\sum_{i=m^{*}}^{m^{'}-1}x_{i}+1-\varepsilon\right)R^{*}<\overline{C}_{m^{'}}$.
Therefore, we can strictly increase $R$ without violating any constraints.
Hence, at the optimal solution $\left(\mathbf{x}^{*},R^{*}\right)$,
the condition in (\ref{eq:Requ}) must be satisfied.

In the following, we show that the solution to the conditions in Theorem
\ref{thm:Optimality-Condition-of} is unique, which implies that these
conditions are also sufficient for $\left(\mathbf{x}^{*},R^{*}\right)$
to be the optimal solution of Problem (\ref{eq:cachePmain-1}). Specifically,
it can be shown that 
\begin{align}
L_{m}\left(\overline{C}_{m}\right) & >L_{m}\left(\overline{C}_{m+1}\right)=L_{m+1}\left(\overline{C}_{m+1}\right)\nonumber \\
L_{m}\left(\overline{C}_{m}\right) & =L_{m-1}\left(\overline{C}_{m}\right)<L_{m-2}\left(\overline{C}_{m-1}\right).\label{eq:Lminequ}
\end{align}
If $m^{*}$ is the solution to the optimality conditions, we must
have $L_{m^{*}}\left(\overline{C}_{m^{*}+1}\right)<L_{C}$ and $L_{m^{*}}\left(\overline{C}_{m^{*}}\right)\geq L_{C}$.
Then it follows from (\ref{eq:Lminequ}) that $L_{m}\left(\overline{C}_{m}\right)<L_{C},\forall m>m^{*}$
and $L_{m}\left(\overline{C}_{m+1}\right)\geq L_{C},\forall m<m^{*}$,
which implies any $m\neq m^{*}$ cannot satisfy the two conditions
$L_{m}\left(\overline{C}_{m+1}\right)<L_{C}$ and $L_{m}\left(\overline{C}_{m}\right)\geq L_{C}$
simultaneously. Therefore, the solution to the conditions in Theorem
\ref{thm:Optimality-Condition-of} is unique.

\subsection{Reformulation of the Problem (\ref{eq:cachePmain}) \label{subsec:Reformulation-of-the}}

We reformulate Problem (\ref{eq:cachePmain}) as a ZOLP for fixed
$R$ as follows. Define the binary variables $\ensuremath{\delta_{m,l}\in\{0,1\}}$
and the $\ensuremath{(M+2)\times L}$ matrix $\boldsymbol{\triangle}\ensuremath{=[\delta_{m,l}]}$
formed as follows: the first row is the all-one vector, that is, $\ensuremath{\delta_{m,l}=1}$
for all $\ensuremath{l=1,\ldots,L}$. The last row is the all-zero
vector, that is, $\ensuremath{\delta_{M+1,l}=0}$ for all $\ensuremath{l=1,\ldots,L}$.
The remaining rows between $m=1$ and $m=M$ satisfy the following
monotonicity conditions on the rows and on the columns:
\begin{align*}
\delta_{m,l} & \leq\delta_{m,l+1},\:m=1,...,M,\:l=1,...,L,\\
\delta_{m,l} & \geq\delta_{m+1,l},\:m=1,...,M,\:l=1,...,L.
\end{align*}
In words, the matrix $\boldsymbol{\triangle}$ has monotonically non-decreasing
rows, and monotonically non-increasing columns. Since the $\ensuremath{\delta}$-variables
are binary, this means that each row of $\boldsymbol{\triangle}$
is formed by a leading block of zeros followed by a block of ones,
and each column of $\boldsymbol{\triangle}$ is formed by a leading
block of ones followed by a block of zeros. 

\textbf{Observation 1:} For $\ensuremath{m=0,\ldots,M}$, we let 
\[
x_{m}=\sum_{l=1}^{L}\delta_{m,l}-\sum_{l=1}^{L}\delta_{m+1,l}.
\]
Hence, the link capacity constraint (\ref{eq:rateconmain}) can be
written as a linear constraint with respect to the variables $\delta_{m,l}$
for fixed $\ensuremath{R}$ as follows
\[
\boldsymbol{\triangle}\mathbf{p}R\leq\mathbf{c},
\]
where $\mathbf{p}$ is the $L\times1$ vectors containing the file
request distribution $\{p_{l}\}$, and $\mathbf{c}$ is a $\ensuremath{(M+2)\times1}$
vector with the first element $\ensuremath{c_{0}=R}$, the last element
$\ensuremath{c_{M+1}=0}$, and elements $\ensuremath{m=1,\ldots,M}$
equal to $\ensuremath{c_{m}=C_{m}4^{-(m-1)}}$. 

\textbf{Observation 2:} We define row-differential matrix of dimensions
$\ensuremath{(M+1)\times(M+2)}$ given by
\[
\mathbf{D}=\left[\begin{array}{cccccc}
1 & -1 & 0 & \cdots &  & 0\\
0 & 1 & -1 & 0 & \cdots & 0\\
\vdots &  & \ddots & \ddots &  & \vdots\\
0 & \cdots & 0 & 1 & -1 & 0\\
0 &  & \cdots & 0 & 1 & -1
\end{array}\right].
\]
Then, it is not difficult to see that the cache size constraint (\ref{eq:cachesizeconmain})
can be rewritten as
\[
\mathbf{w}^{T}\mathbf{D}\boldsymbol{\triangle}\boldsymbol{1}\leq L_{C},
\]
where $\mathbf{w}$ is a $\ensuremath{(M+1)\times1}$ weight vector
with elements $w_{m}=4^{-m}$ and $\boldsymbol{1}$ is the all-one
column vector of dimension $L\times1$. 

\textbf{Observation 3:} The constraint (\ref{eq:cacheallLmain}) with
this new parameterization of the problem becomes irrelevant since
it is automatically imposed by the side of the matrix $\boldsymbol{\triangle}$. 

It follows that the problem re-parameterized in the binary variables
$\ensuremath{\delta_{m,l}}$ can be written as
\begin{align}
\max_{R,\boldsymbol{\triangle}}\: & R\label{eq:ZOLP}\\
\text{s.t.}\: & \boldsymbol{\triangle}\mathbf{p}R\leq\mathbf{c},\nonumber \\
 & \mathbf{w}^{T}\mathbf{D}\boldsymbol{\triangle}\boldsymbol{1}\leq L_{C},\nonumber \\
 & \delta_{m,l}\leq\delta_{m,l+1},\:m=1,...,M,\:l=1,...,L,\nonumber \\
 & \delta_{m,l}\geq\delta_{m+1,l},\:m=1,...,M,\:l=1,...,L,\nonumber \\
 & \delta_{0,l}=1,\:l=1,...,L,\nonumber \\
 & \delta_{m+1,l}=0,\:l=1,...,L.\nonumber 
\end{align}
This is a ZOLP feasibility problem for any fixed value of $R$. Therefore,
it is possible to use standard ZOLP solvers for fixed $\ensuremath{R}$,
and perform a bisection search over $R\in[0,R_{\max}]$, where $R_{\max}$
is some upper bound on the per node throughput that can be found by,
e.g., relaxing the binary constraint on $\ensuremath{\delta_{m,l}}$
to $\ensuremath{\delta_{m,l}}\in\left[0,1\right]$. 

\subsection{Proof of Lemma \ref{lem:Performance-loss-Integer} \label{subsec:Proof-of-Lemmaloss}}

After steps 2a and 2b of Algorithm \ref{alg:cachebisec}, at the $m^{*}$
level, we have $x_{m^{*}}^{o}=\left\lfloor x_{m^{*}}^{*}\right\rfloor <x_{m^{*}}^{*}$,
and thus
\begin{align*}
\sum_{i=m^{*}}^{M}x_{i}^{\circ} & =\sum_{i=m^{*}}^{M}x_{i}^{*}=L,\\
\sum_{i=m^{*}+1}^{M}x_{i}^{\circ} & =L-\left\lfloor x_{m^{*}}^{*}\right\rfloor <\sum_{i=m^{*}+1}^{M}x_{i}^{*}+1.
\end{align*}
At the $m^{*}+1$ level, we have $x_{m^{*}+1}^{o}=\left\lfloor x_{m*+1}^{*}+b_{m^{*}}4^{m^{*}+1}\right\rfloor >\left\lfloor x_{m*+1}^{*}+4\left(x_{m^{*}}^{*}-x_{m^{*}}^{o}\right)\right\rfloor $,
and thus
\begin{align*}
\sum_{i=m^{*}+2}^{M}x_{i}^{\circ} & =L-\left\lfloor x_{m*+1}^{*}+4\left(x_{m^{*}}^{*}-x_{m^{*}}^{o}\right)\right\rfloor -\left\lfloor x_{m^{*}}^{*}\right\rfloor \\
 & <\sum_{i=m^{*}+2}^{M}x_{i}^{*}+1.
\end{align*}
Similarly, it can be shown that 
\begin{align}
\sum_{i=m}^{M}x_{i}^{\circ} & <\sum_{i=m}^{M}x_{i}^{*}+1,\forall m=m^{*},...,M.\label{eq:sum0lessumxstar}
\end{align}
 Let $M^{\circ}=\textrm{argmax}_{m}x_{m}^{\circ},\:\textrm{s.t. }x_{m}^{\circ}>0$
and $m^{\circ}=\textrm{argmin}_{m}x_{m}^{\circ},\:\textrm{s.t. }x_{m}^{\circ}\geq1$,
where $x_{m}^{\circ}$ is the cache content placement parameter after
steps 2a and 2b. It follows from (\ref{eq:sum0lessumxstar}) that
$\sum_{i=m^{\circ}}^{M}x_{i}^{*}>\sum_{i=m^{\circ}}^{M}x_{i}^{\circ}-1=L-1$,
and thus
\begin{equation}
R^{*}=\overline{C}_{m^{\circ}}/f\left(\sum_{i=m^{*}}^{m^{\circ}-1}x_{i}^{*}+1\right)\leq\overline{C}_{m^{\circ}}/f\left(2\right).\label{eq:Rtareq}
\end{equation}
There are three cases as follows:

\subsubsection*{Case 1: $M^{\circ}-m^{\circ}=0$}

In this case, the achievable throughput after steps 2a and 2b is
\[
R^{\circ}=\overline{C}_{M^{\circ}}/f\left(1\right).
\]
It follows from (\ref{eq:sum0lessumxstar}) that $\sum_{i=M^{\circ}}^{M}x_{i}^{*}>\sum_{i=M^{\circ}}^{M}x_{i}^{\circ}-1=L-1$,
and thus
\[
R^{*}=\overline{C}_{M^{\circ}}/f\left(\sum_{i=m^{*}}^{M^{\circ}-1}x_{i}^{*}+1\right)\leq\overline{C}_{M^{\circ}}/f\left(2\right).
\]
Therefore, 
\begin{equation}
R^{\circ}/R^{*}=f\left(2\right)/f\left(1\right)\geq1/\left(1+2^{\tau}\right).\label{eq:Roratio}
\end{equation}

\subsubsection*{Case 2: $M^{\circ}-m^{\circ}=1$}

In this case, the achievable throughput after steps 2a and 2b is 
\begin{align}
R^{\circ} & =\min\left(\frac{\overline{C}_{m^{\circ}}}{f\left(1\right)},\frac{\overline{C}_{m^{\circ}+1}}{f\left(\sum_{i=m^{*}}^{m^{\circ}}x_{i}^{\circ}+1\right)}\right)\nonumber \\
 & \geq\frac{\overline{C}_{m^{\circ}+1}}{f\left(1\right)}.\label{eq:Romin1}
\end{align}
From (\ref{eq:Rtareq}), we have 
\begin{align*}
\frac{\overline{C}_{m^{\circ}}}{f\left(1\right)R^{*}} & \geq\frac{f\left(2\right)}{f\left(1\right)}\geq1/\left(1+2^{\tau}\right).
\end{align*}
If $x_{m^{\circ}}^{\circ}<L-1$, we have $\sum_{i=m^{*}}^{m^{\circ}}x_{i}^{\circ}+2\leq L$,
and thus
\begin{align*}
\frac{\overline{C}_{m^{\circ}+1}}{f\left(\sum_{i=m^{*}}^{m^{\circ}}x_{i}^{\circ}+1\right)R^{*}} & =\frac{f\left(\sum_{i=m^{*}}^{m^{\circ}}x_{i}^{*}+1\right)}{f\left(\sum_{i=m^{*}}^{m^{\circ}}x_{i}^{\circ}+1\right)}\\
 & \geq\frac{f\left(\sum_{i=m^{*}}^{m^{\circ}}x_{i}^{\circ}+2\right)}{f\left(\sum_{i=m^{*}}^{m^{\circ}}x_{i}^{\circ}+1\right)}\geq\frac{1}{1+2^{\tau}}.
\end{align*}
If $x_{m^{\circ}}^{\circ}=L-1$, 
\begin{align*}
\frac{\overline{C}_{m^{\circ}+1}}{f\left(\sum_{i=m^{*}}^{m^{\circ}}x_{i}^{\circ}+1\right)R^{*}} & =\frac{f\left(2\right)}{4f\left(L\right)}\geq\frac{1}{1+2^{\tau}},
\end{align*}
for $L\geq2$. From the above analysis, we have
\begin{equation}
R^{\circ}/R^{*}=\frac{1}{4}f\left(2\right)/f\left(1\right)\geq1/\left(1+2^{\tau}\right).\label{eq:Roration2}
\end{equation}

\subsubsection*{Case 3: $M^{\circ}-m^{\circ}>1$}

In this case, after Step 2f is performed for the first time, the achievable
throughput under the cache content placement parameter $\mathbf{x}^{'}$
is given by
\[
R^{'}=\min_{m^{\circ}\leq m\leq M^{\circ}}\left(\frac{\overline{C}_{m}}{f\left(\sum_{i=m^{*}}^{m-1}x_{i}^{'}+1\right)}\right).
\]
If $x_{M^{\circ}}^{\circ}<3$, we have $x_{M^{\circ}}^{\circ}=0$,
and thus $\frac{\overline{C}_{M^{\circ}}}{f\left(\sum_{i=m^{*}}^{M^{\circ}-1}x_{i}^{'}+1\right)}=\frac{\overline{C}_{M^{\circ}}}{f\left(L+1\right)}=+\infty$.
Otherwise, $x_{M^{\circ}}^{\circ}\geq3$ and 
\begin{align}
\frac{\overline{C}_{M^{\circ}}}{f\left(\sum_{i=m^{*}}^{M^{\circ}-1}x_{i}^{'}+1\right)R^{*}} & \geq\frac{\overline{C}_{M^{\circ}}}{f\left(\sum_{i=m^{*}}^{M^{\circ}-1}x_{i}^{\circ}+1\right)R^{*}}\nonumber \\
 & =\frac{f\left(\sum_{i=m^{*}}^{M^{\circ}-1}x_{i}^{*}+1\right)}{f\left(\sum_{i=m^{*}}^{M^{\circ}-1}x_{i}^{\circ}+1\right)}\nonumber \\
 & \geq\frac{f\left(L-x_{M^{\circ}}^{\circ}+2\right)}{f\left(L-x_{M^{\circ}}^{\circ}+1\right)}\geq\frac{1}{1+2^{\tau}}.\label{eq:Moratio}
\end{align}
For $m=m^{\circ}+1$, we have
\begin{align}
\frac{\overline{C}_{m^{\circ}+1}}{f\left(\sum_{i=m^{*}}^{m^{\circ}}x_{i}^{'}+1\right)R^{*}} & =\frac{\overline{C}_{m^{\circ}+1}}{f\left(\sum_{i=m^{*}}^{m^{\circ}}x_{i}^{\circ}\right)R^{*}}\nonumber \\
 & =\frac{f\left(\sum_{i=m^{*}}^{m^{\circ}}x_{i}^{*}+1\right)}{f\left(\sum_{i=m^{*}}^{m^{\circ}}x_{i}^{\circ}\right)}\nonumber \\
 & \geq\frac{f\left(\sum_{i=m^{*}}^{m^{\circ}}x_{i}^{\circ}+2\right)}{f\left(\sum_{i=m^{*}}^{m^{\circ}}x_{i}^{\circ}\right)}\geq\frac{1}{1+2^{\tau}}.\label{eq:moration}
\end{align}
Similarly, it can be shown that 
\[
\frac{\overline{C}_{m}}{f\left(\sum_{i=m^{*}}^{m-1}x_{i}^{'}+1\right)}>\frac{1}{1+2^{\tau}},m=m^{\circ},...,M^{\circ},
\]
from which it follows that $R^{'}/R^{*}\geq\frac{1}{1+2^{\tau}}$. 

Since steps 2c to 2g do not decrease the achievable throughput, it
follows from the above analysis that $R^{\circ}/R^{*}\geq1/\left(1+2^{\tau}\right)$
also holds after the termination of the algorithm. Finally, the additional
factor of $\frac{1}{M}$ is because $\frac{1}{M}\overline{C}_{m}4^{\left(m-1\right)}\leq C_{m}\leq\overline{C}_{m}4^{\left(m-1\right)}$
and we have used the upper bound $\overline{C}_{m}4^{\left(m-1\right)}$
in the relaxed cache content placement optimization problem in (\ref{eq:cachePmain-1}).

\subsection{Proof of Theorem \ref{thm:thp-bounds} \label{subsec:Proof-of-Theoremtpbounds}}

We first give some useful lemmas. The following lemma follows immediately
from the optimality condition in Theorem \ref{thm:Optimality-Condition-of}. 
\begin{lem}
\label{lem:Mainlemmafbar}Let $0\leq\overline{f}_{L}\left(R,m^{*}\right)\leq L_{m^{*}}\left(R\right),R\in\left(\overline{C}_{m^{*}+1},\overline{C}_{m^{*}}\right]$
and $\overline{f}_{U}\left(R,m^{*}\right)\geq L_{m^{*}}\left(R\right),R\in\left(\overline{C}_{m^{*}+1},\overline{C}_{m^{*}}\right]$
be some lower bound and upper bound of $L_{m^{*}}\left(R\right)$,
respectively. For any $R_{U}$ that satisfies $\overline{f}_{L}\left(R_{U},m^{*}\right)\geq L_{C},R_{U}\in\left(\overline{C}_{m^{*}+1},\overline{C}_{m^{*}}\right]$
for some $m^{*}\in\mathbb{Z}_{+}$, we have $R_{U}\geq R^{*}$, where
$R^{*}$ is the optimal solution of the relaxed cache content placement
optimization problem in (\ref{eq:cachePmain-1}). And for any $R_{L}$
that satisfies $\overline{f}_{U}\left(R_{L},m^{*}\right)\leq L_{C},R_{L}\in\left(\overline{C}_{m^{*}+1},\overline{C}_{m^{*}}\right]$
for some $m^{*}\in\mathbb{Z}_{+}$, we have $R_{L}\leq R^{*}$. 
\end{lem}

The following lemma gives closed-form bounds for $f^{-1}\left(\frac{\overline{C}_{m}}{R}\right)$.
\begin{lem}
\label{lem:finvbounds}For different regions of $\tau$, $f^{-1}\left(\frac{\overline{C}_{m}}{R}\right)$
can be lower bounded as
\begin{align*}
f^{-1}\left(\frac{\overline{C}_{m}}{R}\right) & \geq\left(1-\frac{\overline{C}_{m}}{\left(1-\tau\right)R}\right)^{+},\tau\in\left[0,1\right)\\
f^{-1}\left(\frac{\overline{C}_{m}}{R}\right) & \geq e^{-1}L^{1-\frac{\overline{C}_{m}}{R}},\tau=1\\
f^{-1}\left(\frac{\overline{C}_{m}}{R}\right) & \geq2^{1-\tau}\min\left(\left(\frac{\overline{C}_{m}\tau}{R}\right)^{\frac{1}{1-\tau}},L+1\right),\tau>1.
\end{align*}
and upper bounded as
\begin{align*}
f^{-1}\left(\frac{\overline{C}_{m}}{R}\right) & \leq1-\frac{\overline{C}_{m}}{R},\tau\in\left[0,1\right)\\
f^{-1}\left(\frac{\overline{C}_{m}}{R}\right) & \leq e^{2}L,\tau=1\\
f^{-1}\left(\frac{\overline{C}_{m}}{R}\right) & \leq\min\left(\left(\frac{\overline{C}_{m}}{\tau R}\right)^{\frac{1}{1-\tau}},L+1\right)+1,\tau>1.
\end{align*}
\end{lem}

\begin{IEEEproof}
The upper bound follows from the fact that $f\left(x\right)\geq\frac{\int_{x}^{L+1}z^{-\tau}dz}{\int_{1}^{L}z^{-\tau}dz+1}$
and the lower bound follows from the fact that $f\left(x\right)\leq\frac{\int_{\left\lfloor x\right\rfloor }^{L+1}z^{-\tau}dz+\left\lfloor x\right\rfloor ^{-1}}{\int_{1}^{L+1}z^{-\tau}dz}$.
The detailed calculations are omitted for conciseness.
\end{IEEEproof}

With the above lemmas, we are ready to prove Theorem \ref{thm:thp-bounds}.
The proof contains five cases depending on the value of $\tau$.

\subsubsection*{Case 1: $\tau\in\left[0,1\right)$}

We first prove the lower bound. Replace $f^{-1}\left(\frac{\overline{C}_{m}}{R}\right)$
in $L_{m^{*}}\left(R\right)$ with the upper bound of $f^{-1}\left(\frac{\overline{C}_{m}}{R}\right)$
for $\tau\in\left[0,1\right)$ in Lemma \ref{lem:finvbounds}, and
we obtain an upper bound of $L_{m^{*}}\left(R\right)$ as 
\[
\overline{f}_{U}\left(R,m^{*}\right)=L\left(4-\frac{3c_{n}4^{\gamma_{n}+1}4^{-\gamma_{n}\left(m^{*}+1\right)}}{R\left(4^{\gamma_{n}+1}-1\right)}\right)4^{-\left(m^{*}+1\right)}.
\]
If $\frac{L_{C}}{L}>1-\frac{3\cdot4^{\gamma_{n}}}{4^{\gamma_{n}+1}-1}$,
it can be verified that $R_{L}^{a}=\frac{3c_{n}}{\left(4^{\gamma_{n}+1}-1\right)\left(1-\frac{L_{C}}{L}\right)}$
and $m^{*}=0$ satisfies $\overline{f}_{U}\left(R_{L}^{a},m^{*}\right)\leq L_{C},R_{L}^{a}\in\left(\overline{C}_{m^{*}+1},\overline{C}_{m^{*}}\right]$.
On the other hand, if $\frac{L_{C}}{L}\leq1-\frac{3\cdot4^{\gamma_{n}}}{4^{\gamma_{n}+1}-1}$,
it can be verified that $R_{L}^{b}=c_{n}\left(4-\frac{12}{4^{\gamma_{n}+1}-1}\right)^{-\gamma_{n}}\left(\frac{L_{C}}{L}\right)^{\gamma_{n}}$
and $m^{*}=\left\lfloor \frac{1}{\gamma_{n}}\log_{4}\frac{c_{n}}{R_{L}}\right\rfloor $
satisfies $\overline{f}_{U}\left(R_{L}^{b},m^{*}\right)\leq L_{C},R_{L}^{b}\in\left(\overline{C}_{m^{*}+1},\overline{C}_{m^{*}}\right]$.
Then from Lemma \ref{lem:Mainlemmafbar}, the lower bound given in
Theorem \ref{thm:thp-bounds} is valid for $\tau\in\left[0,1\right)$.

Then we prove the upper bound. Replace $f^{-1}\left(\frac{\overline{C}_{m}}{R}\right)$
in $L_{m^{*}}\left(R\right)$ with the lower bound of $f^{-1}\left(\frac{\overline{C}_{m}}{R}\right)$
for $\tau\in\left[0,1\right)$ in Lemma \ref{lem:finvbounds}, and
we obtain a lower bound of $L_{m^{*}}\left(R\right)$ as
\begin{align*}
\overline{f}_{L}\left(R,m^{*}\right) & =L\left(1-\frac{3\cdot4^{\gamma_{n}}}{\left(4^{\gamma_{n}+1}-1\right)}\right)\left(1-\tau\right)^{\frac{1}{\gamma_{n}}}\left(\frac{R}{c_{n}}\right)^{\frac{1}{\gamma_{n}}}\\
 & \geq L\left(1-\frac{3\cdot4^{\gamma_{n}}}{\left(4^{\gamma_{n}+1}-1\right)}\right)\left(1-\tau\right)^{\frac{1}{\gamma_{n}}}\left(\frac{R}{c_{n}}\right)^{\frac{1}{\gamma_{n}}},
\end{align*}
where the last inequality follows from $4^{-m^{*}}<4\left(\frac{R}{c_{n}}\right)^{\frac{1}{\gamma_{n}}}$
since $R>\overline{C}_{m^{*}+1}$. Let $L\left(1-\frac{3\cdot4^{\gamma_{n}}}{\left(4^{\gamma_{n}+1}-1\right)}\right)\left(1-\tau\right)^{\frac{1}{\gamma_{n}}}\left(\frac{R_{U}}{c_{n}}\right)^{\frac{1}{\gamma_{n}}}=L_{C}$,
and we have
\[
R_{U}=\frac{c_{n}}{1-\tau}\left(1-\frac{3\cdot4^{\gamma_{n}}}{\left(4^{\gamma_{n}+1}-1\right)}\right)^{-\gamma_{n}}\left(\frac{L_{C}}{L}\right)^{\gamma_{n}}.
\]
Clearly, the above $R_{U}$ and $m^{*}=\left\lfloor \frac{1}{\gamma_{n}}\log_{4}\frac{c_{n}}{R_{U}}\right\rfloor ^{+}$
satisfy $\overline{f}_{L}\left(R_{U},m^{*}\right)>L_{C},R_{U}\in\left(\overline{C}_{m^{*}+1},\overline{C}_{m^{*}}\right]$.
Then from Lemma \ref{lem:Mainlemmafbar}, the upper bound $R_{U}$
given in Theorem \ref{thm:thp-bounds} is valid for $\tau\in\left[0,1\right)$.

\subsubsection*{Case 2: $\tau=1$}

We first prove the lower bound. Replace $f^{-1}\left(\frac{\overline{C}_{m}}{R}\right)$
in $L_{m^{*}}\left(R\right)$ with the upper bound of $f^{-1}\left(\frac{\overline{C}_{m}}{R}\right)$
for $\tau=1$ in Lemma \ref{lem:finvbounds}, and we obtain an upper
bound of $L_{m^{*}}\left(R\right)$ as
\begin{align*}
\overline{f}_{U}\left(R,m^{*}\right) & =\left(e^{2}L-1\right)4^{-m^{*}}-\left(e^{2}L-L-1\right)4^{-M}\\
 & \leq\left(e^{2}L-1\right)4\left(\frac{R}{c_{n}}\right)^{\frac{1}{\gamma_{n}}}-\left(e^{2}L-L-1\right)4^{-M},
\end{align*}
where the last inequality follows from $4^{-m^{*}}<4\left(\frac{R}{c_{n}}\right)^{\frac{1}{\gamma_{n}}}$.
Let $\left(e^{2}L-1\right)4\left(\frac{R_{L}}{c_{n}}\right)^{\frac{1}{\gamma_{n}}}-\left(e^{2}L-L-1\right)4^{-M}=L_{C}$,
and we have
\[
R_{L}=c_{n}\left(\frac{\left(e^{2}L-L-1\right)4^{-M}+L_{C}}{4\left(e^{2}L-1\right)}\right)^{\gamma_{n}}.
\]
Clearly, the above $R_{L}$ and $m^{*}=\left\lfloor \frac{1}{\gamma_{n}}\log_{4}\frac{c_{n}}{R_{L}}\right\rfloor ^{+}$
satisfy $\overline{f}_{U}\left(R_{L},m^{*}\right)\leq L_{C},R_{L}\in\left(\overline{C}_{m^{*}+1},\overline{C}_{m^{*}}\right]$.
Then from Lemma \ref{lem:Mainlemmafbar}, the lower bound $\frac{1}{M\left(1+2^{\tau}\right)}R_{L}$
given in Theorem \ref{thm:thp-bounds} is valid for $\tau=1$.

Then we prove the upper bound. Replace $f^{-1}\left(\frac{\overline{C}_{m}}{R}\right)$
in $L_{m^{*}}\left(R\right)$ with the lower bound of $f^{-1}\left(\frac{\overline{C}_{m}}{R}\right)$
for $\tau=1$ in Lemma \ref{lem:finvbounds}, and we obtain a lower
bound of $L_{m^{*}}\left(R\right)$ as
\begin{align}
\overline{f}_{L}\left(R,m^{*}\right) & =\frac{3L}{e}\sum_{m=m^{*}+1}^{M}\frac{L^{-\frac{c_{n}}{R}4^{-m\gamma_{n}}}}{4^{m}}-\left(L+1\right)4^{-M}\nonumber \\
 & \geq\frac{3L}{e}\left(\frac{R}{c_{n}}\right)^{\frac{1}{\gamma_{n}}}\sum_{m=1}^{M-m}4^{-m}L^{-4^{-\gamma_{n}\left(m-1\right)}}-\left(L+1\right)4^{-M},\label{fbarLtau2}
\end{align}
where the last inequality follows from $\frac{c_{n}}{R}<4^{\left(m^{*}+1\right)\gamma_{n}}$
and $4^{-m^{*}}\geq\left(\frac{R}{c_{n}}\right)^{\frac{1}{\gamma_{n}}}$
since $R\in\left(\overline{C}_{m^{*}+1},\overline{C}_{m^{*}}\right]$.
It can be shown that $\max_{m}4^{-m}L^{-4^{-\gamma_{n}\left(m-1\right)}}\geq\frac{\left(\ln L\right)^{-\frac{1}{\gamma_{n}}}L^{-\frac{1}{\ln L}}}{4}$,
from which it follows that 
\begin{equation}
\overline{f}_{L}\left(R,m^{*}\right)\geq\frac{3L}{e}\left(\frac{R}{c_{n}}\right)^{\frac{1}{\gamma_{n}}}\frac{\left(\ln L\right)^{-\frac{1}{\gamma_{n}}}L^{-\frac{1}{\ln L}}}{4}-\left(L+1\right)4^{-M}.\label{eq:fbartau2a}
\end{equation}
Let $\frac{3L}{e}\left(\frac{R_{U}}{c_{n}}\right)^{\frac{1}{\gamma_{n}}}\frac{\left(\ln L\right)^{-\frac{1}{\gamma_{n}}}L^{-\frac{1}{\ln L}}}{4}-\left(L+1\right)4^{-M}=L_{C}$,
and we have 
\[
R_{U}=c_{n}\left(\frac{4e}{3}\right)^{\gamma_{n}}L^{\frac{\gamma_{n}}{\ln L}}\ln L\left(\frac{L_{C}}{L}+\frac{1}{4^{M}}\left(1+\frac{1}{L}\right)\right)^{\gamma_{n}}.
\]
Clearly, the above $R_{U}$ and $m^{*}=\left\lfloor \frac{1}{\gamma_{n}}\log_{4}\frac{c_{n}}{R_{U}}\right\rfloor ^{+}$
satisfy $\overline{f}_{L}\left(R_{U},m^{*}\right)>L_{C},R_{U}\in\left(\overline{C}_{m^{*}+1},\overline{C}_{m^{*}}\right]$.
Then from Lemma \ref{lem:Mainlemmafbar}, the upper bound $R_{U}$
given in Theorem \ref{thm:thp-bounds} is valid for $\tau=1$.

\subsubsection*{Case 3: $\tau\in\left(1,\gamma_{n}+1\right)$}

We first prove the lower bound. Replace $f^{-1}\left(\frac{\overline{C}_{m}}{R}\right)$
in $L_{m^{*}}\left(R\right)$ with the upper bound of $f^{-1}\left(\frac{\overline{C}_{m}}{R}\right)$
for $\tau>1$ in Lemma \ref{lem:finvbounds}, and we obtain an upper
bound of $L_{m^{*}}\left(R\right)$ as
\[
\overline{f}_{U}\left(R,m^{*}\right)=\frac{\left(4^{\frac{\gamma_{n}+\tau-1}{\tau-1}}-4\right)L^{\frac{1+\gamma_{n}-\tau}{\gamma_{n}}}\left(\frac{\tau R}{c_{n}}\right)^{\frac{1}{\gamma_{n}}}}{\left(4^{\frac{1+\gamma_{n}-\tau}{\tau-1}}-1\right)}.
\]
Let $\overline{f}_{U}\left(R_{L},m^{*}\right)=L_{C}$, and we have
\[
R_{L}=\frac{c_{n}}{\tau}\left(\frac{4^{\frac{1+\gamma_{n}-\tau}{\tau-1}}-1}{4^{\frac{\gamma_{n}+\tau-1}{\tau-1}}-4}\right)^{\gamma_{n}}L_{C}^{\gamma_{n}}L^{\tau-1-\gamma_{n}}.
\]
Clearly, the above $R_{L}$ and $m^{*}=\left\lfloor \frac{1}{\gamma_{n}}\log_{4}\frac{c_{n}}{R_{L}}\right\rfloor ^{+}$
satisfy $\overline{f}_{U}\left(R_{L},m^{*}\right)\leq L_{C},R_{L}\in\left(\overline{C}_{m^{*}+1},\overline{C}_{m^{*}}\right]$.
Then from Lemma \ref{lem:Mainlemmafbar}, the lower bound $\frac{1}{M\left(1+2^{\tau}\right)}R_{L}$
given in Theorem \ref{thm:thp-bounds} is valid for $\tau\in\left(1,\gamma_{n}+1\right)$.

Then we prove the upper bound. Replace $f^{-1}\left(\frac{\overline{C}_{m}}{R}\right)$
in $L_{m^{*}}\left(R\right)$ with the lower bound of $f^{-1}\left(\frac{\overline{C}_{m}}{R}\right)$
for $\tau>1$ in Lemma \ref{lem:finvbounds}, and we obtain a lower
bound of $L_{m^{*}}\left(R\right)$ as
\begin{align*}
\overline{f}_{L}\left(R,m^{*}\right) & =\frac{L^{\frac{1+\gamma_{n}-\tau}{\gamma_{n}}}\left(\frac{\tau R}{c_{n}}\right)^{\frac{1}{\gamma_{n}}}}{2^{\tau-1}}-4^{-m^{*}}\\
 & \geq\frac{L^{\frac{1+\gamma_{n}-\tau}{\gamma_{n}}}\left(\frac{\tau R}{c_{n}}\right)^{\frac{1}{\gamma_{n}}}}{2^{\tau-1}}-4\left(\frac{R}{c_{n}}\right)^{\frac{1}{\gamma_{n}}},
\end{align*}
where the last inequality follows from $4^{-m^{*}}<4\left(\frac{R}{c_{n}}\right)^{\frac{1}{\gamma_{n}}}$
since $R>\overline{C}_{m^{*}+1}$. Let $\frac{L^{\frac{1+\gamma_{n}-\tau}{\gamma_{n}}}\left(\frac{\tau R_{U}}{c_{n}}\right)^{\frac{1}{\gamma_{n}}}}{2^{\tau-1}}-4\left(\frac{R_{U}}{c_{n}}\right)^{\frac{1}{\gamma_{n}}}=L_{C}$,
and we have
\[
R_{U}=L_{C}^{\gamma_{n}}\left(\frac{L^{\frac{1+\gamma_{n}-\tau}{\gamma_{n}}}\left(\frac{\tau}{c_{n}}\right)^{\frac{1}{\gamma_{n}}}}{2^{\tau-1}}-4c_{n}^{-\frac{1}{\gamma_{n}}}\right)^{-\gamma_{n}}.
\]
Clearly, the above $R_{U}$ and $m^{*}=\left\lfloor \frac{1}{\gamma_{n}}\log_{4}\frac{c_{n}}{R_{U}}\right\rfloor ^{+}$
satisfy $\overline{f}_{L}\left(R_{U},m^{*}\right)>L_{C},R_{U}\in\left(\overline{C}_{m^{*}+1},\overline{C}_{m^{*}}\right]$.
Then from Lemma \ref{lem:Mainlemmafbar}, the upper bound $R_{U}$
given in Theorem \ref{thm:thp-bounds} is valid for $\tau\in\left(1,\gamma_{n}+1\right)$.

\subsubsection*{Case 4: $\tau=\gamma_{n}+1$}

Replace $f^{-1}\left(\frac{\overline{C}_{m}}{R}\right)$ in $L_{m^{*}}\left(R\right)$
with the upper bound of $f^{-1}\left(\frac{\overline{C}_{m}}{R}\right)$
for $\tau>1$ in Lemma \ref{lem:finvbounds}, and we obtain an upper
bound of $L_{m^{*}}\left(R\right)$ as
\[
\overline{f}_{U}\left(R,m^{*}\right)=\left(3\log_{4}L+4\right)\left(\frac{\tau R}{c_{n}}\right)^{\frac{1}{\gamma_{n}}}.
\]
Let $\overline{f}_{U}\left(R_{L},m^{*}\right)=L_{C}$, and we have
\[
R_{L}=\left(3\log_{4}L+4\right)^{-\gamma_{n}}\frac{c_{n}}{\tau}L_{C}^{\gamma_{n}}.
\]
Clearly, the above $R_{L}$ and $m^{*}=\left\lfloor \frac{1}{\gamma_{n}}\log_{4}\frac{c_{n}}{R_{L}}\right\rfloor ^{+}$
satisfy $\overline{f}_{U}\left(R_{L},m^{*}\right)\leq L_{C},R_{L}\in\left(\overline{C}_{m^{*}+1},\overline{C}_{m^{*}}\right]$.
Then from Lemma \ref{lem:Mainlemmafbar}, the lower bound $\frac{1}{M\left(1+2^{\tau}\right)}R_{L}$
given in Theorem \ref{thm:thp-bounds} is valid for $\tau=\gamma_{n}+1$.

For the throughput upper bound, consider a scheme which caches the
most popular $L_{C}$ files at the $0$-th level and the remaining
$L-L_{C}$ files at the $1$-th level. Clearly, the throughput achieved
by such a scheme must be larger than $R^{*}$ and is given by
\[
\frac{c_{n}4^{-\gamma_{n}}}{f\left(L-L_{C}+1\right)}\leq\frac{\left(\tau-L^{1-\tau}\right)c_{n}4^{-\gamma_{n}}}{\left(L_{C}+1\right)^{1-\tau}-\left(L+1\right)^{1-\tau}}.
\]

\subsubsection*{Case 5: $\tau>\gamma_{n}+1$}

Replace $f^{-1}\left(\frac{\overline{C}_{m}}{R}\right)$ in $L_{m^{*}}\left(R\right)$
with the upper bound of $f^{-1}\left(\frac{\overline{C}_{m}}{R}\right)$
for $\tau>1$ in Lemma \ref{lem:finvbounds}, and we obtain an upper
bound of $L_{m^{*}}\left(R\right)$ for $R\leq c_{n}L^{\tau-1}$ as
\[
\overline{f}_{U}\left(R,m^{*}\right)=\left(\frac{3\tau^{\frac{1}{\tau-1}}4^{\frac{\gamma_{n}+1-\tau}{\tau-1}}}{1-4^{\frac{\gamma_{n}+1-\tau}{\tau-1}}}+4\tau^{\frac{1}{\gamma_{n}}}\right)\left(\frac{R}{c_{n}}\right)^{\frac{1}{\tau-1}}.
\]
Let $\overline{f}_{U}\left(R_{L},m^{*}\right)=L_{C}$, and we have
\[
R_{L}=c_{n}\left(\frac{3\tau^{\frac{1}{\tau-1}}4^{\frac{\gamma_{n}+1-\tau}{\tau-1}}}{1-4^{\frac{\gamma_{n}+1-\tau}{\tau-1}}}+4\tau^{\frac{1}{\gamma_{n}}}\right)^{1-\tau}L_{C}^{\tau-1}.
\]
Clearly, the above $R_{L}$ and $m^{*}=\left\lfloor \frac{1}{\gamma_{n}}\log_{4}\frac{c_{n}}{R_{L}}\right\rfloor ^{+}$
satisfy $\overline{f}_{U}\left(R_{L},m^{*}\right)\leq L_{C},R_{L}\in\left(\overline{C}_{m^{*}+1},\overline{C}_{m^{*}}\right]$.
Since $R_{L}\leq c_{n}L^{\tau-1}$, from Lemma \ref{lem:Mainlemmafbar},
the lower bound $\frac{1}{M\left(1+2^{\tau}\right)}R_{L}$ given in
Theorem \ref{thm:thp-bounds} is valid for $\tau>\gamma_{n}+1$.

The throughput upper bound is the same as in case 4. This completes
the proof.

\subsection{Proof of Lemma \ref{lem:The-sum-capacity-c}\label{subsec:Proof-of-Lemmagamc}}

Lemma \ref{lem:The-sum-capacity-c} can be proved using similar a
technique to that in the proof of Theorem 5.2 in \cite{Tse_IT07_CapscalingHMIMO}.
With some bounded per node power constraint $P$, the sum capacity
of the MIMO channel between the $\mathcal{S}_{c}$ and $\mathcal{D}_{c}$
is
\begin{align}
\Gamma_{c} & =\max_{\begin{array}{c}
\boldsymbol{Q}\left(\boldsymbol{H}\right)\succeq\boldsymbol{0}\\
\mathbb{E}\left(\boldsymbol{Q}_{j,j}\left(\boldsymbol{H}\right)\right)\leq P,\forall j\in\mathcal{S}_{c}
\end{array}}\mathbb{E}\left(\log\left|\boldsymbol{I}+\boldsymbol{H}\boldsymbol{Q}\left(\boldsymbol{H}\right)\boldsymbol{H}^{H}\right|\right),\label{eq:gammac}
\end{align}
where $\boldsymbol{H}=\left[h_{i,j}\right]_{i\in\mathcal{D}_{c},,j\in\mathcal{S}_{c}}$.
Let $\overline{V}_{c}$ denote the set of nodes inside the square
at the center of the network with area $\left(\sqrt{n_{c}}-2\right)^{2}$,
where $n_{c}=n^{\beta_{1}-\beta_{2}}$, and let $V_{c}=\mathcal{D}_{c}\backslash\overline{V}_{c}$.
By the generalized Hadamard\textquoteright s inequality, we have
\begin{align*}
\log\left|\boldsymbol{I}+\boldsymbol{H}\boldsymbol{Q}\left(\boldsymbol{H}\right)\boldsymbol{H}^{H}\right| & \leq\log\left|\boldsymbol{I}+\boldsymbol{H}^{(1)}\boldsymbol{Q}\left(\boldsymbol{H}\right)\boldsymbol{H}^{(1)H}\right|\\
 & +\log\left|\boldsymbol{I}+\boldsymbol{H}^{(2)}\boldsymbol{Q}\left(\boldsymbol{H}\right)\boldsymbol{H}^{(2)H}\right|,
\end{align*}
where $\boldsymbol{H}^{(1)}=\left[h_{i,j}\right]_{i\in V_{c},,j\in\mathcal{S}_{c}}$
is the channel between the $\mathcal{S}_{c}$ and $V_{c}$, and $\boldsymbol{H}^{(2)}=\left[h_{i,j}\right]_{i\in\overline{V}_{c},,j\in\mathcal{S}_{c}}$
is the channel between the $\mathcal{S}_{c}$ and $\overline{V}_{c}$,
and thus (\ref{eq:gammac}) is bounded above by
\begin{align}
\Gamma_{c} & \leq\max_{\begin{array}{c}
\boldsymbol{Q}\left(\boldsymbol{H}^{(1)}\right)\succeq\boldsymbol{0}\\
\mathbb{E}\left(\boldsymbol{Q}_{j,j}\left(\boldsymbol{H}^{(1)}\right)\right)\leq P,\forall j\in\mathcal{S}_{c}
\end{array}}\mathbb{E}\left(\log\left|\boldsymbol{I}+\boldsymbol{H}^{(1)}\boldsymbol{Q}\left(\boldsymbol{H}^{(1)}\right)\boldsymbol{H}^{(1)H}\right|\right)\nonumber \\
 & +\max_{\begin{array}{c}
\boldsymbol{Q}\left(\boldsymbol{H}^{(2)}\right)\succeq\boldsymbol{0}\\
\mathbb{E}\left(\boldsymbol{Q}_{j,j}\left(\boldsymbol{H}^{(2)}\right)\right)\leq P,\forall j\in\mathcal{S}_{c}
\end{array}}\mathbb{E}\left(\log\left|\boldsymbol{I}+\boldsymbol{H}^{(2)}\boldsymbol{Q}\left(\boldsymbol{H}^{(2)}\right)\boldsymbol{H}^{(2)H}\right|\right).\label{eq:gammac12}
\end{align}
Applying Hadamard\textquoteright s inequality once more, the first
term in (\ref{eq:gammac12}) can be upper-bounded by the sum of the
capacities of the individual MISO channels between nodes in $\mathcal{S}_{c}$
and each node in $V_{c}$. Following a similar analysis to that in
the proof of Theorem 5.2 in \cite{Tse_IT07_CapscalingHMIMO}, the
first term in (\ref{eq:gammac12}) is upper bounded by $K^{'}\sqrt{n_{c}}\left(\log n\right)^{2}$,
where $K^{'}$ is a constant independent of $n$.

To bound the second term in (\ref{eq:gammac12}), we introduce the
concept of the total power received by all the nodes in $\overline{V}_{c}$,
when the nodes in $\mathcal{S}_{c}$ are transmitting independent
signals with power $P$. Specifically, let $P_{j}$ denote the total
received power in $\overline{V}_{c}$ of the signal sent by $j\in\mathcal{S}_{c}$:
$P_{j}=P\sum_{i\in\overline{V}_{c}}r_{i,j}^{-\alpha}.$ Let $P_{tot}\left(n_{c}\right)=\sum_{j\in\mathcal{S}_{c}}P_{j}$
and define $\tilde{\boldsymbol{H}}=\left[h_{i,j}/\sqrt{d_{j}}\right]_{i\in\overline{V}_{c},,j\in\mathcal{S}_{c}}$,
where $d_{j}=\sum_{i\in\overline{V}_{c}}r_{i,j}^{-\alpha}$. Then
the second term is equal to
\begin{align}
 & \max_{\begin{array}{c}
\boldsymbol{Q}\left(\tilde{\boldsymbol{H}}\right)\succeq\boldsymbol{0}\\
\mathbb{E}\left(\tilde{\boldsymbol{Q}}_{j,j}\left(\tilde{\boldsymbol{H}}\right)\right)\leq P_{j},\forall j\in\mathcal{S}_{c}
\end{array}}\mathbb{E}\left(\log\left|\boldsymbol{I}+\tilde{\boldsymbol{H}}\tilde{\boldsymbol{Q}}\left(\tilde{\boldsymbol{H}}\right)\tilde{\boldsymbol{H}}^{H}\right|\right)\nonumber \\
\leq & \max_{\begin{array}{c}
\boldsymbol{Q}\left(\tilde{\boldsymbol{H}}\right)\succeq\boldsymbol{0}\\
\mathbb{E}\left(\textrm{Tr}\left(\tilde{\boldsymbol{Q}}\left(\tilde{\boldsymbol{H}}\right)\right)\right)\leq P_{tot}\left(n_{c}\right)
\end{array}}\mathbb{E}\left(\log\left|\boldsymbol{I}+\tilde{\boldsymbol{H}}\tilde{\boldsymbol{Q}}\left(\tilde{\boldsymbol{H}}\right)\tilde{\boldsymbol{H}}^{H}\right|\right)\nonumber \\
\leq & \max_{\begin{array}{c}
\boldsymbol{Q}\left(\tilde{\boldsymbol{H}}\right)\succeq\boldsymbol{0}\\
\mathbb{E}\left(\textrm{Tr}\left(\tilde{\boldsymbol{Q}}\left(\tilde{\boldsymbol{H}}\right)\right)\right)\leq P_{tot}\left(n_{c}\right)
\end{array}}\mathbb{E}\left(\log\left|\boldsymbol{I}+\tilde{\boldsymbol{H}}\tilde{\boldsymbol{Q}}\left(\tilde{\boldsymbol{H}}\right)\tilde{\boldsymbol{H}}^{H}\right|1_{B_{n,\epsilon}}\right)\nonumber \\
+ & \max_{\begin{array}{c}
\boldsymbol{Q}\left(\tilde{\boldsymbol{H}}\right)\succeq\boldsymbol{0}\\
\mathbb{E}\left(\textrm{Tr}\left(\tilde{\boldsymbol{Q}}\left(\tilde{\boldsymbol{H}}\right)\right)\right)\leq P_{tot}\left(n_{c}\right)
\end{array}}\mathbb{E}\left(\textrm{Tr}\left(\tilde{\boldsymbol{H}}\tilde{\boldsymbol{Q}}\left(\tilde{\boldsymbol{H}}\right)\tilde{\boldsymbol{H}}^{H}\right)1_{B_{n,\epsilon}^{c}}\right),\label{eq:captutbound}
\end{align}
where the set $B_{n,\epsilon}=\left\{ \left\Vert \tilde{\boldsymbol{H}}\right\Vert ^{2}>n^{\epsilon}\right\} $.

For the first term in (\ref{eq:captutbound}), denoted as $C_{B_{n,\epsilon}}$,
using a similar analysis to Equation (11) in \cite{Tse_IT07_CapscalingHMIMO},
it can be shown that 
\begin{equation}
C_{B_{n,\epsilon}}\leq\frac{1}{2}n^{\beta_{1}-\beta_{2}}\log\left(1+\frac{nP_{tot}\left(n_{c}\right)}{\Pr\left(B_{n,\epsilon}\right)}\right)\Pr\left(B_{n,\epsilon}\right).\label{eq:CBnepson}
\end{equation}
Furthermore, following a similar analysis to Lemma 5.3 in \cite{Tse_IT07_CapscalingHMIMO},
it can be shown that for any $\epsilon>0$ and $p\geq1$, there exists
$K_{1}^{'}>0$ such that for all $n$,
\begin{equation}
\Pr\left(B_{n,\epsilon}\right)\leq\frac{K_{1}^{'}}{n^{p}}.\label{eq:PrBn}
\end{equation}
It follows from (\ref{eq:CBnepson}), (\ref{eq:PrBn}), and $P_{tot}\left(n_{c}\right)\leq Pn^{2}$
that 
\begin{equation}
C_{B_{n,\epsilon}}\leq K_{1}^{'}n^{\beta_{1}-\beta_{2}-p}\log\left(1+\frac{n^{3+p}}{K_{1}^{'}}\right),\label{eq:Cbepson1}
\end{equation}
which decays to zero with an arbitrary exponent as $n$ tends to infinity.

For the second term in (\ref{eq:captutbound}), denoted as $C_{B_{n,\epsilon}^{c}}$,
we have
\begin{align}
C_{B_{n,\epsilon}^{c}} & \leq\max_{\begin{array}{c}
\boldsymbol{Q}\left(\tilde{\boldsymbol{H}}\right)\succeq\boldsymbol{0}\\
\mathbb{E}\left(\textrm{Tr}\left(\tilde{\boldsymbol{Q}}\left(\tilde{\boldsymbol{H}}\right)\right)\right)\leq P_{tot}\left(n_{c}\right)
\end{array}}\mathbb{E}\left(\left\Vert \tilde{\boldsymbol{H}}\right\Vert ^{2}\textrm{Tr}\left(\tilde{\boldsymbol{Q}}\left(\tilde{\boldsymbol{H}}\right)\right)1_{B_{n,\epsilon}^{c}}\right)\nonumber \\
 & \leq n^{\epsilon}P_{tot}\left(n_{c}\right).\label{eq:Cbepson2}
\end{align}
Moreover, it can be verified that
\begin{equation}
P_{tot}\left(n_{c}\right)=\begin{cases}
K^{'}n_{c}^{2-\alpha/2}, & \alpha\in\left(2,3\right)\\
K^{'}\sqrt{n_{c}}, & \alpha\geq3
\end{cases}.\label{eq:Ptot}
\end{equation}
Finally, (\ref{eq:Cbepson1}-\ref{eq:Ptot}) complete the Proof of
Lemma \ref{lem:The-sum-capacity-c}.



\begin{thebibliography}{10}
\providecommand{\url}[1]{#1}
\csname url@samestyle\endcsname
\providecommand{\newblock}{\relax}
\providecommand{\bibinfo}[2]{#2}
\providecommand{\BIBentrySTDinterwordspacing}{\spaceskip=0pt\relax}
\providecommand{\BIBentryALTinterwordstretchfactor}{4}
\providecommand{\BIBentryALTinterwordspacing}{\spaceskip=\fontdimen2\font plus
\BIBentryALTinterwordstretchfactor\fontdimen3\font minus
  \fontdimen4\font\relax}
\providecommand{\BIBforeignlanguage}[2]{{%
\expandafter\ifx\csname l@#1\endcsname\relax
\typeout{** WARNING: IEEEtran.bst: No hyphenation pattern has been}%
\typeout{** loaded for the language `#1'. Using the pattern for}%
\typeout{** the default language instead.}%
\else
\language=\csname l@#1\endcsname
\fi
#2}}
\providecommand{\BIBdecl}{\relax}
\BIBdecl

\bibitem{Caire_INFOCOM12_femtocache}
N.~Golrezaei, K.~Shanmugam, A.~Dimakis, A.~Molisch, and G.~Caire,
  ``Femtocaching: Wireless video content delivery through distributed caching
  helpers,'' \emph{in Proc. IEEE INFOCOM}, pp. 1107--1115, 2012.

\bibitem{Niesen_TIT12_FLcaching}
M.~Maddah-Ali and U.~Niesen, ``Fundamental limits of caching,'' \emph{IEEE
  Trans. Info. Theory}, vol.~60, no.~5, pp. 2856--2867, May 2014.

\bibitem{Liu_TWC15arxiv_adhoccaching}
A.~Liu and V.~K.~N. Lau, ``Asymptotic scaling laws of wireless ad hoc network
  with physical layer caching,'' \emph{IEEE Trans. Wireless Commun.}, vol.~15,
  no.~3, pp. 1657--1664, March 2016.

\bibitem{Caire_arxiv13_D2Dcaching}
\BIBentryALTinterwordspacing
M.~Ji, G.~Caire, and A.~Molisch, ``Fundamental limits of distributed caching in
  {D2D} wireless networks,'' 2013. [Online]. Available:
  \url{http://arxiv.org/abs/1304.5856}
\BIBentrySTDinterwordspacing

\bibitem{Caire_arxiv13_d2dcachingtradeoff}
M.~Ji, G.~Caire, and A.~F. Molisch, ``The throughput-outage tradeoff of
  wireless one-hop caching networks,'' \emph{IEEE Trans. Info. Theory},
  vol.~61, no.~12, pp. 6833--6859, Dec 2015.

\bibitem{Altieri_arxiv14_d2dcaching}
A.~Altieri, P.~Piantanida, L.~R. Vega, and C.~G. Galarza, ``On fundamental
  trade-offs of device-to-device communications in large wireless networks,''
  \emph{IEEE Trans. Wireless Commun.}, vol.~14, no.~9, pp. 4958--4971, Sept
  2015.

\bibitem{Jeon_ICC15_D2Dcaching}
S.-W. Jeon, S.-N. Hong, M.~Ji, and G.~Caire, ``Caching in wireless multihop
  device-to-device networks,'' in \emph{Proc. IEEE ICC 2015}, June 2015, pp.
  6732--6737.

\bibitem{GuptaKumar}
P.~Gupta and P.~Kumar, ``The capacity of wireless networks,'' \emph{IEEE Trans.
  Info. Theory}, vol.~46, no.~2, pp. 388--404, Mar 2000.

\bibitem{Xie_liangliang_IT04_network_information_capacity}
X.~Liang-Liang and P.~R. Kumar, ``A network information theory for wireless
  communication: scaling laws and optimal operation,'' \emph{IEEE Trans. Info.
  Theory}, vol.~50, no.~5, pp. 748--767, 2004.

\bibitem{Jovicic_TIT2004_TCahoc}
A.~Jovicic, P.~Viswanath, and S.~Kulkarni, ``Upper bounds to transport capacity
  of wireless networks,'' \emph{IEEE Trans. Info. Theory}, vol.~50, no.~11, pp.
  2555--2565, Nov 2004.

\bibitem{Kumar_TIT06_TCadhoc}
L.-L. Xie and P.~Kumar, ``On the path-loss attenuation regime for positive cost
  and linear scaling of transport capacity in wireless networks,'' \emph{IEEE
  Trans. Info. Theory}, vol.~52, no.~6, pp. 2313--2328, June 2006.

\bibitem{Tse_IT07_CapscalingHMIMO}
A.~Ozgur, O.~Leveque, and D.~Tse, ``Hierarchical cooperation achieves optimal
  capacity scaling in ad hoc networks,'' \emph{IEEE Trans. Info. Theory},
  vol.~53, no.~10, pp. 3549--3572, Oct 2007.

\bibitem{Niesen_TIT09_CSadhoc}
U.~Niesen, P.~Gupta, and D.~Shah, ``On capacity scaling in arbitrary wireless
  networks,'' \emph{IEEE Trans. Info. Theory}, vol.~55, no.~9, pp. 3959--3982,
  Sept 2009.

\bibitem{Niesen_TIT10_CSadhoc}
------, ``The balanced unicast and multicast capacity regions of large wireless
  networks,'' \emph{IEEE Trans. Info. Theory}, vol.~56, no.~5, pp. 2249--2271,
  May 2010.

\bibitem{Franceschetti_TIT09_Maxwellscaling}
M.~Franceschetti, M.~D. Migliore, and P.~Minero, ``The capacity of wireless
  networks: Information-theoretic and physical limits,'' \emph{IEEE Trans.
  Info. Theory}, vol.~55, no.~8, pp. 3413--3424, Aug 2009.

\bibitem{Caire_TIT2015_HMIMOimp}
S.~N. Hong and G.~Caire, ``Beyond scaling laws: On the rate performance of
  dense device-to-device wireless networks,'' \emph{IEEE Trans. Info. Theory},
  vol.~61, no.~9, pp. 4735--4750, Sept 2015.

\bibitem{Gitzenis_TIT13_wirelesscache}
S.~Gitzenis, G.~Paschos, and L.~Tassiulas, ``Asymptotic laws for joint content
  replication and delivery in wireless networks,'' \emph{IEEE Trans. Info.
  Theory}, vol.~59, no.~5, pp. 2760--2776, May 2013.

\bibitem{Yamakami_PDCAT06_Zipflaw}
T.~Yamakami, ``A zipf-like distribution of popularity and hits in the mobile
  web pages with short life time,'' in \emph{Proc. Parallel Distrib. Comput.,
  Appl. Technol., Taipei, Taiwan}, Dec 2006, pp. 240--243.

\bibitem{Caire_GlobalSIP2014_codedcaching}
M.~Ji, A.~M. Tulino, J.~Llorca, and G.~Caire, ``Caching and coded multicasting:
  Multiple groupcast index coding,'' in \emph{proc. 2014 IEEE GlobalSIP}, Dec
  2014, pp. 881--885.

\bibitem{Hachem_ISIT14_Codedcaching}
J.~Hachem, N.~Karamchandani, and S.~Diggavi, ``Multi-level coded caching,'' in
  \emph{Proc. IEEE ISIT}, Jun. 2014.

\bibitem{Niesen_TIT2016_codedcaching}
N.~Karamchandani, U.~Niesen, M.~A. Maddah-Ali, and S.~N. Diggavi,
  ``Hierarchical coded caching,'' \emph{IEEE Trans. Info. Theory}, vol.~62,
  no.~6, pp. 3212--3229, June 2016.

\bibitem{Liu_TSP14_CacheRelay}
A.~Liu and V.~Lau, ``Cache-enabled opportunistic cooperative {MIMO} for video
  streaming in wireless systems,'' \emph{IEEE Trans. Signal Processing},
  vol.~62, no.~2, pp. 390--402, Jan 2014.

\bibitem{Liu_TSP13_CacheIFN}
------, ``Mixed-timescale precoding and cache control in cached {MIMO}
  interference network,'' \emph{IEEE Trans. Signal Processing}, vol.~61,
  no.~24, pp. 6320--6332, Dec 2013.

\bibitem{Wei_TSP15_cacheDoFrelay}
W.~Han, A.~Liu, and V.~Lau, ``Degrees of freedom in cached mimo relay
  networks,'' \emph{IEEE Trans. Signal Processing}, vol.~63, no.~15, pp.
  3986--3997, Aug 2015.

\bibitem{Wei_Globecom15_DoFcachedIFC}
W.~Han, A.~Liu, and V.~K.~N. Lau, ``Improving the degrees of freedom in {MIMO}
  interference network via {PHY} caching,'' in \emph{proc. 2015 IEEE GLOBECOM},
  Dec 2015, pp. 1--6.

\bibitem{naderializadeh2016fundamental}
N.~Naderializadeh, M.~A. Maddah-Ali, and A.~S. Avestimehr, ``Fundamental limits
  of cache-aided interference management,'' \emph{arXiv preprint
  arXiv:1602.04207}, 2016.

\bibitem{Zhang_2016fundamentalCSITFeedback_arxiv}
J.~Zhang and P.~Elia, ``Fundamental limits of cache-aided wireless {BC}:
  Interplay of coded-caching and {CSIT} feedback,'' \emph{arXiv preprint
  arXiv:1511.03961}, 2016.

\bibitem{Liu_ToN16_capscalingCaching}
A.~Liu and V.~K.~N. Lau, ``How much cache is needed to achieve linear capacity
  scaling in backhaul-limited dense wireless networks?'' \emph{IEEE/ACM
  Transactions on Networking}, vol.~PP, no.~99, pp. 1--10, 2016.

\bibitem{Ozgur_TIT2010_HMIMOimp}
A.~Ozgur and O.~Leveque, ``Throughput-delay tradeoff for hierarchical
  cooperation in ad hoc wireless networks,'' \emph{IEEE Trans. Info. Theory},
  vol.~56, no.~3, pp. 1369--1377, March 2010.

\bibitem{Shen_TIT2009_HMIMO}
J.~Ghaderi, L.~L. Xie, and X.~Shen, ``Hierarchical cooperation in ad hoc
  networks: Optimal clustering and achievable throughput,'' \emph{IEEE
  Transactions on Information Theory}, vol.~55, no.~8, pp. 3425--3436, Aug
  2009.

\bibitem{Niesen_IT12_caching}
U.~Niesen, D.~Shah, and G.~W. Wornell, ``Caching in wireless networks,''
  \emph{IEEE Trans. Info. Theory}, vol.~58, no.~10, pp. 6524--6540, Oct 2012.

\bibitem{Chung_TIT12_CapScaling}
S.~H. Lee and S.~Y. Chung, ``Capacity scaling of wireless ad hoc networks:
  Shannon meets maxwell,'' \emph{IEEE Trans. Info. Theory}, vol.~58, no.~3, pp.
  1702--1715, March 2012.

\end{thebibliography}
\end{document}